\documentclass[%
 preprint,
 tightenlines,
 amsmath,amssymb,
 aps,
prb,
]{revtex4-2}

\usepackage{graphicx}
\usepackage{dcolumn}
\usepackage{bm}
\usepackage{mathrsfs}
\usepackage{xfrac}
\usepackage{braket}
\usepackage{hyperref}
\usepackage{comment}
\usepackage{wrapfig}
\usepackage{mathtools}
\usepackage{hhline}

\usepackage{xcolor}
\usepackage{soul}

\DeclareMathAlphabet{\mathpzc}{OT1}{pzc}{m}{it}

\makeatletter
\renewcommand{\p@subsection}{}
\renewcommand{\p@subsubsection}{}
\makeatother

\begin{document}
\setcitestyle{super}

\title{Quantitative infrared nanoscopy:  Probe-cavity eigenmodes and nano-gap polaritons for strongly coupled nanoscale optics}

\author{Benjamin Allen}
\altaffiliation{These authors contributed equally to this work.}
\affiliation{School of Physics and Astronomy, University of Minnesota Twin Cities
}
\author{Brayden Lukaskawcez}
\affiliation{School of Physics and Astronomy, University of Minnesota Twin Cities
}
\author{Alyssa Bragg}
\affiliation{School of Physics and Astronomy, University of Minnesota Twin Cities
}
\author{Nitzan Hirshberg}
\affiliation{School of Physics and Astronomy, University of Minnesota Twin Cities
}
\author{Michael Fogler}
\affiliation{Dept. of Physics, University of California San Diego
}
\author{Stephanie N. Gilbert Corder}
\affiliation{Advanced Light Source, Lawrence Berkeley National Laboratory
}
\author{Hans A. Bechtel}
\affiliation{Advanced Light Source, Lawrence Berkeley National Laboratory
}
\author{Dimitri N. Basov}
\affiliation{Dept. of Physics, Columbia University
}
\author{Alexander S. McLeod}
 \thanks{These authors contributed equally to this work.}
\affiliation{School of Physics and Astronomy, University of Minnesota Twin Cities}
 \email{mcleoda@umn.edu}

\date{\today}

\begin{abstract}
Optical nanoscopies including near-field optical microscopy and spectroscopy circumvent the diffraction limit of conventional optics thanks to the nanoscale light focus emerging at the apex of a sharp irradiated probe.  However, while strong optical coupling between the apex and its dielectric environment affords both enhanced nanoscopic measurement sensitivity and potentially ultra-strong fields within a nano-gap cavity, the conditions for this coupling remain poorly quantified by prevailing analytic models. Here we present a robust formalism of probe-cavity eigenmodes that fully describes how mutual near-field interactions between probe and environment produce a composite response to external fields qualitatively distinct from that of its distinct components.  This \textit{EigenProbe} model identifies the fundamental excitations of realistic optical nanoscopies as nano-gap polaritons, which provide an elegant basis to accurately predict near-field microscopy and spectroscopy experiments especially when probe-sample interactions are non-perturbative.  Through comparison to carefully controlled nanoscopies of polar phonons and molecular vibrations alike, we show how nano-gap polaritons are both realized and utilized for reliable and rapid "inversion" of local optical constants.  This advance involves both qualifying how probe geometry influences its optical response through quantitative calibration, and applying an efficient semi-analytic description of cavity eigenmode scattering at the probe apex.  Our EigenProbe formalism sets the stage for maturing diverse and proliferating optical nanoscopies into precision metrologies of nano-scale optical environments, and guides future use of nano-gap cavities to manipulate local excitations of quantum materials and to achieve strong coupling over photonic emitters.
\end{abstract}

\maketitle

\tableofcontents

\listoffigures

\section{Introduction}

Today’s increasingly diverse methods for nano-optical microscopy, which we broadly label as nanoscopies, provide researchers and metrologists profound tools to reveal materials and optical phenomena at scales previously regarded physically impossible.  Broadly stated, the diffraction limit of light forbids synthesizing an optical focus smaller in dimension than the wavelength of a photon (or a plane wave of the electromagnetic field) from purely propagating electromagnetic fields.  To construct few nanometer-sized (nano-scale) light fields then demands synthesis using non-propagating or “evanescent” photons \cite{novotny_near-field_2006,novotny2012principles}.  These comprise the electromagnetic near-field around and at spatial scales set by distributions of electric charge or current, as might be driven to the sharp apex of a metallic tip.  The possibility to use such tip apex as a “near-field optical probe” was suggested many decades before first realized \cite{novotny_history_2007,pohl_scanning_1988,hillenbrand_complex_2000,Raschke2003}, but has in recent decades become essential to nanoscopies like THz-resolved scanning tunneling microscopy (THz-STM) \cite{cocker_ultrafast_2013}, photo-induced force microscopy (PIFM) \cite{shcherbakov_photo-induced_2025}, scattering-based (or apertureless) scanning near-field optical microscopy (s-SNOM) \cite{hillenbrand_visible--thz_2025,chen_modern_2019,atkin_nano-optical_2012}, nanometer-scale Fourier transform infrared spectroscopy (nano-FTIR) \cite{huth_nano-ftir_2012}, and synchrotron-based infrared nano-spectroscopy (SINS) \cite{BECHTEL2020100493,khatib_far_2018}.  

Unlike conventional microscopies, in these nanoscopies the imaging instrument must be maintained in such close proximity to the target  (via distance-feedback with concurrent scanned probe modalities like atomic force microscopy) as to both {\bf i)} supply a nano-focus of light and {\bf ii)} receive influence from the local response of the environment.  Profoundly, by temporal reciprocity, if such a sharp probe confines near-fields into interaction with an analyte, it can equally deconfine those near-fields that arise in response, and in this process the electromagnetic state (e.g. polarization) of the probe is modified, thus forming an “active microscopy”.  Were this the whole story, detection of scattering from the probe (or of other metrics to report its polarization, like photo-induced force) would already suffice to resolve sub-diffraction optical contrasts that are linear in the environment response.  However, re-polarization of the probe and its subsequent re-illumination of the analyte go hand in hand, forming a self-consistent cycle that both complicates the simple picture of “linear microscopy”, but also underlies exciting opportunities to fundamentally alter landscapes of optical phenomena within the probe environment, including e.g. “strong coupling” to a nanoscale system.  These opportunities broadly intersect with the field of “cavity quantum-electrodynamics”, whereby tailoring optical excitations within an optical cavity (“coupled optical system”) can modify behavior of materials within its volume \cite{schlawin_cavity_2022}.

Active microscopy is then by nature non-perturbative, and fortunately so.  However, a unified theoretical and quantitative understanding of how both effects realistically impact optical nanoscopies is woefully lacking, despite the broad commonalities inherent to most optical nanoscopies (comprising “pointed metal probe over planar surface”).  This void owes to the awkward gulf between simple nanoscale models \cite{keilmann_near-field_2004,ocelic_subwavelength-scale_2004,Raschke2003} and multi-scale electromagnetic simulations \cite{amarie_broadband-infrared_2011,dai_polarization-engineered_2026,chen_scattering_2018,mooshammer_quantifying_2020}, and largely follows from the separation of spatial scales intrinsic to near-fields that mediate interaction and the far-fields that trigger and/or report it.  Treatments of the nanoscopy problem at increasing complexity bridge this gap at varying levels of algebraic or analytic simplification \cite{cvitkovic_analytical_2007,hauer_quasi-analytical_2012} or computational expense \cite{McLeod2014,chen_validity_2021}.  Ideal would be a framework that simultaneously rationalizes active, non-perturbative, and realistic features of optical nanoscopy without sacrificing simplicity and speed of prediction that are so necessary for high-throughput and “inverse” nanoscopy.  Building upon insightful treatment of the optical nanoscopy measurement problem using a quasi-normal basis of “eigenmodes” not unlike those within an optical cavity \cite{Jiang2016}, our work presents a so-called \textit{EigenProbe} formalism that marries experimental realism with an economical treatment of the physics problem underlying a probe-coupled optical response.

Before introducing this new formalism, Sec. \ref{sec:coupledresponsefunction} lays out a general language of response functions that can reliably account for the multifold near- and far-field interactions operative in the full nanoscopy problem.  Secs. \ref{sec:configurationalresonance}-\ref{sec:strongcoupling} review how fundamental features of interest, including probe-enhancement of active microscopy, configurational resonance, and strong coupling generically arise from our formalism, and motivates our search for appropriate parameters that can approach description of realistic experiments.  Secs. \ref{sec:multimodalscattering}-\ref{sec:eigenmodesbasis} describe our multi-modal \textit{EigenProbe} description of probe-sample interaction that is both economical and in principle exact, and explore how its quantitative parameters connect to measurables of optical nanoscopy.   Secs. \ref{sec:confinementandscattering}-\ref{sec:probegapcavitypolariton} investigate how these parameters sensitively follow from configurations of the experiment and geometry of the utilized nano-probe, and lead to appropriately quantitative descriptions of (in particular) probe-mediated electromagnetic scattering as recorded by s-SNOM.  Here we present evidence for predicted nano-gap polaritons when probe-sample coupling is non-perturbative  Sec. \ref{sec:imaging} describes a straightforward generalization of our formalism that quantitatively describes formation of imaging contrast from inhomogeneous media and nonlocal optical excitations, like plasmon polaritons.  Secs. \ref{sec:calibration}-\ref{sec:energydependence} compare predictions by our formalism against experimental energy-resolved nanoscopies of standard materials like gold, silicon, and silicon carbide (SiC).  Furthermore, here we propose that \textit{EigenProbe} provides a facile route to “probe calibration” that quantifies its optical response when well isolated from unknowns of its optical  environment.  This calibration is essential to reliable application of any interpretation model, including our own, and with this tool empower both remote- \cite{amarie_broadband-infrared_2011,huth_nano-ftir_2012} and self-normalized \cite{McLeod2021,chen_near-field_2022,mester_high-fidelity_2022} nano-spectroscopies as quantitative tools for optical metrology of materials.  Sec. \ref{sec:farfieldfactor} applies the \textit{EigenProbe} formalism to rationalize and semi-quantify how undesired far-field effects easily intrude upon remote-normalized nano-spectroscopies, while proposing practical remedies in concurrence with and advancing upon recent work \cite{mester_high-fidelity_2022}.  Sec. \ref{sec:inversion} applies the balanced speed and accuracy of our formalism to the inverse problem of optical metrology, demonstrating faithful extraction of optical constants from polar crystals and polymers both within and without the non-perturbative regime of probe-sample coupling.  Once quantitative confidence in our model is established, Sec. \ref{sec:STO} revisits opportunities supplied by strong coupling of a probe to its environment, proposing a practical experiment with promise to manipulate polar excitations and order in the quantum-paraelectric SrTiO$_3$ \cite{hameed_enhanced_2022,latini_ferroelectric_2021,gastiasoro_superconductivity_2020}.  Finally, Sec. \ref{sec:discussion} discusses some implications of our findings and avenues for future advancement in the practice and interpretation of nano-optical metrology. 

To minimize the burden of formalism within the main text, we reserve detailed mathematical derivations for the Appendices, the most of important of which is Appendix \ref{app:eigenmodeproperties}, which details properties of probe-cavity eigenmodes that are essential to our findings and predictions.   At the same time, our text retains a bare minimum of mathematical detail needed to accurately formulate our EigenProbe solution to the problem of actively coupled nanoscopy.  For concreteness, we apply our theoretical formalism to interpret experimental findings chiefly by s-SNOM, in which light scattered from a sharp metallic nano-probe (as positioned on a micro-cantilever by an atomic force microscope, or AFM), in response to illumination by focused radiation, provides the nano-resolved contrast-forming signal \cite{ocelic_pseudoheterodyne_2006,chen_modern_2019,atkin_nano-optical_2012}.  Fig. \ref{fig1}a sketches the layout and mechanisms of this nanoscopy in a fashion suited to our formalism.  Moreover though, as exemplified in Appendix \ref{app:eigenmodeproperties}, the EigenProbe formalism is equally suited to quantitative description and interpretation of scanned probe nanoscopies ranging from THz-STM to PIFM and beyond \cite{cocker_ultrafast_2013,shcherbakov_photo-induced_2025}.  To enable application and extension of the findings presented here, the authors have made a software implementation of EigenProbe publicly available for use by researchers and metrologists alike.\cite{mcleod_eigenprobe_2026}

\section{Fundamentals of a coupled response function}

\subsection{An emergent response function for active microscopy} \label{sec:coupledresponsefunction}

Scattering-type near-field microscopy is unique among microscopies not only for the novel spatial resolution afforded, but also for its principle of operation, which makes the microscope probe itself an active participant in the optical scattering process of the analyte (or sample / environment; we will utilize these terms interchangeably).  The mandate of such "active microscopy" can be described as follows.  Suppose we have some control over the scattering properties of the ``probe", including for instance its size, shape, optical permittivity, and most importantly, position.  We place this probe into an environment treated as our analyte and thus perturb the scattering properties of the combined system.   Heuristically, our goal in measuring these modified scattering properties is to infer the response function of the original analyte and, in turn, its optical properties, where these are otherwise inaddressible through ``passive microscopies" like \textit{e.g.} optical reflectance that do not invoke active participation of a probe object.  To refine our distinction between ``active" and ``passive" microscopies and to rationalize the former requires establishing a general but formal framework to articulate both the ``known" response function of the isolated probe, the response from the analyte, and the composite response emerging from the interaction.

We will formalize the scenario of active microscopy in the language of a composite response (scattering) function for the probe-analyte system.  Dyadic Green's functions provide a helpful way to describe the complete scattering properties of a system by supplying the total field arising in response to arbitrary sources of free current ``external" to that system.  To be concrete, in linear media described by inhomogeneous permittivity $\varepsilon(\bm{r})$, Maxwell's equations prescribe that electric fields $\bm{E}$ sourced by external ``drive" currents $\bm{j}_\mathrm{ext}$  satisfy the
\begin{eqnarray}
    \text{\textit{Maxwell wave equation:}} &\quad
    \bm{j}_\mathrm{ext}(\bm{r}) & = \hat{\mathcal{M}}\, \bm{E}(\bm{r}), \label{eq:maxwelloperator} \\
    \text{where}&\quad \hat{\mathcal{M}} & \equiv \frac{1}{4\pi i\omega} \left[ \left(\nabla \times \nabla \times \right) -\omega^2 \varepsilon(\bm{r}) \right] \nonumber
\end{eqnarray}

\noindent is often termed the \textit{Maxwell operator}.  Up to a homogeneous solution $\bm{E}_0$ for which $\hat{\mathcal{M}} \,\bm{E}_0=0$, a particular solution to Eq. \ref{eq:maxwelloperator} is formally expressed by $\bm{E}(\bm{r}) = \hat{\mathcal{M}}^{-1}\, \bm{j}_\mathrm{ext}(\bm{r}) \equiv \hat{\mathcal{G}}\, \bm{j}_\mathrm{ext}(\bm{r})$, where $\hat{\mathcal{G}}$ is the Green (dyadic) function associated with $\varepsilon(\bm{r})$ and the formal inverse of the Maxwell operator \cite{novotny2012principles,jackson1999classical}.  A bra-ket notation is also convenient in cases where $\hat{\mathcal{G}}$ can be obtained through expansion in some set of orthonormal basis functions for the electric field, as provided for instance in frameworks like the quasi-normal mode (QNM) or generalized orthogonal mode expansions (GENOME) \cite{lalanne_light_2018,alpeggiani_quasinormal-mode_2017,rao_generalized_2022}:

\begin{align}
    |E) &= \hat{\mathcal{G}} |j_\mathrm{ext}) 
    \quad \text{with} \quad \hat{\mathcal{G}} \equiv \sum_\nu g_\nu(\ldots) |E_\nu)(E_\nu| \label{eq:GreensExpansion} \\
    &=\hat{G} |E_\mathrm{ext})
    \quad \text{with} \quad \hat{G} \equiv \sum_\nu g_\nu(\ldots) |E_\nu)(j_\nu|. \label{eq:Gdef}
\end{align}

\noindent Here $|E_\nu)$ denote members of the relevant (quasi-)orthonormal set, and $g_\nu(\ldots)$ denote associated coefficients depending variously on the light frequency \cite{lalanne_light_2018}, the scatterer permittivity \cite{alpeggiani_quasinormal-mode_2017}, or other ``adjustable" parameters of the scatterer for which the ``quasi-normal modes" $|E_\nu)$ are otherwise a fixture of its configuration.  The inner product is defined by $(E_\nu|j_\mu)\equiv \int dV\, \bm{E}_\nu \cdot \bm{j}_\mu \equiv (j_\mu|E_\nu)$, where $\bm{j}_{\mu}$ is the current sourcing the field $\bm{E}_{\mu}$.  (Notably, since the Maxwell and Green function operators are symmetric but not Hermitian, our parenthetical bra-ket notation denotes a non-Hermitian inner product.) Defining $|E_\mathrm{ext})$ as the field ``sourced" by $|j_\mathrm{ext})$ in free space, Eq. \ref{eq:Gdef} follows from Lorentz reciprocity for modes satisfying outgoing radiation conditions (Appendix \ref{app:reciprocity}), generating a complimentary Green's function $\hat{G}$ that is more convenient to our application.  Whereas $\hat{\mathcal{G}} |j_\mathrm{ext})$ describes the total field, $\left(\hat{G} -1\right)|E_\mathrm{ext})$ generates the scattered field in response to the external one.  We henceforth define a scattering response function $\hat{S}\equiv \hat{G} -1$, where Eqs. \ref{eq:GreensExpansion}-\ref{eq:Gdef} formally connect it to the Green dyadic function.  Field modes are selected in particular to satisfy a useful orthonormality condition $(E_\mu|\hat{O}|E_\nu)=\delta_{\mu\nu}$ with $\hat{O}$ a framework-specific metric operator. For instance, in the QNM expansion $\hat{O}\equiv\omega_\mu\, \varepsilon(\omega_\mu,\bm{r})-\omega_\nu\, \varepsilon(\omega_\nu,\bm{r})$ describes the permittivity-weighted difference in mode frequencies, whereas in GENOME $\hat{O}\equiv \hat{\theta}_{\Omega_S}$ is a projector restricting to the scatterer volume $\Omega_S$.\cite{rao_generalized_2022}  In such a modal basis, $\hat{\mathcal{G}}$ is simply a diagonal matrix acting on the vector of elements $(E_\nu|j_\mathrm{ext})$.  

On the other hand, to describe the response of a composite system poses a challenge even when response functions of its constituents are known.  This is the scenario encountered by an optical nanoscopy experiment, in which the (nominally ``known") response function of a probe interfaces with that (``unknown") of its environment, and the composite response (scattered field) is measured.  The following formalism handles all bookkeeping of this scenario:  Consider an $N$-part system of subsystems indexed by $n$, together specifying $\varepsilon = \sum_n \varepsilon_n$.  We are interested in the composite scattering operator $\hat{S}_+$ defined by $|E_\mathrm{scat})=\sum_n^N |E_n) \equiv \hat{S}_+ |E_\mathrm{ext})$.  For the $N=2$ composite system of a probe interacting with its environment (each described by ``bare" response functions $\hat{S}_{P,E}$, respectively), self-consistency (see Appendix \ref{app:compositeresponsefunction}) supplies the

\begin{equation} \label{eq:gcomposite2}
    \text{\textit{Composite response function:}}\quad
    \hat{S}_+ = \hat{S}_E+\left(\hat{S}_E+1\right) \frac{\hat{S}_P}{1-\hat{S}_P \hat{S}_E} \left(1+\hat{S}_E\right).
\end{equation}

\begin{figure}[tbp]
\includegraphics[width=\columnwidth]{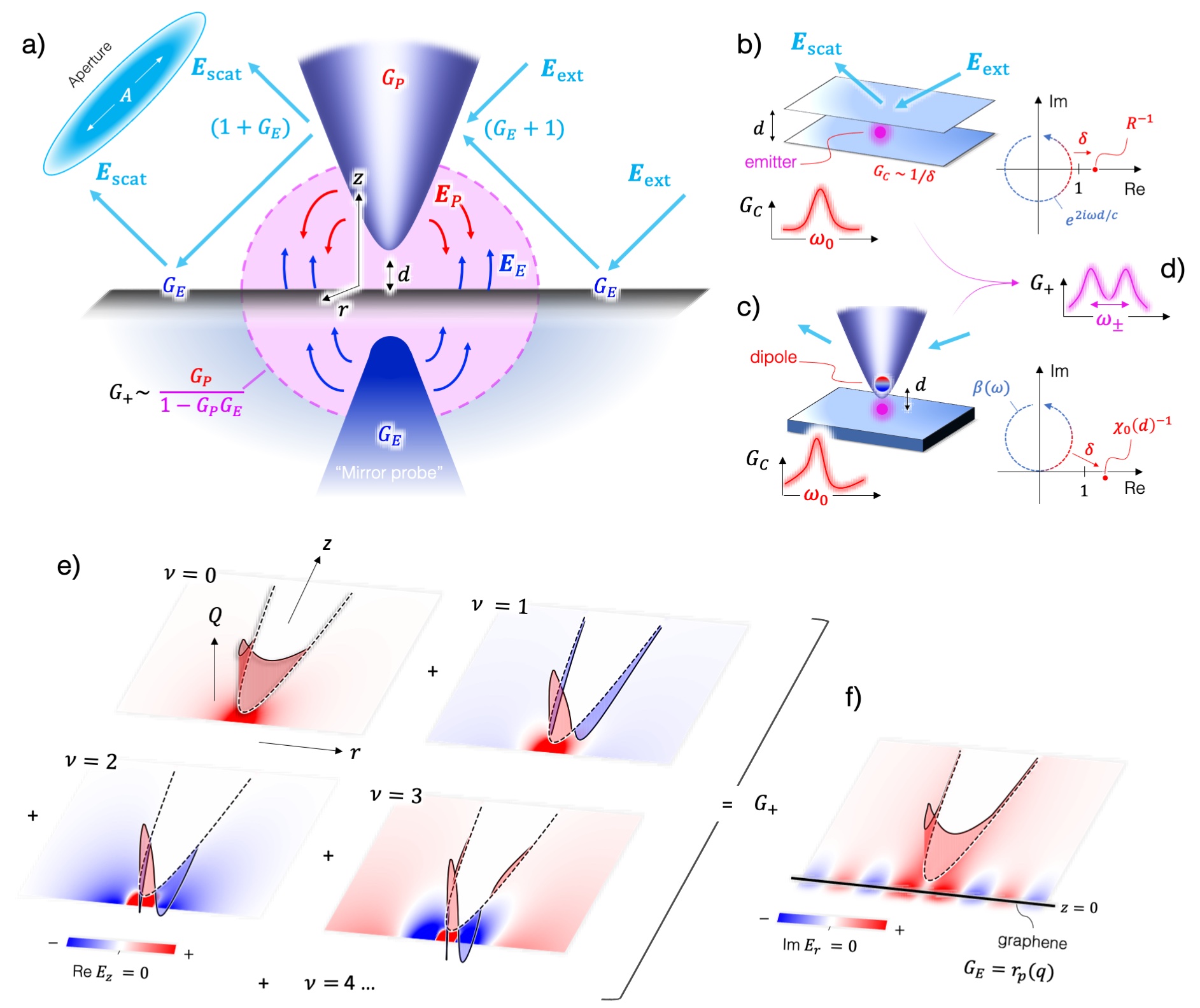}
\caption[Probe nano-gap as an optical cavity.]{\label{fig1}  {\bf Probe nano-gap as an optical cavity.} a) Conceptual scenario of two optical subsystems (“probe” and “environment”) interacting to transform distinct response functions $G_P$ and $G_E$ into a composite response $G_+$ (Eq. \ref{eq:gcomposite2}), which fully describes scattering ($E_\mathrm{scat}$) of an external excitation ($E_\mathrm{ext}$) even when both systems strongly couple.  Symbols are defined in text.  b-c) Optical cavities where $G_+\sim\delta(\omega)^{-1}\equiv G_C$ produces a configurational resonance, and possibility of strong coupling to an emitter is contingent on minimum distance $\delta(\omega_0)$ to a pole in the lower complex plane (inset axes). b) Etalon cavity and c) nano-gap cavity formed between a near-field probe and resonantly reflective surface. d) Emergence of fundamentally new poles $\omega_\pm$ in the composite emitter-cavity response $G_+$ is the hallmark of strong coupling, which underlies emitter control through Purcell-enhanced emission or electromagnetically-induced transparency.  e) Discrete nano-gap polariton modes fully describe the composite response function of the nano-gap cavity and quantify proximity to configurational resonance.  Shown: z-polarized electric field and associated probe surface charge $Q_\nu$ for the four most physically realizable (smallest eigenvalue $\rho_\nu$) polaritons.  f) A superposition of these polaritons describes excitations within the nanogap even over dispersive environments (as described by reflectivity $r_p(q)$) such as e.g. surface plasmon-resonant graphene.}
\end{figure}

\noindent As we will show, this seemingly formal expression in fact provides a practical route to concrete expressions for the composite response function.  Although interchanging labels $P,E$ necessarily yields a mathematically equivalent response function, the present form of Eq. \ref{eq:gcomposite2} brings focus to $\hat{S}_P$, whereby its operator quotient should be regarded as the probe's ``dressed" response function.  In the case of a near-field probe over a semi-infinite half-space environment as sketched in Fig. \ref{fig1}a, $\hat{S}_E$ denotes reflection from the half-space, $\hat{S}_P$ denotes scattering by the probe, and the composite terms signify multiple scattering processes.

In Sec. \ref{sec:farfieldfactor} we will contend with consequences of the leading and trailing factors of $(1+\hat{S}_E)$ for optical nanoscopy experiments.  For now, we focus on the operator quotient in which $\hat{S}_P \hat{S}_E$ ``enhances"  scattering from the region of the environment where characteristic fields from the probe and environment overlap.  We distinguish two regimes of probe-sample coupling: {\bf i)} \textit{active coupling}, in which the amplitude of probe-coupled response exceeds that of the probe alone ($|\hat{S}_P \hat{S}_E|>|\hat{S}_P|$, whose meaning depends on the nature of excitation), and {\bf ii)} \textit{non-perturbative coupling} in which the probe and environment becomes ``inseparable" and a perturbative treatment of Eq. \ref{eq:gcomposite2} will fail ($|\hat{S}_P \hat{S}_E|>1$).  In the latter case, ``circulation" of scattered fields within the probe-sample gap will dominate the response function; we denote this phenomenon a \textit{nano-gap polariton}.  The present work obtains an explicit form for Eq. \ref{eq:gcomposite2} that is both conceptually simple and broadly applicable to quantitative predictions in near-field microscopy.  But first, we recapitulate some general consequences of a composite response function that emerge mutually in two of its simplest realizations -- the Fabry-P\'{e}rot interferometer (or \textit{etalon}), and a polarizable dipole over a dielectric surface.  Whereas the latter (the ``point-dipole model") supplies an elementary understanding of near-field microscopy, its formal similarity to the etalon will demonstrate also how the probe-sample system forms a functional cavity.  Two essential features of bipartite optical cavities generally follow from Eq. \ref{eq:gcomposite2}:  i) \textit{configurational resonance} \cite{keller_configurational_1993} (defined in Sec. \ref{sec:configurationalresonance}) and ii) opportunities for \textit{strong coupling} \cite{simpkins_control_2023,flick_strong_2018} to (and thus control over) optical emitters positioned within the cavity.\cite{park_tip-enhanced_2019}  As we will show, these features are inherited and fully quantified within our realistic treatment of Eq. \ref{eq:gcomposite2} for optical nanoscopy, which affords a definitive description of the probe-sample gap as a functional cavity.  

\subsection{Configurational resonance and formation of an optical cavity} \label{sec:configurationalresonance}
Since Eq. \ref{eq:gcomposite2} is fundamental to the linear response of any pair of interacting systems, we can anticipate resonant phenomena that are generic to pairs of interacting oscillators.  Here we define the \textit{configurational resonance} of a response function as that whose resonant energy $\omega_0$ and oscillator strength $f$ (defined as in Eq. \ref{eq:resonance}) are both modifiable through the geometric configuration of the bipartite system. To see how a composite response function produces this phenomenon, here we apply the ``machinery" of Eq.  \ref{eq:gcomposite2} establish familiar response functions for 1) the simple Fabry-P\'{e}rot cavity (etalon) and for 2) a polarizable point dipole over a dielectric surface, shown in Fig. \ref{fig1}b-c, respectively.

First, the etalon comprises two parallel infinitesimally thin partial reflectors separated by vacuum over a gap distance $d$.  For simplicity we consider only $p$-polarized (with respect to the reflectors) drive and response fields of in-plane momentum $q$ within the gap; generalization to arbitrary field modes is a formal exercise.  Numbering bottom and top reflectors with index $\alpha=1,2$, we construct $\hat{S}_1 \hat{S}_2$ from each:
\begin{equation} \label{eq:gsinglemode}
    \text{\textit{Elementary response function:}} \quad 
    \hat{S}_\alpha=|E_\alpha^q)r_\alpha(j^q_\alpha|.
\end{equation}
\noindent Here the scalars $r_\alpha=r_{1,2}$ denote the respective interior reflectivities, $|E_\alpha^q)$ denote plane waves with in-plane momentum $q$ and longitudinal momenta $\pm k_z=\sqrt{k^2-q^2}$ ``emitted" by (reflected from) the respective reflectors ($k\equiv\omega/c$), and $|j^q_\alpha)$ are modal currents defined by  $(j^q_1|E^q_2)=(j^q_2|E^q_1)=e^{i k_z d}$, which describe retardance of the two modes $|E_{1,2})$ counter-propagating from each surface to the other (we hence suppress the extraneous mode label $q$).  These ingredients are sufficient to observe the product $\hat{S}_1 \hat{S}_2 \hat{S}_1 = r_1 r_2 |E_1)(j_1|E_2)(j_2|E_1)r_1(j_1|$  ``circulates" the two counter-propagating modes, whence \cite{hecht_optics_2017}:
\begin{equation} \label{eq:Airy}
    \text{\textit{Fabry-P\'{e}rot:}}\quad (1-\hat{S}_1 \hat{S}_2)^{-1} \hat{S}_1 = |E_1) r_1 \mathcal{A}(\omega) (j_1|
    \quad \text{where} \quad
    \mathcal{A}(\omega)\equiv \left(1-r_1 r_2 e^{2i k_z d}\right)^{-1}
\end{equation}
\noindent is the familiar Airy distribution.  It is also a formal exercise to apply the totality of $\hat{S}_+$ obtained through Eq. \ref{eq:gcomposite2} to predict response to an external ``drive" $|E_\mathrm{ext})=|E_1)$ launched towards reflector 2 (and not 1), whence we recover the more familiar result $|E)\equiv|E_\mathrm{scat})+|E_\mathrm{ext})=(1+\hat{S}_+)|E_1)=\mathcal{A}(\omega)|E_1)$.  As elaborated in Appendix \ref{app:fabryperotandpointdipole}, $\mathcal{A}(\omega)\propto \delta(\omega)^{-1}$ describes resonance, where $\delta(\omega)$ as sketched in Fig. \ref{fig1}b quantifies the dimensionless ``distance" to the nearest simple pole, and the cavity finesse controls its minimum value.  The resonance frequencies $\omega_\mathrm{FP}$ and oscillator strength $f_\mathrm{FP}$ that describe $\mathcal{A}(\omega)$ are elaborated in Appendix \ref{app:fabryperotandpointdipole} in accord with standard texts \cite{hecht_optics_2017}.  All depend sensitively on the cavity's geometric configuration through both $d$ (owing to spatial structure of the cavity mode) and $r_1 r_2$; thus $\delta(\omega)$ for the etalon evidently describes a configurational resonance that is famously tunable and would be lost were its subsystems ``broken apart".

Second, the polarizable point dipole over a dielectric half-space of (for now, scalar) reflectivity $\beta$ provides a celebrated minimal model (the ``point dipole model") for optical nanoscopy experiments.  Here we consider the dipole with polarizability $\alpha_0$ positioned a height $d$ over the surface with reflectivity $\beta(\omega)$.  Resemblance of the point dipole model to an optical cavity remains underappreciated, and its configurational resonance \cite{keller_configurational_1993} remains poorly quantified, especially in experimentally realistic extensions of the model \cite{McLeod2014,cvitkovic_analytical_2007,hauer_quasi-analytical_2012}.  We can first obtain its familiar composite response function \cite{keilmann_near-field_2004,Raschke2003,ocelic_subwavelength-scale_2004} by taking $\hat{S}_{P,E}$ (Eq. \ref{eq:gcomposite2}) to describe the polarizable dipole and dielectric surface, respectively. As shown in Appendix \ref{app:fabryperotandpointdipole}, manipulations like those leading to Eq. \ref{eq:Airy} show that:
\begin{gather} \label{eq:pdcompositeresponse}
    \text{\textit{Point dipole model:}}\quad
    (1-\hat{S}_P \hat{S}_E)^{-1} \hat{S}_P = |E_\mathrm{pd}) \alpha_\mathrm{pd} (j_\mathrm{pd}| \\
    \text{where} \quad
     \alpha_\mathrm{pd}(\omega)\equiv \frac{\alpha_0}{1-\beta(\omega) \chi_0(2d)} \quad \text{with} \quad \chi_0(2d)\equiv \alpha_0\cdot 2/\left(2d\right)^3. \nonumber
\end{gather}
\noindent As elaborated in Appendix \ref{app:fabryperotandpointdipole}, here $|j_\mathrm{pd})$ is a localized $z$-unit vector, whereby $(j_\mathrm{pd}|E_\mathrm{ext})$ ``samples" the external field at the polarizable dipole.  Meanwhile, $\chi_0$ is a susceptibility not unlike $r_{1,2}$ in the etalon, and encodes the spatial (distance $d$) dependence of the modal field $|E_\mathrm{pd})$ from a $z$-oriented point dipole.  Unlike the etalon, the point-dipole model miniaturizes $d$ to spatial scale much smaller than the free-space light wavelength; thus, without retardation, another source of $\omega$-dispersion is necessary to generate configurational resonance in $\alpha_\mathrm{pd}(\omega)\propto \delta(\omega)^{-1}$.  We can consider the familiar case \cite{keilmann_near-field_2004,Raschke2003} of a substrate with surface (\textit{e.g.} optical phonon, exciton) resonance approximated by a single pole $\beta\approx f_\beta \omega_\beta/(\omega_\beta-\omega-i\gamma)$, where $\omega_\beta$ denotes the surface resonance energy with quality factor $Q_\beta=\omega_\beta/\gamma$.  The inset of Fig. \ref{fig1}c provides graphical representation of evolution in $\beta$ with increasing energy.  As shown in Appendix \ref{app:fabryperotandpointdipole}, the minimum proximity $\delta_0=\chi_0^{-1}-\beta(\omega_\mathrm{pd})$ describes the configurational resonance with emergent frequency and oscillator strength given by:
\begin{equation} \label{eq:pdoscparams}
    \frac{\omega_\mathrm{pd}}{\omega_\beta}=1-f_\beta \chi_0(2d)
    \quad \text{and} \quad
    \frac{f_\mathrm{pd}}{f_\beta} =  \frac{\chi_0(d) \chi_0(2d)}{1-f_\beta \chi_0(2d)}.
\end{equation}
\noindent Here $f_\mathrm{pd}/f_\beta$ measures the dipole-emitted field in the gap $d$ on the dielectric surface; it can vastly exceed unity at $d$ where $\chi_0(d)\sim f_\beta^{-1}$.  Notably, $\alpha_\mathrm{pd}(\omega)$ now obtains a Fano lineshape described by negative Fano parameter $q$:
\begin{equation} \label{eq:pdfano}
    |\alpha_\mathrm{pd}(\omega)|^2 = \frac{\gamma_\beta^2 + (\omega-\omega_\mathrm{pd}+q \gamma_\beta )^2}{\gamma_\beta^2+(\omega-\omega_\mathrm{pd})^2} \alpha_0^2.
    \quad \text{with} \quad
    q = -\chi_0(2d) f_\beta Q_\beta.
\end{equation}
We can trace this behavior to interaction between the ``discrete" $\beta$-resonance and the continuum susceptibility $\chi_0$ of the probe.  As we will show also in later results, this asymmetric Fano lineshape red-shifted relative to $\omega_\beta$ is indeed characteristically observed in near-field optical spectroscopies of \textit{e.g.} surface optical phonons detected by nonresonance probes\cite{hillenbrand_complex_2000,amarie_broadband-infrared_2011,lewin_nanospectroscopy_2018,McLeod2014}.  Since, as we will show, our sophisticated treatment of probe-sample interactions in realizable optical nanoscopy experiments preserves the structure of Eq. \ref{eq:pdcompositeresponse}, the ``optical cavity enhancement" (Eq. \ref{eq:pdoscparams}) and its Fano resonance (Eq. \ref{eq:pdfano}) are both genuine.   This prospect compels us to refine the notion of $\chi_0(d)$ from a single mode towards a quantitative description of all cavity modes conceivably sustained within the gap between a realistic probe and material surface (the ``probe gap cavity").

\subsection{Strong-coupling within a nano-gap optical cavity} \label{sec:strongcoupling}

We have established a configurational resonance generically emerges from the composite response function of adjustibly coupled elements, including the system of an optical nanoscopy probe over a surface.  Notably, a controllable resonance is also a key tool for establishing the celebrated phenomenon of ``strong coupling" \cite{flick_strong_2018,dolfo_damping_2018,simpkins_control_2023} to an additional (tertiary) system, like an optical emitter.   Strong coupling can be regarded as a tool to re-engineer the excited states of a coupled system, with applications spanning enhanced sensing \cite{dolado_remote_2022,bylinkin_-chip_2024}, manipulation of single-photon emitters \cite{maragkou_controlling_2012,park_tip-enhanced_2019}, and even proposed stabilization of unique ground states in quantum materials\cite{latini_ferroelectric_2021,schlawin_cavity_2022}.  Our analysis in Sec. \ref{sec:configurationalresonance} quantified resonance frequency and oscillator strength emerging from configurational cavities, including the gap between a nanoscopy probe and resonant sample.  For brevity, we now denote the bipartite system described by Eq. \ref{eq:gcomposite2} in the former section as the ``optical cavity" and a tertiary system with response function $G_e$ as the ``emitter".

To quantify the necessary ingredients for strong coupling between reconfigurable optical cavity and emitter, we envision iterating Eq. \ref{eq:gcomposite2} once more, now taking the cavity as our ``probe" with composite response denoted by $G_C$, whereas for the ``environment" response $G_E$ we substitute the emitter response $G_e$.  We address the question: When both $G_C$ and $G_e$ are each reasonably described by a single-pole oscillator (Eq. \ref{eq:resonance}; with resonant frequencies $\omega_{C,e}$, oscillator strengths $f_{C,e}$, and linewidths $\gamma_{C,e}$), under what conditions does $(1-\hat{S}_e \hat{S}_C)^{-1}$ generate fundamentally new resonant poles $\omega_\pm$? Strong coupling is defined where the renormalized mode energies differ from the ``bare" ones by substantially more than their net linewidth \cite{dolfo_damping_2018,simpkins_control_2023}:
\begin{gather} \label{eq:strongcouplingcondition}
   (\Delta \omega)^2 > \bar{\gamma}^2 + (\delta \omega)^2 \\
   \text{where}\quad \Delta \omega \equiv \omega_+-\omega_-,
   \quad \bar{\gamma}\equiv (\gamma_C+\gamma_e)/2,
   \quad \text{and} \quad \delta \omega = \omega_C-\omega_e. \nonumber
\end{gather}
\noindent Eq. \ref{eq:gcomposite2} shows us that modes $\omega_{\pm}$  emerge where $1-\hat{S}_e\hat{S}_C$ has vanishing determinant (whereby $\hat{S}_+$ is singular).  Clearly this coincides with the regime where $|\hat{S}_C \hat{S}_e| \gtrsim 1$ and a perturbative Born series in $g_P g_e(\omega)$ fails utterly to describe the dressed response function of cavity and emitter.  To identify $\omega_\pm$ it suffices still to consider a single ``cavity mode" $|E_C)$ scattered by $\hat{S}_C$, and a single \textit{e.g.} electric point dipole field $|E_e)$ scattered by $\hat{S}_e$, whence $\hat{S}_{\alpha=C,e}$ can be expressed as in Eq. \ref{eq:gsinglemode} with $g_\alpha(\omega)$ encoding their respective resonances. Defining the emitter's Lorenz-reciprocal current by $(E_e|j_e)$ allows writing the condition for $\omega_\pm$ as:
\begin{equation}\label{eq:vanishingdeterminant}
(j_e|1-\hat{S}_e\hat{S}_C|E_e) = 1-(j_e|E_C)(j_C|E_e)\cdot g_e(\omega_\pm)g_C(\omega_\pm)= 0.
\end{equation}
\noindent Appendix \ref{app:strongcoupling} shows that Eq. \ref{eq:vanishingdeterminant} is equivalent to identifying normal modes of the cavity-emitter system \cite{dolfo_damping_2018}.  The form factors $\epsilon \equiv (j_e|E_C)=(j_C|E_e)$ equate by reciprocity and evaluate the cavity mode at the emitter location.  With these ingredients, Appendix \ref{app:strongcoupling} (Eq. \ref{eq:criteria2}) prescribes the oscillator strengths $f_C$ and $f_e$ necessary to establish strong coupling between a cavity and emitter, respectively, when brought into mutual resonance at $\omega$:
\begin{equation} \label{eq:strongcouplingsimple}
    \epsilon^2 f_C f_e\, \omega^2 = (\Delta\omega)^2-(\delta\omega)^2 \equiv \Omega^2 \gtrsim \langle \gamma^2 \rangle.
\end{equation}
\noindent Here $\Omega$ denotes the \textit{Rabi frequency} of the coupled system (the rate of energy exchange between Cavity and emitter), $\delta \omega$ is the relative \textit{detuning} of the resonances in isolation, and $\langle \gamma^2 \rangle$ denotes the average squared linewidth of the bipartite system.  Most importantly, as shown in Appendix \ref{app:strongcoupling}, applying this condition to our simplified probe-sample ``optical cavity" predicts that configurational resonance in the oscillator strength $f_\mathrm{pd}$ enhances strong coupling to an emitter.  With the latter placed a distance $d_e$ from the probe, $d$ should be chosen so that:
\begin{equation} \label{eq:pdstrongcouplingmain}
    2 \chi_0(d_e)^2 \cdot \frac{f_\beta \chi_0(2d)}{1-f_\beta \chi_0(2d)}  \left(\frac{\omega_\mathrm{pd}}{\gamma_\beta}\right)^2 > f_e^{-1}.
\end{equation}

\noindent Note that these criteria apply symmetrically to the case where the probe presents resonance in $\chi_0(\omega)$ and the sample showcases non-resonant $\beta$, in which case Eq. \ref{eq:pdstrongcouplingmain} requires simply interchanging $f_\beta$ with an oscillator strength $f_\chi$ and $\chi_0$ with $\beta$.  

When can Eq. \ref{eq:pdstrongcouplingmain} be realistically satisfied?  Considering the probable case of a uniform material surface with $f_\beta \sim 1$ or smaller, the strong-coupling threshold is practically unattainable without sufficiently high probe susceptibility $\chi_0$, which saturates \textit{e.g.} near $1$ for the point dipole system at the minimum physically achievable gap dimension $d=a$ (Appendix \ref{app:fabryperotandpointdipole}).  An actionable answer must proceed beyond this pedagogical but crude point dipole model towards a realistic and quantitative description of near-field optical probes and the probe-cavity.  This task raises important questions:  Is Eq. \ref{eq:pdcompositeresponse} a salvageable description of the composite response that ultimately generates observables of optical nanoscopy experiments?  If so, how should we understand and quantify the probe response $\chi$, and can such knowledge allow inference (``inversion") of the unknown sample response function or its optical constants? \cite{keilmann_near-field_2004,McLeod2014,chen_validity_2021,govyadinov_quantitative_2013,govyadinov_recovery_2014}  How are the conditions of strong coupling to an emitter (Eq. \ref{eq:pdstrongcoupling}) sensitive to experimentally controllable properties of the probe and the gap size $d$?  To answer these questions requires a robust treatment of the non-perturbative regime of the composite response function (Eq. \ref{eq:gcomposite2}) where $|\hat{S}_P \hat{G_E}|\gtrsim 1$ is the precondition for both i) probe-enhanced sensitivity to the analyte and ii) sufficient configurational resonance within the probe-cavity to manipulate a small emitter (Eq. \ref{eq:pdstrongcoupling}).  Motivated by these concerns, the present work generalizes the simple scalars $\chi_0$ and $\beta$ that describe a single ``cavity mode" towards a rigorous ``multi-mode" two-system response function in the non-perturbative regime of optical nanoscopy.  The formalism of ``probe cavity eigenmodes" (or ``eigenprobe" formalism, for short) we now present will reveal added degrees of freedom with enhanced opportunities for probe-sample interaction and manipulation of optical excitations \textit{e.g.} in emitters.

\section{EigenProbe: The probe-cavity eigenmode expansion}
\subsection{From formal response function to multi-modal scattering matrix} \label{sec:multimodalscattering}
Since the point-dipole model is at best a rudimentary treatment of probe-sample interactions, the foregoing discussion supplies at most a qualitative understanding of the composite response function.  We now rigorously apply Eq. \ref{eq:gcomposite2} to the problem of an optical nanoscopy probe interacting with a half-space optical environment.  In this framework the preceding concepts still apply directly, but across a discrete set of ``cavity eigenmodes" variably localized to the probing volume.  For simplicity, we approximate the probe as a perfect electrical conductor (PEC) defined by a closed 3D surface denoted $\partial \Omega_P$.  The response function of the (isolated) probe $\hat{S}_P$ follows from the PEC boundary condition:
\begin{equation} \label{eq:PECboundarycondition}
     \hat{\mathcal{E}}_P |j_P)+\hat{\theta}_{\partial \Omega_P}|E_\mathrm{ext})=0
    \quad \text{where} \quad  \hat{\mathcal{E}}_P \equiv \hat{\theta}_{\partial \Omega_P} \hat{\mathcal{G}}_0
\end{equation}
\noindent Eq. \ref{eq:PECboundarycondition} holds for the isolated probe in presence of an external drive field $|E_\mathrm{ext})$ describing \textit{e.g.} the illumination of the probe by a coherent light source.  Here $|E_P)=\hat{\mathcal{G}}_0|j_P)=\hat{S}_P|E_\mathrm{ext})$ is the counter-field generated by currents $|j_p)$ flowing along and tangent to the probe surface $\partial\Omega_P$, $\hat{\mathcal{G}}_0$ is the free-space Green dyadic function \cite{novotny2012principles}, and $\hat{\theta}_{\partial \Omega_P}$ is a projector that evaluates fields only at points along and tangent to the surface $\partial\Omega_P$.  For later convenience we have defined $\hat{\mathcal{E}}_P$  that generates the probe-scattered field and evaluates strictly along its surface.  As described in Appendix \ref{app:quasistaticsurface}, Eq. \ref{eq:PECboundarycondition} can be solved by assembling matrix elements of $\hat{\mathcal{E}}_P$ and vector elements of $|E_\mathrm{ext})$ in a basis of local current elements along the probe.  This ``method of moments" renders an ``impedance matrix" for the probe and an ``excitation vector", respectively, from which $|j_P)$ is obtained by a linear solver.\cite{gibson_method_2008,virtanen_scipy_2020}

Next, we expand $|j_P)\equiv \sum_\nu \alpha_\nu |j_\nu)$ by a complete basis of current distributions $|j_\nu)$ defined exclusively on the probe surface and chosen to satisfy an orthogonality relation $(j_\nu|E_\mu)\equiv (j_\nu| \hat{\mathcal{G}}_0|j_\mu)\equiv \rho_\nu \delta_{\mu \nu}$ with $\rho_\nu$ a dimensionless normalization factor.  (Here $\delta_{\mu\nu}$ is the Kronecker delta.)  Applying this orthogonality relation to the boundary condition reveals that $\alpha_\nu=-\chi_\nu (j_\nu|E_\mathrm{ext})$, where $\chi_\nu \equiv \rho_\nu^{-1}$ and the inner product is evaluated only on $\partial \Omega_P$ where $|j_\nu)$ has support.  The resulting scattered field supplies the sought response function $\hat{S}_P=-\sum_\nu \chi_\nu |E_\nu)(j_\nu|$ where $|E_\nu)\equiv \hat{\mathcal{G}}_0 |j_\nu)$.  The choice of $|j_\nu)$ is not unique, since eigenfunctions of the operator $\hat{\mathcal{E}}_P$ can be chosen $\hat{\mathcal{E}}_P$-orthonormal, and any orthogonal combination thereof remain $\hat{\mathcal{E}}_P$-orthonormal.  However, the novelty of our approach lies in a unique selection that also nearly diagonalizes the response function of the probe's environment $\hat{S}_E$.  The field from such a set of $|j_\nu)$ might be said to ``circulate" within the probe gap cavity with minimal distortion.  Clearly, these sought ``cavity eigenmodes" will depend manifestly on the probe geometry (\textit{viz.} $\partial \Omega_P$).  Previous work has described these modes as ``eigenoscillations", explored their analytic forms suited to simple probe geometries $\partial \Omega_P$ like the spheroid, and combined them in a ``generalized spectral" description of the probe-sample response function \cite{Jiang2016}.  Next we rearticulate the needed formalism before extending it towards realistic probe geometries.  For the first time, this will allow us to quantitatively compare predictions of this ``eigenprobe" formalism against new experimental results.

As noted in previous work \cite{govyadinov_quantitative_2013,esslinger_reciprocity_2012}, the probe-environment scattering operator $\hat{S}_{PE}\equiv (1-\hat{S}_P \hat{S}_E)^{-1} \hat{S}_P$ in Eq. \ref{eq:gcomposite2} heuristically represents an infinite sequence of scattering events beginning and ending at the probe that may or may not converge in a perturbation series, but we seek a solution operable in the non-perturbative regime.  Our set $|E_\nu)$ spans the image of this operator, which together with the orthogonality relations $(j_\mu|E_\nu)=\rho_\nu(j_\mu|\hat{S}_P|E_\nu)= \rho_\nu \delta_{\mu\nu} $ allows to expand its action on $|E_\mathrm{ext})$ in a combination of eigenmodes as $\hat{S}_{PE} |E_\mathrm{ext}) \equiv \sum_\nu e_\nu |E_\nu)$.  Under excitation $|E_\mathrm{ext})$, the amplitudes $e_\nu$ satisfy:
\begin{gather}
   (j_\mu| \sum_\nu e_\nu (1-\hat{S}_P \hat{S}_E) |E_\nu) = (j_\mu|\hat{S}_P |E_\mathrm{ext}) \nonumber \\
    \rho_\nu\, e_\nu\delta_{\mu\nu}+\sum_\nu (j_\mu|\hat{S}_E|E_\nu) e_\nu = -(j_\mu|E_\mathrm{ext}), \nonumber \\
    \text{thus} \quad \hat{S}_{PE} |E_\mathrm{ext}) = \sum_{\nu} |E_\nu) \left[-\sum_\mu \left(\frac{1}{ \bm{\rho}+\bm{S}_E }\right)_{\hspace{-4pt}\nu\mu} (j_\mu|E_\mathrm{ext})\right]. \label{eq:inversematrix}
\end{gather}

\noindent Here $\bm{\rho}$ denotes a diagonal matrix of all $\rho_\nu$,  $\bm{S}_E$ denotes the scattering matrix with elements $(j_\mu|\hat{S}_E|E_\nu)$, and $e_\nu$ is given within the square brackets.  Eq. \ref{eq:inversematrix} thus describes probe-sample interaction with a multi-modal scattering matrix $(\bm{\rho}+\bm{S}_E)^{-1}$ set in an orthogonal basis set of currents $|j_\nu)$ along the probe.  With our basis tending in size towards infinity, the inverse matrix in Eq. \ref{eq:inversematrix} is well defined provided that $\rho_\nu$ increases sufficiently more rapidly with $\nu$ than the matrix elements $(j_\nu|E_\mathrm{ext})$, and our choice will satisfy this criterion.  Most importantly, a basis in which the scattering matrix is sparse or diagonal will reduce Eq. \ref{eq:inversematrix} to a convergent analytic (Laurent) series that might be truncated to a finite number of terms with little loss of accuracy.  Together with a protocol for realistically accurate calculation of the needed eigenmodes, this prescription comprises the eigenprobe expansion.

\subsection{Probe-cavity eigenmodes as a sparse basis for optical nanoscopy} \label{sec:sparseeigenmodes}

To generate an eigenmode basis where the scattering matrix is sparse would seem to demand specific knowledge not only of the probe response via $\bm \rho$, but also the response function of the environment $\bm g_E$, which seems \textit{a priori} unavailable.  On the other hand, layered environments with in-plane translational and rotational invariance form a class of common analytes for optical nanoscopy inclusive of dielectric half-spaces, 2D materials, and heterolayers of van der Waals materials even with uniaxial anisotropy.  Their optical response  $\hat{S}_E$ is well described by momentum-(and frequency-)resolved Fresnel coefficients $r_{s,p}(\omega,q)$ for incident $s$- and $p-$polarized fields, respectively.  Hereafter we approximate the environment $\hat{S}_E$ by such a surface scattering function $\hat{S}_S$.

Although strongly $q-$dependent (\textit{spatially dispersive} or \textit{nonlocal}) reflectivities signify propagating polariton modes \cite{basov_polaritonic_2025}, reflectivities at $q \gg k\equiv \omega/c$ tend toward a local ($q$-independent) quasi-electrostatic (QS) limit often dubbed $\beta$.  Taking $z=0$ as the surface plane with the most proximal point on the probe at $z=d>0$, in this limit $(j_\mu|\hat{S}_S|E_\nu) \approx (j_\mu|\delta \hat{S}_{S,q}|E_\nu) -\beta(j_\mu|\tilde{E}_\nu^\mathrm{\,QS})$, where the reflected field $|\tilde{E}_\nu^\mathrm{\,QS})=\hat{M}_z |E_\nu^\mathrm{\,QS})$ denotes a version of $|E_\nu^\mathrm{\,QS})$ simply mirrored across the $z=0$ plane (in both its spatial profile and the $z$-component of polarization) by an inversion operator $\hat{M}_z$, $|E_\nu^\mathrm{\,QS})$ is the quasi-electrostatic field generated by the QS part of $\hat{\mathcal{G}}_0|j_\nu)$, and $\delta \hat{S}_{S,q}$ denotes the nonlocal part of the sample surface response.  When the nonlocal response vanishes, $|E_S)\approx -\beta |\tilde{E}_\nu^\mathrm{QS})$ scattered by the sample is that described by the ``method of images" for quasi-electrostatic problems, as generated from a ``mirror probe" as sketched in Fig. \ref{fig1}a. (These definitions realize $|E_\nu)+|E_S)$ with vanishing in-plane polarization along the $z=0$ plane in the case of a quasi-static perfectly metallic surface: $\beta=1$.) Considering that nonlocal reflectance should be in principle ever more weakly excited by fields from nanoscopy probes with ever smaller (sharper apex) dimension, we choose $|j_\nu)$ such that $(j_\nu|\tilde{E}_\mu^\mathrm{\,QS})=\delta_{\nu\mu}$ whereby the matrix elements $\left( \bm{\rho}+\bm{g}_S \right)^{-1}_{\mu\nu}$ tend asymptotically to $\left( \rho_\nu-\beta \right)^{-1} \delta_{\nu\mu}=\chi_\nu/(1-\beta \chi_\nu)\delta_{\nu \mu}$.  Together with the earlier $\hat{\mathcal{E}}_P$-orthogonality, this selection identifies $|j_\nu)$ as solutions to the

\begin{equation} \label{eq:generalizedeigenvalue}
    \text{\textit{Eigenvalue problem:}} \quad 
   \left(\hat{\mathcal{E}}_P - \rho_\nu\, \hat{\mathcal{E}}_S^{\mathrm{QS}} \right) |j_\nu)=0 \quad \text{with} \quad \hat{\mathcal{E}}_S^{\mathrm{QS}} \equiv \hat{\theta}_{\partial \Omega_P} \hat{M}_z \hat{\mathcal{G}}_0^\mathrm{QS}
\end{equation}

\noindent in which $\rho_\nu$ and $|j_\nu)$ are eigen-pairs, $\hat{\mathcal{G}}_0$ is again the free-space Green's function at finite $\omega$, and $\hat{\mathcal{G}}_0^\mathrm{\,QS}$ is its quasi-electrostatic counterpart obtained in the limit $\omega \rightarrow 0$.  Since $\hat{\mathcal{E}}_S^{\mathrm{QS}} \equiv \hat{\theta}_{\partial \Omega_P} \hat{M}_z \hat{\mathcal{G}}_0^\mathrm{\,QS}$ is a symmetric operator, solutions $|j_\nu)$ can be chosen simultaneously $\hat{\mathcal{E}}_P$- and $\hat{\mathcal{E}}_S^{\mathrm{QS}}$-orthogonal with $(j_\mu|\hat{\mathcal{E}}_P|j_\nu)=\rho_\nu (j_\mu|\hat{\mathcal{E}}_S^{\mathrm{QS}}|j_\nu)= \rho_\nu \delta_{\mu\nu}$, where the last equality signifies our choice of normalization, consistent with that in Sec. \ref{sec:multimodalscattering}.  Just as Eq. \ref{eq:PECboundarycondition} is solved by a method of moments after casting $\hat{\mathcal{E}}_P$ as a probe impedance matrix, we similarly cast $\hat{\mathcal{E}}_S^{\mathrm{QS}}$ as a ``surface impedance matrix" (see Appendix \ref{app:quasistaticsurface} for details), whence solutions to Eq. \ref{eq:generalizedeigenvalue} are obtained with any generalized eigenvalue solver.\cite{virtanen_scipy_2020,gibson_method_2008,mcleod_eigenprobe_2026}

This eigenvalue problem describes a scenario where the surface scatters fields $|E_\nu)$ quasistatically with a reflectivity $\rho_\nu$ sufficiently large that re-scattering from the PEC probe will reproduce the same field.   In this hypothetical case $\rho_\nu$ signify the discrete ``eigenreflectivites" whereby fields $|E_\nu^\mathrm{\,tot})=(1+\hat{S}_S) |E_\nu)\approx |E_\nu)-\rho_\nu |\tilde{E}_\nu)$ are self-sustaining in the open cavity of the probe-sample gap and zero along the probe surface; $\hat{\theta}_{\partial \Omega_P}|E_\nu^\mathrm{\,tot})=0$.  Sec. \ref{sec:eigenmodeproperties} addresses the numerical and physical significance of the eigenvalues $\rho_\nu$.  Most critically, these self-sustaining modes convey their resonant character to $\hat{S}_{PE}$ (hereafter denoted $\hat{S}_{PS}$, where $S$ emphasizes a surface-like environment), which as we will show can dominate the probe response, even where nonlocal surface response is important.  A nonlocal reflectivity matrix with $\left[\delta \bm{r}_{S,q}\right]_{\mu\nu} \equiv -(j_\mu|\delta \hat{S}_{S,q}|E_\nu)$ describes the latter case, which could also account for far-field scattering by the sample surface. In this basis Eq. \ref{eq:inversematrix} describes the:
\begin{eqnarray}
    \text{\textit{Probe-sample response:}}&\quad \hat{S}_{PS} &= -\sum_{\mu\nu} |E_\mu) \left[ \frac{1}{\bm{\rho}-\beta(\omega) - \delta \bm{r}_{S,q}} \right]_{\mu\nu} \hspace{-6pt}(j_\nu| \label{eq:PSScattering} \\
    \text{\textit{with local reflectance} $\beta$:}& \quad &\approx -\sum_\nu \frac{|E_{\nu})(j_{\nu}|}{\rho_\nu-\beta(\omega)} 
    = -\sum_\nu |E_{\nu})\frac{\chi_\nu }{1-\beta \chi_\nu}(j_{\nu}|
     \label{eq:PSScatteringlocal} \\ 
   \text{\textit{or as perturbation series:}}&  \quad &\approx \hat{S}_P \sum_{n=0}^\infty \left( \sum_\nu -\chi_\nu |E_\nu)(j_\nu|\, \hat{S}_S \right)^n \\
   & & \approx \hat{S}_P \sum_{n=0}^\infty \sum_\nu \left(\beta \chi_\nu \right)^n.
    \quad \label{eq:PSScatteringPerturbative}
\end{eqnarray}

\noindent These expressions form a central result of our work (and are formally equivalent to Eq. (3) of \textit{Jiang et al.} \cite{Jiang2016}).  We observe that for each eigenmode contribution to Eq. \ref{eq:PSScatteringlocal}, the analogy with the point-dipole model formulation Eq. \ref{eq:pdcompositeresponse} is complete.  In particular, we observe that configurational resonance owing to a single eigenmode is possible wherever $\beta \approx \chi_\nu^{-1}$, with characteristics given identically by Eq. \ref{eq:pdoscparams}.  Because this case describes a surface optical resonance (polariton) dressed by the probe-cavity, we denote such phenomenon a \textit{nano-gap polariton}.  Sec. \ref{sec:probegapcavitypolariton} will compare realistic predictions to an experimental realization of configurational resonance in the nano-gap cavity formed between an optical nanoscopy probe and the phonon-resonant surface of SiC.

Regarding $\hat{S}_S$ as an integral operator, Eq. \ref{eq:PSScatteringPerturbative} presents a conventional Born series expansion sometimes considered a model for optical nanoscopy experiments\cite{esslinger_reciprocity_2012,govyadinov_quantitative_2013} whose criterion for convergence is evidently $|\hat{S}_S|\approx |\beta| \le \mathrm{min} |\rho_\nu|$, where $|\hat{S}_S|$ is measured by its largest eigenvalue in the eigenmode basis.  Bounded rigorously by the smallest probe-cavity mode eigenvalue of the probe-sample geometry, this condition delineates the regime of ``passive" (perturbative) probe-sample coupling from that where probe and analyte become indivisible participants in the response function.  Eq. \ref{eq:PSScatteringlocal} further highlights the manifest dependence of eigenpairs on the probe-sample separation distance $d$ even with otherwise fixed probe geometry.  Finally, we close this section by addressing whether the eigenmode expansion provided by Eqs. \ref{eq:generalizedeigenvalue} \& \ref{eq:PSScattering} is \textit{ad hoc} or generalizable.  Notably, our particular (hereafter dubbed \textit{EigenProbe}) expansion is in fact generated from the ``closest"  scattering response to the family of $\hat{S}_S$ that might be anticipated from analytes of optical nanoscopy experiments.  What does such description leave out?  We note that perturbative corrections to Eq. \ref{eq:PSScatteringlocal} from \textit{e.g.} far-field interactions between probe and sample scale as $\left| (j_\nu|\delta \hat{S}_{S,
\mathrm{FF}}|E_\nu)/(\rho_\nu-\beta) \right| \ll 1$, where $\delta \hat{S}_{S,
\mathrm{FF}}$ arises from the ``far-field part" of $\hat{\mathcal{G}}_0$.  The inequality is ensured by the fact that, as we later discuss, $|\rho_\nu-\beta|$ is at least as large as the eigenvalue's imaginary part, which arises from $|(j_\nu|\delta \hat{S}_{P,
\mathrm{FF}}|E_\nu)|>|(j_\nu|\delta \hat{S}_{S,
\mathrm{FF}}|E_\nu)|$.  Meanwhile, the \textit{EigenProbe} solution may converge more slowly (\textit{viz.} may demand more terms $\nu$ in Eq. \ref{eq:PSScattering}) for increasingly nonlocal $\hat{S}_S$ for which emission and reflection of the eigenmode field $\hat{\theta}_{\partial \Omega_P}\, \hat{S}_S \hat{\mathcal{G}}_0|j_\nu) \not\propto \hat{\mathcal{E}_S^\mathrm{QS}}|j_\nu)$. In such cases, the surface response is ``distant" from the quasi-electrostatic operator $\hat{\mathcal{E}}_S^\mathrm{QS}$, whence an alternative eigenfield expansion analogous to Eq. \ref{eq:generalizedeigenvalue} could be better suited to those applications of active microscopy.

While we defer details to Sec. \ref{sec:realisticcalculations}, Fig. \ref{fig1}e presents exemplary (four lowest order $\nu$) eigenmodes calculated by Eq. \ref{eq:generalizedeigenvalue} for a realistic conical probe geometry  that is later revisited in Fig. \ref{fig:3}.  Here the probe gap $d$ is chosen equal to the apex radius of curvature $a$.  Surface charge density $Q_\nu$ along the probe is shown along with the $z$-component of the associated nano-gap polariton field that a superposition of that sourced by $Q_\nu$ as well as that reflected from a surface at $z=0$ with reflectivity $\beta=\rho_\nu$.  The total field satisfies the boundary condition of Eq. \ref{eq:PECboundarycondition} even without an external drive field, substantiating these polaritons as self-sustaining modes, albeit demanding a particular (as we will see, nonphysical) choice of $\beta$. Appropriate composition of these eigenmodes will solve any particular scattering problem like that shown in Fig. \ref{fig1}f, where an environment with strongly nonlocal reflectivity ($\hat{S}_E \sim r_p(q)$) sustains propagating surface plasmons.  Here, the resulting localized surface plasmon excitation is itself a particular superposition of probe-cavity eigenmodes described precisely by Eq. \ref{eq:inversematrix}.  It remains to be shown how such nonlocal surface scattering is practically evaluated in the basis of eigenmodes; Sec. \ref{sec:eigenmodeproperties} next addresses this question.

\begin{figure}[tbp]
\centering
\includegraphics[width=0.8 \columnwidth]{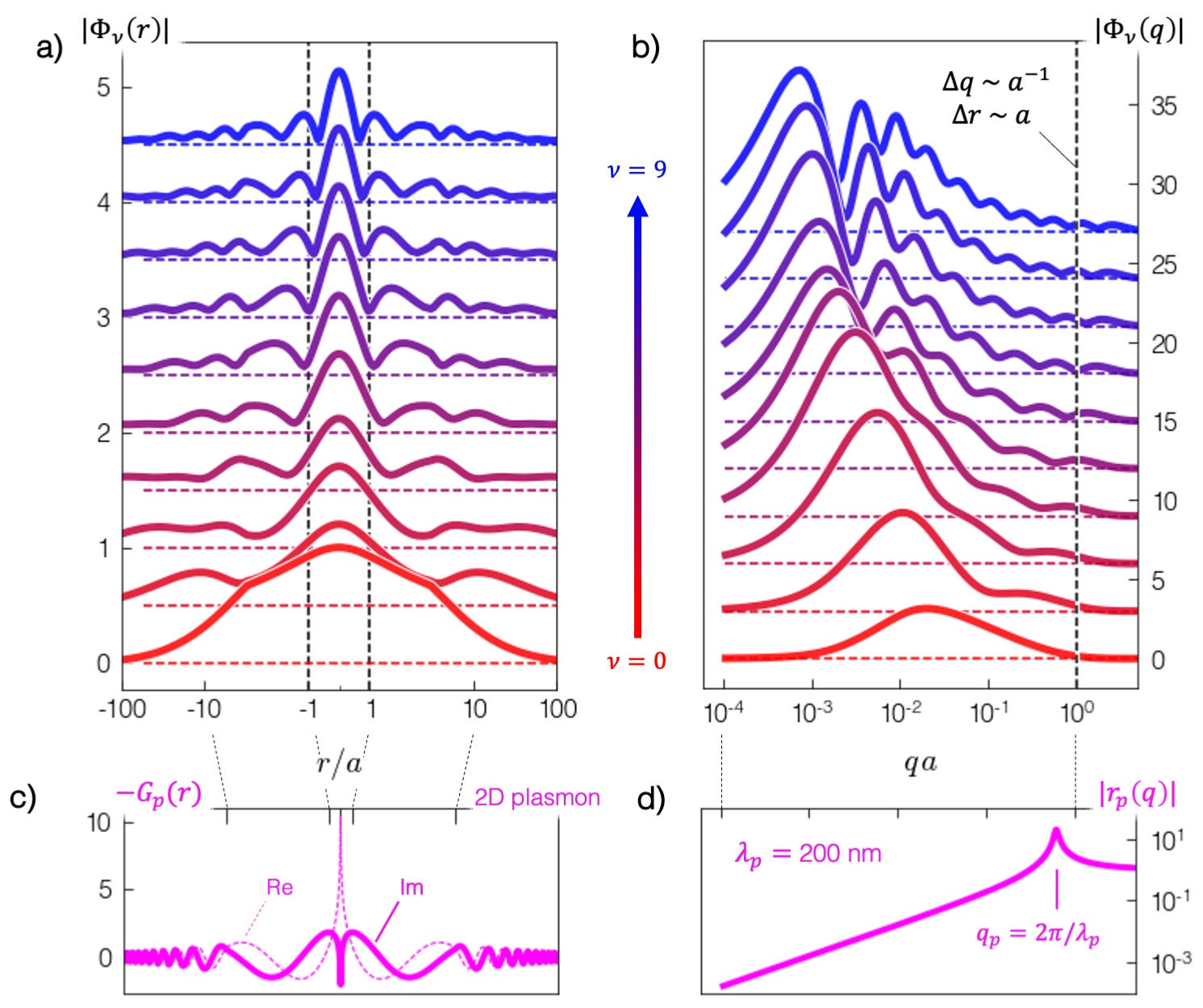}
\caption[Eigenmodes in real- and momentum-space.]{\label{fig:2} {\bf Eigenmodes in real- and momentum-space.} a) Cross-sectional view of the scalar potential $\Phi_\nu$ associated with the first ten eigenmodes (colored red-to-blue while incrementing $\nu=0-9$) of the conical probe pictured in Fig. 3b when positioned one tip radius ($a$) above the material half-space.  b) Hankel transform (from $\Phi_{\nu}(r)$ to) $\Phi_{\nu}(q)$. In a-b), curves are offset for clarity (by 0.5 and 3, respectively); dashed curves indicate $\Phi_\nu=0$. c) Real-space cross-sectional view of the plasmonic Green function for graphene with plasmon wavelength $\lambda_p=200$ nm and a quality factor $Q=100$; position axis is shared with a) (but in expanded view) depicting the case of a “sharp” probe radius $a=20$ nm.  d) The momentum space Green function $(1-q_p/q)^{-1}$ shows overlap with higher order eigenmodes that is key to describing the modal reflectivity of graphene (Eq. \ref{eq:genreflectanceq}). }
\end{figure}

\subsection{Properties of the probe-cavity eigenmodes} \label{sec:eigenmodeproperties}

Although the \textit{probe-cavity eigenmodes} described here have been previously explored in the context of quasistatic probe-sample interactions \cite{Jiang2016}, our treatment shows they remain well defined and can be obtained from Eq. \ref{eq:generalizedeigenvalue} even when $\hat{S}_P$ is treated without approximation in the electrodynamic context, with rich implications for far-field scattering by a near-field probe.  Before exploring the implications of specific eigenmodes $|j_\nu)$ corresponding to realistic probe geometries and their potential to accurately describe optical nanoscopy experiments, we summarize some formal properties required for their physical interpretation.  The following properties are derived in Appendix \ref{app:eigenmodeproperties}.

First, since a local reflectance is often a poor description of the environmental response $\hat{S}_E$ around a nanoscopy probe, it is essential to generalize the surface reflectivity factor $\beta$ appearing in Eq. \ref{eq:PSScattering} to a full \textit{modal reflectivity} matrix $\bm \beta$ described by elements $\beta_{\mu\nu}\equiv-(j_\mu|\hat{S}_S \hat{\mathcal{G}}_0|j_\nu)$.  In terms of this modal reflectivity, the inverse matrix shown (in square brackets) in Eq. \ref{eq:PSScattering} can be written as $\left(\bm{\rho}-\bm{\beta}\right)^{-1}$. Elements of the modal reflectivity are remarkably given by integrals over the electromagnetic potential generated by $|j_\nu)$ in a plane infinitesimally above the surface at $z=0^+$.  Among the explicit expressions obtained for $\bm \beta$, the following two within the QS-approximation are most useful:
\begin{align}
    \text{\textit{Multi-modal reflectance:}} \quad
    \beta_{\mu\nu} & \approx \frac{i\omega}{4\pi} \int_{z=0^+} \hspace{-6pt} dA\,\left( \Phi_\mu(\bm{r}) \partial_z  \tilde{\Phi}_\nu (\bm{r})- \partial_z \Phi_\mu(\bm{r}) \tilde{\Phi}_\nu(\bm{r}) \right) \label{eq:genreflectancer} \\
    & \approx i\omega \int_{0}^\infty dq \,q^2\, \cdot r_p(q) \, \Phi_\mu(q)\, \Phi_\nu(q) \label{eq:genreflectanceq} \\
    & = \beta\, \delta_{\mu\nu} \quad \text{when} \quad r_p(q)=\beta. \label{eq:genreflectanceorthonorm}
\end{align}
\noindent  Above, $\Phi_{\mu,\nu}$ are the scalar potentials emitted by eigenmodes $|j_{\mu,\nu})$, respectively, and $\tilde{\Phi}_\nu$ is the latter potential following its reflection from the nearby surface ($=-\beta \hat{M}_z \Phi$, \textit{e.g.} after reflection by a local reflectivity $\beta$).  As addressed in Appendix \ref{app:eigenmodeproperties}, the overall factor of $i\omega$ follows only from the specific normalization condition for $|j_\nu)$ introduced in Sec. \ref{sec:sparseeigenmodes}, and is non-universal.  These expressions can be visualized with aid of the representative eigenfields shown in Fig. \ref{fig:2}.  Here $\Phi_\nu(\bm r)$ denotes the scalar potential emitted from eigencurrent $|j_\nu)$ evaluated in the $z=0$ plane, and $\Phi_\nu(q)$ denotes its Hankel transform.  Eq. \ref{eq:genreflectanceq} describes the momentum decomposition of each eigenfield $\nu$ and its scattering via Fresnel coefficient $r_p(q)$ into alternate modes $\mu$. Eq. \ref{eq:genreflectancer} evaluates the spatial overlap between $\Phi_\mu$ and surface-reflected $\tilde{\Phi}_\nu \neq \Phi_\nu$ scalar potentials evaluated for $\bm r$ in the $z=0$ plane, where $\partial_z$ evaluates the associated surface-normal electric field.  This form is convenient for real-space descriptions of probe-sample interactions, essential to predictions introduced in Sec. \ref{sec:imaging}.  A form of Eq. \ref{eq:genreflectancer} has been used (with $\mu=\nu$ describing a single dipolar field mode) in former work to predict real-space optical nanoscopy contrasts with the point-dipole model \cite{jing_terahertz_2021,jing_phase-resolved_2023,rizzo_nanometer-scale_2022,xu_deep_2021}, in which case $\beta(\bm r)$ was shown equal to the photonic density of states at $z=d$ over a heterogeneous surface.  More generally, Eq. \ref{eq:genreflectancer} composites this photonic density of states over the available probe-cavity eigenmodes for a probe positioned at $d$.  Meanwhile, Eq. \ref{eq:genreflectanceorthonorm} follows from the $\hat{\mathcal{E}}_S^\mathrm{QS}$-orthonormality of $|j_\nu)$, recovering a scalar (diagonal) description of scattering when the surface reflectivity is local.  Hereafter, for simplicity, we adopt an approximately \textit{axisymmetric description} of the probe-sample interaction and the surface modal reflectivity, whence the probe and all $|j_\nu)$ are considered dependent only on the radial and axial coordinates $r$ and $z$, respectively.

Second, the spatial confinement of an eigenmode $|E_\nu)\equiv \hat{\mathcal{G}}_0|j_\nu)$ is well quantified by the ratio of i) interaction energy between probe and surface when $\beta=1$ and a single eigenmode is excited, and ii) the electrostatic energy of the eigenmode when the probe is brought into isolation.  This ratio is given by $1/\mathrm{Re}\,\rho_\nu$, whence increasing $\mathrm{Re}\,\rho_\nu|$ signifies a decreasing proportion of the quasi-static fields emitted by the ``mirror probe" (pursuant to the electrostatic method of images) is capable of interacting with the physical probe.  This scenario would demand a proportionally increasing hypothetical surface reflectivity $\beta\sim \rho_\nu$ to sustain the associated nano-gap polariton.  Conversely, when reducing the probe-sample gap $d$ we expect $\mathrm{Re}\,\rho_\nu(d)$ to reduce in proportion to increasing confinement of $d$-dependent eigenfields.  Intuitively, a pointed metallic probe with apex brought sufficiently close to $d=0$ is locally indistinguishable from a uniform mirror of unit reflectance, whence $\lim_{d\rightarrow 0}\chi_\nu(d)=1$ and Eq. \ref{eq:PSScatteringlocal} (akin to Eq. \ref{eq:pdcompositeresponse}) is indistinguishable from a multi-mode etalon (without retardance).  This limit is demonstrated in \textit{Jiang et al.} \cite{Jiang2016} and consistent with realistic calculations introduced in Sec. \ref{sec:realisticcalculations}.

Third, the integrated radiant flux from an eigenmode $|j_\nu)$ excited on the probe when isolated in free space (disregarding for now scattering by $\hat{S}_E$) is proportional to $-\mathrm{Im}\rho_\nu>0$.  Consequently, energy outflow by radiation limits the achievable amplitude ($\sim |\rho_\nu-\beta|^{-1}$) of nano-gap polaritons excited by a probe in the same way as absorption ($\sim \mathrm{Im}\,\beta)$ by the local surface.  This finding confirms a proposal proffered in \textit{Jiang et al.}\cite{Jiang2016} that the poles of Eq. \ref{eq:PSScatteringlocal} are addressable in real optical nanoscopy experiments only to the extent that $|\rho_\nu-\beta|$ is rendered small; thus configurational resonance is ultimately hindered by cumulative losses of the probe-sample system proportional to $\mathrm{Im}(\beta-\rho_\nu)$.  The consequential inclusion of  realistic loss in our \textit{EigenProbe} expansion (\textit{viz.} realistic probe-cavity eigenmodes $|j_\nu)$) ensures  the predicted response function Eq. \ref{eq:PSScatteringlocal} remains unconditionally finite.  We later demonstrate and discuss the minimum achievable value of $\mathrm{Im}(\beta(\omega)-\rho_\nu(z))$, variable in both frequency $\omega$ and probe-sample gap separation $d$.  A key physical realization is the case of near-field spectroscopy over polar crystals where $\beta(\omega) \equiv (\varepsilon(\omega)-1)/(\varepsilon(\omega)+1)$ is dominated by optical phonons at mid-infrared energies. After presenting $\rho_\nu(d)$ for a realistic probe geometry in Sec. \ref{sec:realisticcalculations}, Sec. \ref{sec:probegapcavitypolariton} compares this predicted regime with nano-gap polaritons experimentally observed over the surface of bulk silicon carbide which supports a strong surface optical phonon resonance.  Having established that any eigenmode can only be excited finitely, its lifetime is thus limited before its energy is radiated by the probe.  As shown in Appendix \ref{app:eigenmodeproperties}, the ratio of eigenmode energy to its integrated radiant flux provides its lifetime as $\tau_{\nu,\mathrm{rad}} = -2\pi/\omega \cdot \mathrm{Re}\,\rho_\nu/\mathrm{Im}\,\rho_\nu$ which, as Sec. \ref{sec:realisticcalculations} will bear out, grows shorter for more delocalized (greater $\nu$) modes.

Fourth, since the probe response function $\hat{S}_P$ combines its geometric structure with the free-space Green dyadic function, it encodes explicit $\omega$-dependence and, for slender geometries, even pronounced antenna resonances, such as have been observed in purpose-built nanoscopy experiments \cite{huth_resonant_2013,mooshammer_quantifying_2020}.  When the probe is isolated ($\hat{S}_E=0$), Eq. \ref{eq:PSScattering} describes the bare probe response, whence $\hat{S}_P=\sum_\nu |E_\nu) g_\nu(\omega)(j_\nu|$ with $g_\nu(\omega)=-\rho_\nu(\omega)^{-1}$.  Since for any independent component $g_\nu(\omega)$ describes the spectral density of a causal response, $\rho_\nu(\omega)^{-1}$ is necessarily a Kramers-Kronig-compatible function, which in principle allows recovery of \textit{e.g.} $\mathrm{Im}\,\rho_\nu(\omega)$ from knowledge of $\mathrm{Re}\,\rho_\nu(\omega)$.\cite{kuzmenko_kramerskronig_2005}  Stated conversely, antenna-enhanced radiance from eigenmodes at particular $\omega$ that is encoded on $\mathrm{Im}\,\rho_\nu(\omega)$ will necessarily influence both the confinement and excitation capacity of the corresponding nano-gap polariton through its connection to $\mathrm{Re}\,\rho_\nu(\omega)$.  Sec. \ref{sec:realisticcalculations} will show this phenomenon can be highly pronounced for slenderized probe geometries, and Sec. \ref{sec:STO} proposes how it might be harnessed to mediate strong-coupling between an antenna-enhanced resonant nano-gap polariton and the surface of a quantum material.

Lastly, the \textit{EigenProbe} expansion predicts both excitation of an optical nanoscopy probe and capacity for detection of probe-scattered fields.  Both are necessary processes for most s-SNOM experiments.  We consider an experimental geometry relevant to numerous s-SNOM experiments,\cite{chen_modern_2019,atkin_nano-optical_2012,hillenbrand_visible--thz_2025} where a single aperture $A$ (shown schematically in \ref{fig1}a) both delivers excitation $|E_\mathrm{ext})$ and recovers back-scattered radiation $|E_P)\sim \hat{S}_{PS}|E_\mathrm{ext})$ to a detector.  As shown in Appendix \ref{app:reciprocity}, the amplitude of probe-radiated electric fields recovered at a point on the aperture is proportional to the matrix element $(j_P|E_\mathrm{PW})$, where $|E_\mathrm{PW})$ denotes a plane wave field propagating from an aperture point at polar angle $\theta$ to the probe.  When the probe is likewise illuminated by such a plane wave, and the surface reflectivity is local, the far-field scattered by the probe is described by the
\begin{equation} \label{eq:scatteredamplitude}
    \text{\textit{Scattering amplitude:}} \quad 
    E_\mathrm{scat}(\theta) \propto -\sum_\nu \frac{B_\nu(\theta)^2}{\rho_\nu-\beta}
    \quad \text{with} \quad B_\nu(\theta) \equiv (j_\nu|E_\mathrm{PW,\theta}).
\end{equation}

\noindent Here for brevity we have used Eq. \ref{eq:PSScatteringlocal}, and $\theta$ labels the polar angle of the aperture on a unit sphere.  (We suppress azimuthal dependence by again assuming cylindrical symmetry about the $z$-axis.)  We define by $B_\nu(\theta)$ the relative ``angular brightness" of $|j_\nu)$ whose orientation dependence uniquely defines the emitted radiation pattern (see Appendix \ref{app:reciprocity}).  The special case $\beta=0$ reduces to a description of $\hat{S}_P$ alone, which cannot depend on the probe-sample gap $d$ chosen for the eigenvalue problem in Eq. \ref{eq:generalizedeigenvalue}.  Hence, as previously noted \cite{Jiang2016} $\sum_\nu B_\nu(\theta)/\rho_\nu$ should not depend on $d$, which implies that obligatory partial summation on $\nu$ in Eq. \ref{eq:scatteredamplitude} also requires subtracting the $\beta$-independent part from $E_\mathrm{scat}(\theta)$ in order to correctly preserve $d$-independence in the limit where $\beta=0$.  The result of this subtraction we will denote the ``sample-mediated" scattering amplitude $\delta E_\mathrm{scat}$.

Notably, Eq. \ref{eq:scatteredamplitude} well describes a practical near-field microscopy experiment utilizing an illumination/collection aperture $A$ only when $B_\nu(\theta)$ is replaced by $\int_A d\Omega B_\nu(\theta)$ with $d\Omega \propto \sin\theta d\theta$, which emulates excitation by a (\textit{e.g.} Gaussian) focused beam of radiation.  Even when the aperture $A$ encompasses the entire sphere of solid angles, the integrated far-field brightness (\textit{viz.} amplitude) is bounded by $\int_A d\Omega (B_\nu(\theta))^2 \le \int_A d\Omega |B_\nu(\theta)|^2$.  The latter is proportional to the integrated radiant flux described by $-\mathrm{Im} \rho_\nu$, as elaborated in Appendix \ref{app:eigenmodeproperties}.  Thus the numerator of Eq. \ref{eq:scatteredamplitude} might be approximately replaced by $-\mathrm{Im} \rho_\nu$ only when explicitly computed eigenmodes $|j_\nu)$ are unavailable.  However, since typical s-SNOM experiments utilize a restricted numerical aperture ($\lesssim 0.6$) for probe illumination, our forthcoming predictions dispense with this approximation and instead integrate $B_\nu(\theta)$ across an (axisymmetric) aperture spanning $45^\circ \le \theta \le 90^\circ$ relative to the $z$-axis.  This approach thus approximates the excitation field by a Gaussian focus (in $z$) comprised by a superposition of cylindrical Bessel beams $E_z\sim J_0(q r) e^{i k_z z}$ defined by cylindrical ($q$) and longitudinal ($k_z$) momenta corresponding to $\tan \theta_q = k_z/q$ with $k_z=\sqrt{\omega^2/c^2-q^2}$.  While not essential, such field profile is likewise economical to quantify ``detection" of the integrated far-field amplitude within the same aperture.  Focusing primarily on s-SNOM optical nanoscopy, we find that our forthcoming predictions are only weakly effected by these choices of $B_\nu$, which affect all eigenmodes in similar (while not identical) fashion.  For brevity, henceforth we will denote integrated brightness $\int_A d\Omega B_\nu(\theta)$ by simply $B_\nu$. 

\section{Quantitative scattering from realistic probe geometries} \label{sec:realisticcalculations}

\subsection{Eigenmodes as a basis for complex near-field interactions} \label{sec:eigenmodesbasis}

To motivate our further exploration of probe-cavity eigenfields as a physical concept and calculation tool, we pause to demonstrate application of this framework to obtain a composite response function relevant to a realistic probe-sample coupled system.  Consider the case of an optical nanoscopy probe interacting with a quasi-infinite conductive sheet endowed with a metallic (\textit{e.g.} plasma, lattice vibrational, excitonic \textit{etc.}) response that supports propagating polaritons as its fundamental optical excitations \cite{koppens_graphene_2011}.  Like other van der Waals materials, few-layer graphene is a rich target of near-field microscopy investigation whose surface plasmon polariton response falls into this category.  The 2D optical response function (\textit{viz.} $\hat{S}_S$) of such media follows from the Fresnel coefficient $r_p(q,\omega)=q/(q-q_p(\omega))$ describing TM modes at momenta $q \approx \mathrm{Re}\,q_p$, where $q_p=i\omega/2\pi \sigma_{2D}$ is the complex-valued plasmon momentum prescribed by the ($\omega$-dependent) sheet conductivity $\sigma_{2D}$ \cite{jing_terahertz_2021,jing_phase-resolved_2023}.  It is straightforward to construct this Fresnel function with a characteristic plasmon wavelength of $\mathrm{Re} (2\pi/q_p)\approx 200\,\mathrm{nm}\approx 10 a$, where $a=20$ nm will describe the apex curvature radius of the near-field probe (\textit{cf.} Fig. \ref{fig1}a) -- a regime matching seminal plasmonic studies of monolayer graphene \cite{Fei2012,chen_optical_2012,rizzo_charge-transfer_2020} . 

Now we discuss parameters of the near-field probe.  For simplicity, in this work we consider exclusively PEC probes with axisymmetric (about the $z$-axis) geometry and response functions $\hat{S}_P$, although solutions to Eq. \ref{eq:generalizedeigenvalue} by our method of moments can also accommodate describing a non-axisymmetric response. Conical probes are characterized by an opening half-angle $\theta$ (identified in Fig. 1b) which we presently take as $\theta=10^o$.  To obtain probe-cavity eigenfields, we treat Eq. \ref{eq:generalizedeigenvalue} as an electric field integral equation (EFIE) and compute the eigenpairs $\rho_\nu$ and $|j_\nu)$ for a prescribed probe geometry $\partial \Omega_P$ and probe-sample distance $z$. This computation first applies a Galerkin method to evaluate self-impedance and quasistatic mirror-impedance matrices corresponding to $\hat{\mathcal{E}}_P$ and $\hat{\mathcal{E}}_S^{\mathrm{QS}}$, respectively, then solves their associated generalized matrix eigenvalue problem.  The probe-scattered fields $|E_\nu)=\hat{\mathcal{G}}_0 |j_\nu)$ are then computed through convolution with the free-space Green's dyadic \cite{novotny2012principles,jackson1999classical}.  For $d=a$, Fig. \ref{fig:2}a presents these eigenfields (namely the $z$-polarized component of $|E_\nu^\mathrm{tot})=|E_\nu)-\rho_\nu |\tilde{E}_\nu^{\mathrm{QS}})$) in false color corresponding to the four eigenvalues $\rho_\nu$ of lowest magnitude (\textit{viz.} highest field confinement).  The charge density associated with $\nu$ ($q\propto \hat{n} \cdot \bm{E}_P$ along $\partial \Omega_P$) is plotted in vertical projection along the probe surface with a symmetric-log scale (to suppress the presentation of singularities).  We observe the eigenfields abide the PEC boundary condition expressed by Eq. \ref{eq:generalizedeigenvalue}, and also that the order $\nu$ of the eigenfield characterizes the number of ``nodes" in the probe surface charge density.  This feature is inherited in turn by the eigenfield itself, and is a natural consequence of the orthogonality among eigencurrents, as previously reported \cite{Jiang2016}.  These eigenfield nodes are accompanied by field deconfinement as $|\rho_\nu|$ grows exponentially with $\nu$.

Standard eigenvalue solvers allow us to obtain up to $N=50$ of these modes (not shown) that showcase increasingly intricate spatial character, affording a capable basis for real-space fields involved in the probe-sample interactions.  Evidencing this utility, we apply Eq. \ref{eq:PSScattering} to compute the field scattered by the probe-sample system following ``standard" illumination by a plane wave at mid-infrared energy (free-space wavelength $\lambda_\mathrm{IR}\equiv 2\pi c/\omega\approx 10$ microns) at oblique incidence ($45^o$ to the surface-normal).  Since dispersion is indispensable to the graphene response, we discard scalar reflectance $\beta$ and explicitly compute matrix elements of the modal reflectivity (Eq. \ref{eq:genreflectanceq}) in momentum space up to $\mu,\nu=10$.  This Sommerfeld integral is readily evaluated by numerical quadrature, provided that poles of $r_p(q)$ (\textit{e.g.} presently, the plasmon polariton) are adequately sampled.  As described in Appendix \ref{app:eigenmodeproperties}, when $r_p\approx \beta$, nonzero values of Eq. \ref{eq:genreflectanceq} are $\beta$ only when $\mu=\nu$ owing to the orthonormality property imposed among eigencurrents.  Fig. \ref{fig:2}a-b help reveal this orthogonality among eigenfields from Fig. \ref{fig1}e, visualized in both real-space and momentum space, respectively.  Proliferating nodes in the real- and momentum-space eigenfield distributions sustain their orthogonality.  Fig. \ref{fig:2}b shows the broad momentum content of eigenmodes peaked below the inverse gap size $q~a^{-1}$.  That peak momenta of $\Phi_{\nu}(q)$ shift lower with increasing $\nu$ indicates the increasing deconfinement of higher order modes.  On the other hand, to localize a gap field at scales below $a$ requires marshaling a superposition of probe-cavity eigenmodes up to higher order $\nu$, akin to a Nyquist criterion.

For comparison, Fig. \ref{fig:2}c-d display the Fresnel coefficient $r_p(q)$ of our plasmonic membrane as well as its inverse Hankel transform, the 2D plasmonic Green's function $G_p(r)$.  The latter displays a logarithmic singularity near $r=0$ responsible for the screening behavior of a metallic membrane.  Meanwhile, outwards-propagating plasma oscillations comprise the distant fields where $G_p(r)\sim H^{(1)}_0(q_p r)$, with $H^{(1)}_0$ an outward Hankel function. The generalized reflection coefficients can be viewed as momentum integrals between pairs of traces in Fig. \ref{fig:2}b and the Fresnel coefficient (Fig. \ref{fig:2}d), or alternatively (\textit{cf.} Eq. \ref{eq:genreflectancer}) as spatial integrals over pairs of $\Phi_\mu(r)$ (Fig. \ref{fig:2}a) with the reflected field supplied by the plasmonic Green's function (Fig. \ref{fig:2}c) as $\tilde{\Phi}_\nu(r) \equiv G_p(r) * \Phi_\nu(r)$.  Here $*$ denotes spatial convolution in the $z=0$ plane.   Finally, Fig. \ref{fig1}f presents the resulting scattered field so computed with Eq. \ref{eq:PSScattering}, which simultaneously satisfies the PEC boundary condition of the probe together with that of the graphene sheet, leading to propagating plasmon polaritons that are especially evident in (shown) $\text{Im}\,E_r$.  This example demonstrates that probe-cavity eigenfields provide a recipe to construct the composite response function for arbitrary probe-analyte combinations.  The \textit{EigenProbe} framework non-perturbatively admixes the probe's response function with that of its environment to faithfully describe the composite response of ``active" optical nanoscopy.

\subsection{Field confinement and scattering efficiency of realistic probe-cavity eigenmodes} \label{sec:confinementandscattering}

In contrast to previous studies of eigenfields for simple probe geometries in the quasistatic limit \cite{Jiang2016}, the demonstration in Fig. \ref{fig1}e-f leverages an eigenfield expansion computed for a conically shaped (precisely, hyperboloidal) near-field optical probe of mesoscopic length comparable to the free-space wavelength of mid-infrared light ($L=20\, \mu\mathrm{m} \sim \lambda_\mathrm{IR}$), characteristic of widely used nanoscopy probes for s-SNOM.\cite{chen_modern_2019,atkin_nano-optical_2012,hillenbrand_visible--thz_2025}  These physical dimensions relative to the wavelength of external excitation places such probes unambiguously in the electrodynamic regime of active optical microscopy.  We emphasize that although probe-sample interactions may be treated reasonably with the quasistatic approximation, there should be no expense spared in accurate dynamic treatment of the probe response function, since $\hat{S}_P$ provides the only solid ``anchor" for interpretation of the composite response function and inferences about $\hat{S}_S$.  Our treatment satisfies this criterion.  Previous work \cite{McLeod2014} considered the implications of electrodynamics for naturally attenuating the severity of long-range electrostatic interactions encountered in the response of large ($L \ge \lambda_\mathrm{IR}$) quasistatic probes \cite{Jiang2016}.  However, any precise implications for the emergent response function that owe to the probe size and overall shape remain underexplored except by large-scale simulations. \cite{amarie_broadband-infrared_2011,huber_optical_2014,mooshammer_quantifying_2020}  Here we address these implications concretely with the \textit{EigenProbe} formalism.

\begin{figure}[b]
\includegraphics[width=\columnwidth]{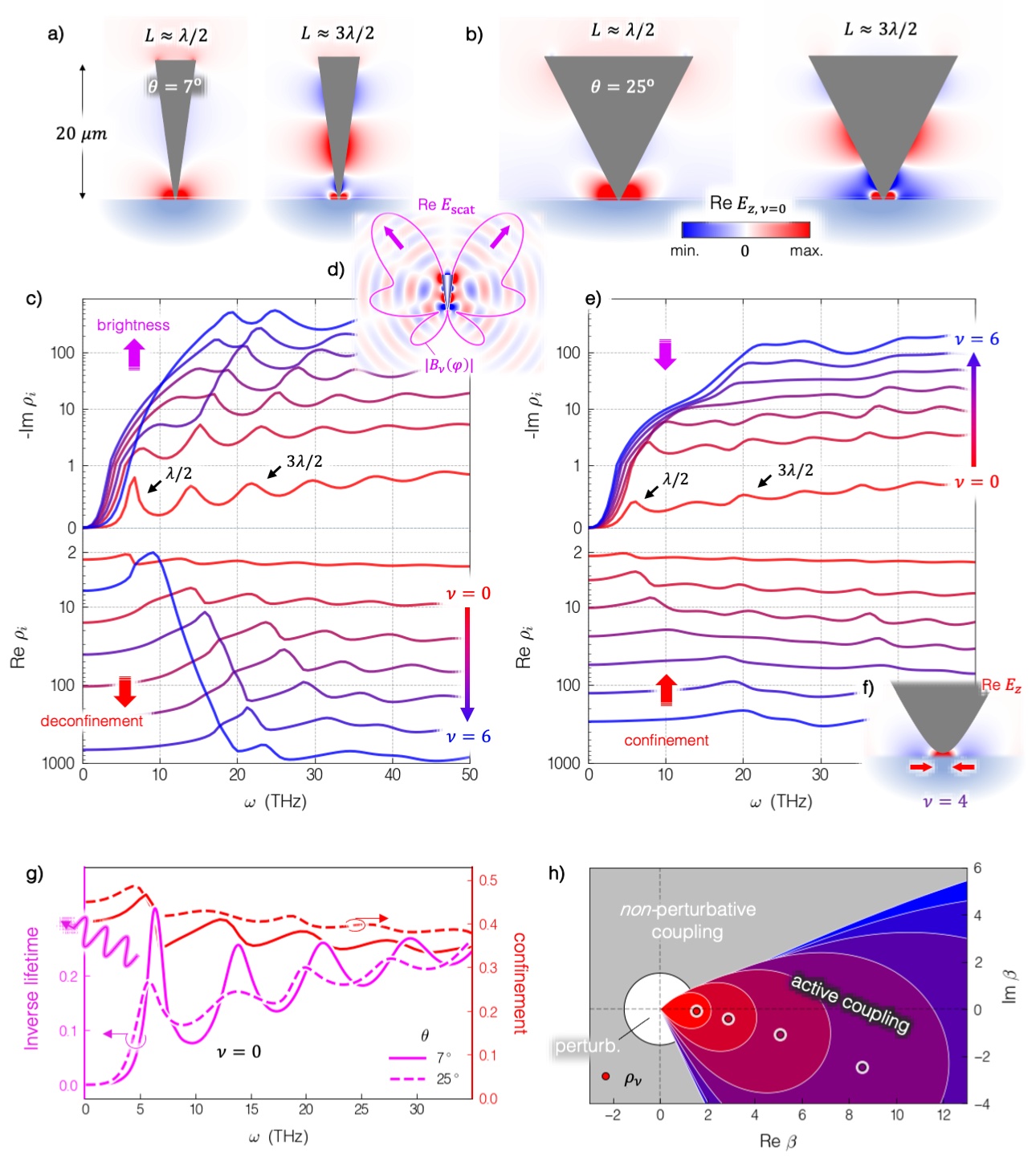}
\end{figure}
\clearpage
\begin{figure}
\caption[Energy dependence of probe-cavity eigenmodes.]{\label{fig:3} {\bf Energy dependence of probe-cavity eigenmodes.}  a) Slender and b) widely tapered probe geometries considered in panels c-d) and in e-f), respectively.  Shown is the z-polarized electric field of the lowest order eigenfield for varied probe lengths $L$ relative to the free-space light wavelength $\lambda=2\pi c/\omega$, revealing that eigenmodes show standing wave character along the probe at energies of “antenna resonance”.  c) The imaginary and real parts of the modal eigenvalues quantify their brightness and deconfinement, respectively.  Both increase exponentially with $\nu$ (coded increasing red-to-blue) and modulate with energy across  antenna resonances.  d) Polar angle ($\phi$)-resolved brightness $B_\nu$ (pink curve) quantifies the field amplitude radiated into $\phi$ by each eigenmode; $\nu=0$ is shown at $\omega=30$ THz.  The real part of the $\phi$-polarized field $E_\mathrm{scat}$ is shown with symmetric negative-positive (blue-red) color scale, revealing “antenna lobes” of radiation. e) Same quantities as panel (c) but for the wide probe. f) Cross-sectional view of an eigenmode at the probe apex. g) Comparison of inverse lifetime (see text) and confinement for the first eigenmode between the two probe geometries; lifetime depends strongly on $\omega$ owing to narrow antenna resonances for the slender probe, whereas broadband brightness of the wide probe lends uniformity in $\omega$. h) In the complex plane of surface reflectance $\beta$, the non-perturbative regime onsets where $|\beta/\rho_\nu|>1$, determined by minimum eigenvalue $\rho_{\nu=0}$.  The regime of active coupling is where the probe response is numerically enhanced through coupling to $\beta$; colored zones identify which eigenvalue (marked by dots) dominates the coupling.}
\end{figure}

Fig. \ref{fig:3}a-b present the lowest order eigenfields computed at 2 distinct light frequencies corresponding to $\lambda=2L$ and $2L/3$ for two probe geometries: a) a slender needle-like probe with $\theta=7^o$, and b) a stout conical probe at $\theta=25^o$, both with $L=20\, \mu\mathrm{m}$ and an apex curvature radius of $a=20$ nm matching the probe-sample distance ($z=a$).  Notably, these eigenmodes are constructed strictly for near-field interaction with the nearby surface.  Thus their far-field reflection from the surface is excluded here but in principal follows according to the surface $\theta$-resolved reflectivity via leading and trailing factors of $(1+\hat{S}_E)$ appearing in Eq. \ref{eq:gcomposite2}; Sec. \ref{sec:farfieldfactor} addresses the practical impact of these ``far-field factors" on nanoscopy experiments.  It is evident for both probes that eigenfields globally oscillate at the corresponding free-space wavelength, and that our selections of $\omega$ are compatible with an antenna resonance condition $L/\lambda=(\frac{1}{2}+n)$ for integer $n$.  Fig. \ref{fig:3}c\&e present the $\omega$-dependent eigenvalues from $\nu=0$ (lowest order, by convention) up to $\nu=6$ (increasing for curves from red to blue) computed for these geometries, illustrating two fundamental concepts.  As described in Sec. \ref{sec:eigenmodeproperties}, $-\mathrm{Im}\,\rho_\nu$ reports the radiant flux from scattered eigenfields, provides an upper bound on the eigenmode brightness, and correspondingly estimates its contribution to the $\omega$-resolved probe scattering cross section.  The slender probe exhibits sharp peaks in $-\mathrm{Im}\,\rho_\nu$ precisely at the antenna resonance conditions that have been experimentally observed \cite{huth_resonant_2013} (arrows mark the energies associated with Fig. \ref{fig:3}a-b).  Fig. \ref{fig:3}d presents the intermediate- and far-eigenfield associated with one such resonance compared with the mode's polar angle($\theta$)-resolved brightness $|B_\nu(\theta)|$ (pink curve) in polar format, showing distinct antenna-enhanced radiation lobes directed favorably above the analyte.  These owes to the conical shape of this antenna and likely favors execution of optical nanoscopies that leverage illumination/detection from above the sample plane, as formerly predicted.\cite{amarie_broadband-infrared_2011}  Meanwhile, for the stout probe these resonances are attenuated at least three-fold, owing presumably to destructive interference from surface currents across the large probe radius ($R\sim \lambda$).  On the other hand, $\mathrm{Re}\rho_\nu$ gauges the eigenfield energy density and its deconfinement as indicated schematically by Fig. \ref{fig:3}f.  These metrics are both manifestly lower by a similar factor as compared to the slender probe.  The higher order eigenvalues ($\nu \ge 1$) essentially replicate these behaviors at much larger scale, with the exception of a notable blue-shift in their resonance frequencies as compared with the lower orders.  Although high-order eigenmodes also satisfy the self-consistency condition of Eq. $\ref{eq:generalizedeigenvalue}$, the quasi-electrostatic approximation leveraged may become increasingly inaccurate for these deconfined modes.  Therefore we surmise that physical excitation of high-order eigenmodes, where possible, may showcase quantitatively different $\omega$-dependence than suggested by Fig. \ref{fig:3}

As shown by Fig. \ref{fig:3}g, the deconfinement for both probe geometries  exhibits a dispersive lineshape near the antenna resonance frequencies that emerges due to the Kramers-Kronig compatibility of susceptibilities $\rho_\nu(\omega)^{-1}$ already discussed in Sec. \ref{sec:eigenmodeproperties}.  We can draw a clear conclusion about the impacts of global probe geometry:  a stout probe geometry (large $\theta$) will help to confine fields near the sample surface, and by virtue of Eq. \ref{eq:PSScatteringlocal} will afford higher scattering ``contrast" (\textit{viz.} relative sensitivity) to analytes of lower reflectivity (\textit{viz.} $\beta$).  On the other hand, the scattering efficiency of such widely tapered non-resonant probes can be several factors lower than that afforded by a slender probe shape excited close to resonance.  (Note the trade-off: enhanced scattering efficiency of slender probes excited near resonance comes at the expense of reduced scattering contrast, that is \textit{e.g.} comparatively more linear in $\beta$, for analytes of all but the highest reflectivities.)  Fig. \ref{fig:3}h presents the exemplary coupling regimes associated with eigenmodes of the stout probe at a close probe-sample separation of $z/a=0.1$ ($\omega=30$ THz). Values of local surface reflectivity are distinguished across the complex-$\beta$ plane based on whether they trigger non-perturbative (gray) or active (colored) probe-sample coupling, where both are simultaneously possible, and simultaneously necessary for conceivable strong-coupling to an emitter within the nano-gap (\textit{cf.} Eq. \ref{eq:pdstrongcouplingmain} with $\chi_0 \le \rho_{\nu=0}^{-1}$).  Lossy materials occupy $\mathrm{Im}\,\beta(\omega)>0$ where nano-gap polaritons can be only finitely excited (\textit{cf.} Eq. \ref{eq:PSScatteringlocal}).  Owing to the minimum achievable values of $\rho_\nu\sim 1$, the range of $\beta$ is quite small for which a perturbative description (Eq. \ref{eq:PSScatteringPerturbative}) is adequate.  Moreover, colored points identify the location of eigenvalues $\rho_\nu$ and each centers over a zone of active coupling.  Namely, the color of each zone indicates which eigenmode is most quantitatively responsible (via its excited eigenmode amplitude) for the coupled probe-sample response. Clearly $\beta$ values with ever increasing ``quality factor" $\mathrm{Re}\,\beta/\mathrm{Im}\,\beta$ (ever decreasing relative absorption) are necessary to significantly excite eigenmodes of increasing deconfinement ($\mathrm{Re}\,\rho_\nu$) and ``brightness" ($-\mathrm{Im}\,\rho_\nu$).  Consequently, material media falling into this regime of active coupling require an increasingly large number of eigenmodes to adequately describe the resulting composite probe-sample response function (Eq. \ref{eq:gcomposite2}).  

We conclude this section by remarking on the complex behavior of $\rho_\nu$ for $\omega L/c \ll 1$.  Sufficiently low frequencies show $-\mathrm{Im}\, \rho_\nu \propto \omega^4$ characteristic of Raleigh scattering of a ``small" probe, whereas $\mathrm{Re}\, \rho_\nu$ tends to a quasistatic limit dictated exclusively by probe geometry \cite{Jiang2016}.  At such low energies, the scattering efficiency of a probe is so comparatively reduced as to render $\hat{S}_{PE}$ a potentially ineffective apparatus for active microscopy, at least when excitation and detection both still occur in the far-field.  At frequencies approaching and exceeding the first antenna resonance of the slender probe, curves describing $\mathrm{Re} \rho_\nu$ evolve rapidly -- this behavior is \textit{bona fide} within our framework, and physical implications may demand further scrutiny.  It is worth noting for both probes that $\mathrm{Re}\,\rho_{\nu}$ is minimized over frequency just below the first antenna resonance, signifying the energy of highest possible scattering contrast towards analytes with low-reflectivity.  Such $\omega$-dependent scattering contrast (by way of field deconfinement) is the underlying origin for imprecise spectroscopic ``normalization" in this energy range, as alluded in ref. \cite{McLeod2014}.  On the other hand, the attenuated antenna resonances and comparitive $\omega$-independent scattering efficiency (especially for $\omega$ above the first) should endow large-$\theta$ probes with more reliable broadband performance, especially for near-field spectroscopy.  Since commercially available near-field probes offer such a fortuitous geometry, many broadband nano-spectroscopy studies likely leverage this feature unknowingly.\cite{BECHTEL2020100493,huth_nano-ftir_2012}

\begin{figure}[tbp]
\centering
\includegraphics[width=\columnwidth]{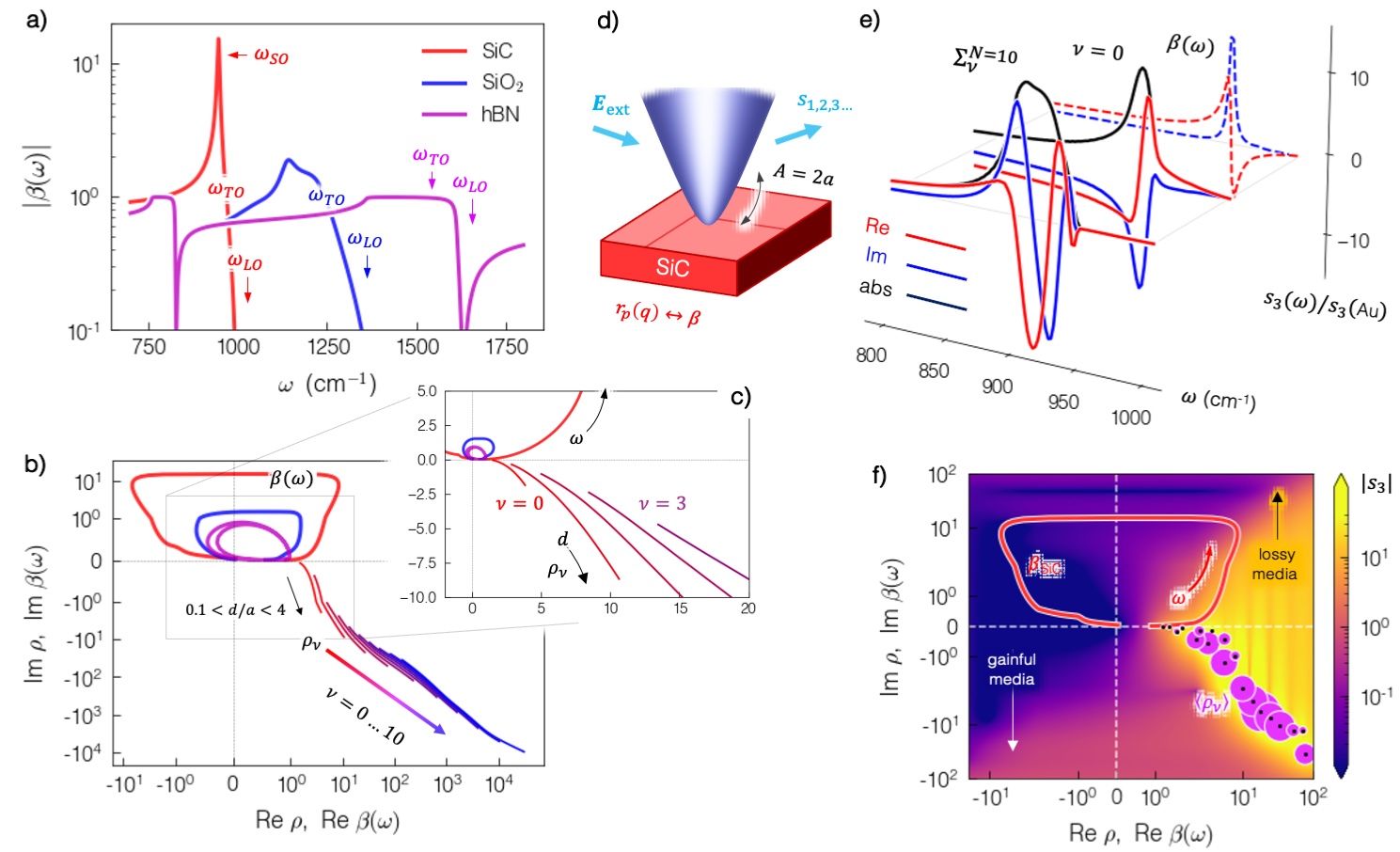}
\caption[Gap-dependent probe-sample coupling.]{\label{fig4} {\bf Gap-dependent probe-sample coupling.} a) Several canonical phonon-resonant materials targeted by near-field infrared microscopy feature quasi-electrostatic surface reflectivities $\beta(\omega)$ with magnitudes exceeding unity at their surface optical phonon frequency $\omega_{SO}$; $\omega_{TO,LO}$ denote the transverse and longitudinal optical phonon frequencies, respectively. b) The energy ($\omega$) dependence of surface reflectivity traces one or more complete circuits in the upper half complex $\beta$ plane, whereas the eigenvalues $\rho_\nu$ corresponding eigenfields of a conical probe of apex radius $a$ disperse with gap size $z$ into the lower half plane. c) View of the complex $\beta$ plane in linear scale shows lower order $\nu$ modes are most dominantly excited over surfaces with modest $\beta$.  d) Experimental configuration corresponding to b-c), in which momentum-dispersive surface reflectivity $r_p(q)$ can be approximated by $\beta$, and tapping of the probe with amplitude $A$ allows sampling the scattered field in a Fourier series $s_n$ corresponding to harmonics $n$ of the tapping frequency. e) Prediction of the demodulated scattered field $s_3$ (real and imaginary parts, and absolute value) from reflectivity $\beta$ becomes more accurate as the $d$-modulated scattered field from further excited eigenmodes $0\le \nu < N$ are summed.  Mid- and foreground spectra correspond to predictions with $N=1$ and $10$ eigenmodes, respectively; the latter changes little for higher $N$.  f) Demodulated scattered field $s_3$ predicted across the complex $\beta$ plane by summing the contribution from a finite number of eigenmodes.  Dots show the position of poles in a rational approximation for $s_3$ with disks sized to denote the relative amplitude of corresponding residues.}
\end{figure}

\subsection{Formation of scattering contrasts with actively coupled  nanoscopy} \label{sec:scatteringcontrasts}

Having clarified the physical behavior of eigenvalues $\rho_i$, we can now further scrutinize their quantitative implications for spetroscpopic scattering contrast.  For conceptual simplicity, we consider the class of ``bulk media" whose Fresnel reflectivity is large chiefly in the quasi-electrostatic regime, justifying an approximation $r_p(q)\rightarrow \beta$ under illumination from a near-field probe.  For demonstration, Fig. \ref{fig4}a presents the quasistatic reflectivity of three representative polar materials: silicon carbide (SiC), amorphous silicon oxide (SiO$_2$), and hexagonal boron nitride (hBN), all presenting one or more optically active phonon resonances in the mid-infrared range from $\omega=700-1800$ cm$^{-1}$.  Here $\omega_\mathrm{SO}<\omega_\mathrm{TO}<\omega_\mathrm{LO}$ denote the surface optical, transverse optical, and longitudinal optical phonon frequencies, respectively, associated with the polar surface.  Associated with the conditions that $\varepsilon_\mathrm{SO,TO,LO}=-1,0,+1$, respectively, we have that $\beta_\mathrm{SO}$ is locally maximized whereas $\beta_\mathrm{TO,LO}=1,0$, respectively.  This phenomenology is essentially replicated among the optical phonons of materials considered here, modulated only by the strength, damping rate, and deviations from Lorentzian character among the oscillators.  (We note also that owing to the hyperbolic uniaxial character of hBN, $\varepsilon_{ab}/\varepsilon_c \lesssim 1$ so that $\mathrm{max} |\beta|=\beta_\mathrm{SO}\sim 1$.\cite{Dai2014}  This phenomenology further generalizes among oscillator permittivities in proximity to wherever the condition $\varepsilon(\omega)\lesssim 0$ produces a surface optical phonon.

\begin{figure}[tbp]
\centering
\includegraphics[width=\textwidth]{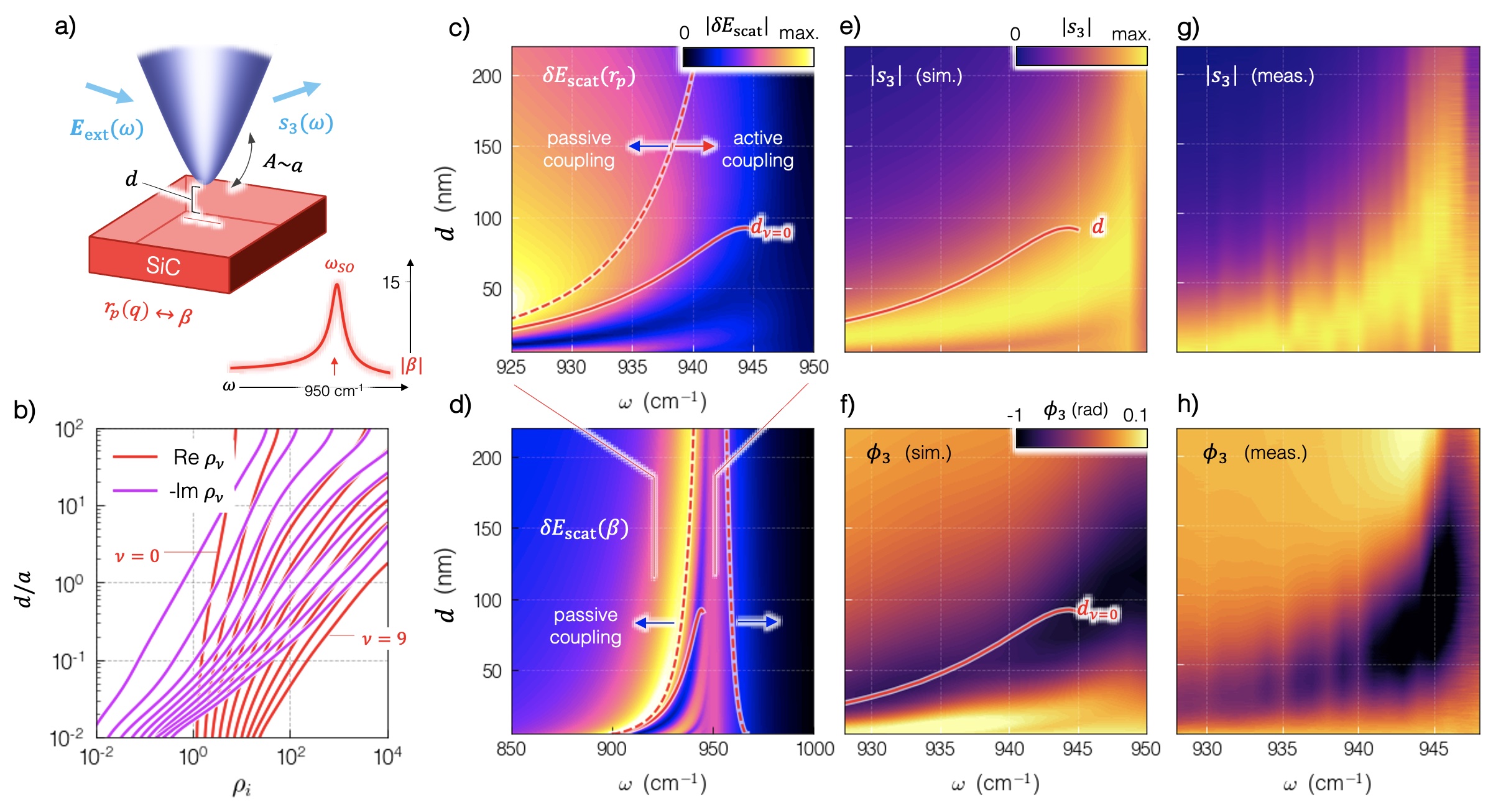}
\caption[Excitation of nano-gap polaritons over SiC.]{\label{fig5} {\bf Excitation of nano-gap polaritons over SiC.} a) Experimental configuration with probe tapping amplitude $A$ comparable to tip radius $a$; inset: quasi-electrostatic reflectivity $\beta$ of the 6H SiC surface is most enhanced at $\omega_{SO}$.  b) Explicit probe-gap $d$-dependence of probe eigenvalues $\rho_\nu$ ($\nu$ increasing left to right; values same as Fig. 4b); these converge to unity for $d\ll a$, the probe radius. c-d) Predicted portion of the probe-scattered field $\delta E_\mathrm{scat}$ induced by coupling to the fully nonlocal response $r_p(q)$ versus to the local one $\beta$, respectively, for $a\approx 50$ nm.  Solid red curves denotes the probe-sample gap $d$ of configurational resonance that minimizes $|\beta(\omega)-\rho_{\nu=0}(d)|$ corresponding to a single nano-gap polariton; nonlocality in $r_p(q)$ shifts and broadens resonance.  Dashed curves distinguish regimes of active coupling near $\omega_{SO}$ (see Sec. \ref{sec:coupledresponsefunction} and Fig. \ref{fig:3}h) from passive coupling where probe and sample weakly couple.  e-f) Simulated amplitude $|s_3|$ and phase $\phi_3$ of the demodulated probe-scattered field versus minimum probe-sample gap show sensitivity to the ``steepest slope" of resonance versus $d$.  g-h) Experimentally measured s-SNOM approach curves corresponding to e-f) evidence configurational resonance of the polariton.}
\end{figure}

The essential implication of this phenomenology for scattering in active microscopy is considered in Fig. \ref{fig4}b, where the representative $\beta(\omega)$ are now rendered with respect to frequency as canonically counterclockwise curves in the complex plane.  The radius for each circulation (of which hBN executes two between its lower and upper Restrahlen bands) serves as a measure of oscillator strength for the surface optical phonon, which varies appreciably enough among these materials to merit a double-log scale.  The eigenvalues of the ``standard" probe considered in Fig. \ref{fig4} are likewise plotted for comparison, each presented as a curve ($\nu$ again increasing from 0 to 9 with color from red to blue) that spans a range of complex values corresponding to its evolution of $\rho_\nu(d)$ with the gap distance $0<d<4a$ between the probe and the analyte surface.  As previously mentioned and later highlighted by Fig. \ref{fig5}b, these eigenvalues drop dramatically in proportion with the confinement of fields in the probe-sample gap, falling superlinearly with $d$ asymptotically towards $\rho_\nu=0$ at $d=0$ \cite{Jiang2016}.  This $d$-evolution manifest in Fig. \ref{fig4}b indicates that proximity to probe-cavity eigenmode resonance or, equivalently, the amplitude of an associated nano-gap polariton, strongly depends on both the size of the probe-sample gap and on the excursion of the surface reflectivity into the first quadrant of the complex $\beta$-plane.  The simultaneously $\omega$- and $d$-dependent resonance proximity is scrutinized in Fig. \ref{fig4}c, where we can observe $\nu \ge 1$ is virtually inaddressible by SiO$_2$ and hBN mid-IR reflectivities, whereas up to $\nu=2$ is proximate to the reflectivity of SiC at frequencies where $\beta$ is positive and mostly real.  We can draw two conclusions from this comparison:  First, probe-sample coupling is strongly non-perturbative, ``active", and potentially resonant where $\beta \gtrsim 1$, and this threshold for non-perturbative coupling is achieved most readily for small probe-sample distances.  Second, in perhaps the majority of cases where this criterion is not satisfied ($\mathrm{Re}\beta<1$), only the lowest order eigenmodes can be expected to interact appreciably with the analyte.

To gauge this selectivity, we predict the scattered field amplitude from a probe realistically executing tapping mode over a SiC surface as shown schematically in Fig. \ref{fig4}d.  Tapping the probe is not only a means to maintain mechanical feedback in near-field microscopy, but an enormous volume of findings demonstrates demodulating the scattered field at higher harmonics of the tapping frequency both excludes the far-field response of the sample surface (the lone $\hat{S}_E$ in Eq. \ref{eq:gcomposite2}) and effectively improves spatial resolution of the measurement even below the scale of $a$.\cite{maissen_probes_2019}  This signal isolation scheme is integral to most s-SNOM imaging and spectroscopy.  In our eigenmode context, it is readily apparent that $\rho_\nu(d)$ varies far more dramatically with $d$ at higher orders $\nu$, so oscillating $d$ at amplitudes $A>a$ can more substantially modulate their contribution to scattering relative to lower order modes. These are precisely the modes conferring spatial resolving power owing to their Fourier bandwidth (Fig. \ref{fig:2}b).  We therefore adapt Eqs. \ref{eq:PSScattering} and \ref{eq:scatteredamplitude} for prediction of the scattering amplitude $s_n$ demodulated at the $n$th harmonic of the (angular) frequency $\Omega$ at which the probe is dithered.  Through the \textit{EigenProbe} formalism, Fig. \ref{fig:2}e predicts the (third order) demodulated scattered field measurable by s-SNOM spectroscopy over SiC. \cite{ocelic_subwavelength-scale_2004,huber_infrared_2009,McLeod2014,amarie_broadband-infrared_2011}  Accumulating the contribution from eigenmodes $\nu$, we find several modes contribute significantly for such surfaces where $|\beta|\gg 1$.  Notably, the excited amplitude each nano-gap polariton and the brightness of each associated cavity eigenmode is complex, resulting in partial interference and a rich measurable spectrum $s_3(\omega)$ (spectra are shown relative to $s_3$ predicted from a medium with $\beta=1$, \textit{e.g.} gold, Au).

The harmonics of a function $E_\mathrm{scat}(d)$ evaluated at an oscillating abscissa $d(t)$ are no more than the coefficients of expansion for that function in a series of Chebyshev polynomials $T_n(x)$.  Provided a function $s(d)$ of $d=A (1+\cos \Omega t)$, our task in computing demodulated $s_n$ up to $n=N$ then is to evaluate the first $N$ expansion coefficients describing $s(d)\approx \sum_n^N s_n\,T_n(d/A-1)$.  Applied to our scattering problem, we have:
\begin{align}
    \textit{\textit{Demodulated scattering:}} \quad
    s_n &= \frac{1}{2\pi} \int_0^{2\pi/ \Omega} \hspace{-10pt} dt\,\cos( n\Omega t) E_\mathrm{scat}\left(d(t),\omega\right) \\
     & \approx \sum_k^N w_k \sum_{\mu,\nu=0}^M B_\mu(d_k)
    \Bigg[
        \underbrace{\frac{1}{\bm{\rho}(d_k)-\bm{\beta}(\omega)}}_{\textstyle
    \begin{gathered}
      \equiv \bm{S}(d_k,\omega)
    \end{gathered}}
        \Bigg]_{\mu\nu}
    \hspace{-6pt}B_\nu(d_k). \label{eq:demodulation}
\end{align}
\noindent Here $k$-summation denotes sampling of the scattering amplitude at $N \gg n$ different $d$-positions selected such that $-1<x_k=d_k/A-1<+1$ are roots of the $N$th order Chebyshev polynomial, and where $w_k=\pi/N $ are (equal) node weights in a Chebyshev-Gauss quadrature.  Note that $d$-dispersion of the eigenvalue matrix $\bm{\rho}$ and eigenmode brightness $B_\nu$ appear explicitly, and the lowest $M$ eigenmodes are retained.  Eq. \ref{eq:demodulation} describes the probe-scattered field at gap $d_k$ as an inner product of ``brightness vectors"  over a (diagonal or sparse) modal scattering matrix $\bm{S}(d_k,\omega)$ describing eigenmode ``mixing" by the surface response $\hat{S}_S$.  It is worth pointing out that, while sinusoidally ``tapping" the probe is a convenient way to conduct optical nanoscopy experiments, the resulting ``Chebyshev transform" of the scattering amplitude $E_\mathrm{scat}(d)$ that results from demodulation is not a particularly ``mathematically natural" description of the underlying $d-$dependent eigenmode excitations (Eq. \ref{eq:scatteredamplitude}), and dictates non-analytic connection between measurement and theory (Eq. \ref{eq:demodulation}).  This source of friction in the conventional practice of s-SNOM might in the future be resolved by ``sampling" the tapping-modulated scattered field with a more purpose-built basis of functions than $\cos(n\Omega t)$, albeit at the expense of (possibly formidable) real-time digital processing. 

In the case where $B_\nu(d_k)$ and eigenvalues $\rho_\nu(d_k)$ are both pre-computed and cached and the sample response $\beta$ is local, Eq. \ref{eq:demodulation} presents a near-instantaneous calculation that reveals inefficiency of previous ``realistic" solutions to the probe scattering problem including the \textit{lightning rod model} \cite{McLeod2014}; it thus forms another central result of this work.  The case where matrix $\bm{\beta}$ is prescribed by nonlocal $r_p(q)$ would seem to demand as many as $M^2\times N$ computations by Eq. \ref{eq:genreflectanceq}.  Still, the \textit{EigenProbe} formalism demands \textit{only} inversion of matrices with low rank ($\nu \lesssim 20$) rather than rank equal to the number of nodes $q$ in the Sommerfeld quadrature of Eq. \ref{eq:genreflectanceq}, as required to implement the \textit{lightning rod model}.  Furthermore, leveraging completeness of eigenmodes $|j_\nu)$ (and associated potentials $\Phi_\nu(\bm{r})$ in the plane of the sample) through the \textit{EigenProbe encoding} provides ultimate economy:  An eigenmode set computed for any $d=d_0$ provides a basis to reconstruct those at other probe-sample gaps according to a linear transformation $\Phi_\nu^d=\sum_\mu \Psi_{\nu\mu}(d) \Phi_\mu^{d_0}$, where matrix $\bm{\Psi} (d)$ is readily precomputed through knowledge of $\Phi_\nu^{d_k}$ (in position or momentum space) and cached at all $d_k$.  (Note, a large quantity of $d_0$-eigenmodes \textit{e.g.} $\nu \lesssim 20$) is required to accurately reconstruct a lesser quantity of more confined eigenmodes at $d<d_0$; so $\bm{\Phi}$ is ideally non-square.)  The modal reflectance at any gap $d_k$ is then a linear transformation of that at $d_0$ whence the modal scattering matrix of Eq. \ref{eq:demodulation} is also given in terms of a single calculation of $\bm{\beta}$ at $d_0$:
\begin{equation}\label{eq:genreflectancetransform}
    \text{\textit{EigenProbe encoding}}:\quad \bm{S}(d_k) \equiv \left[\bm{\rho}-\bm{\Psi}(d_k)^T \bm{\beta}(d_0) \bm{\Psi}(d_k) \right]^{-1}.
\end{equation}

Considering $\beta(\omega)$ as well as the entire $q$-dispersion of the SiC phonon \cite{McLeod2014}, we apply Eqs. \ref{eq:demodulation}-\ref{eq:genreflectancetransform} to predict $s_3(\omega)$ throughout the active coupling regime of the SiC phonon.  Fig. \ref{fig4}e compares $s_3(\omega)$ computed with $M=20$ modes against that computed only with the lowest order eigenmode ($\nu=1$) which primarily effects the active coupling regime.  The complex-valued $\beta(\omega)$ is also shown for comparison.  It is evident even at $\nu=0$ that a scattering peak emerges at $\omega<\omega_\mathrm{SO}$ where $\beta$ is dominantly real and positive.  Inclusion of higher order modes superimpose ``additional" peaks at nearby frequencies to form a continuum that quantitatively matches measured nano-infrared spectra of the SiC phonon \cite{McLeod2014}.  Notably, progression of the complex spectrum through a $\delta \phi_3=4\pi$ phase increment across $\omega_\mathrm{SO}$ \cite{McLeod2014} signifies that $\beta(\omega)$ traverses very near to two ``effective" resonances formed by the $\rho_\nu(z)$ continuum.

To rationalize the possibility of effective ``poles" emerging from the demodulated continuum, we leverage Eq. \ref{eq:demodulation} to rapidly compute $s_3$ at all $\beta$ coordinates shown in the map of Fig. 4f.  Resonant ``hot spots" predictably appear in the lower right quadrant along the trajectories of $\rho_\nu(z)$, but with unexpectedly discrete character. The overall and demodulated fields are greatest for low-loss $\beta$ in the upper right quadrant, whereas hypothetical gainful media in the (lower right quadrant) can offset radiative loss of nano-gap polaritons to sustain diverging amplitude. Since each such polariton contributes a simple pole ``swept" with gap size $d$, $s_n$ sums a continuum of simple poles but can be approximated by a finite number.   We opt to approximate the demodulated scattering amplitude by $s_3(\beta)\approx s_{3,0}+\sum_{\nu=0}^M \frac{R_\nu}{P_\nu-\beta}$ (so-called barycentric rational approximation\cite{AAA_algorithm}) up to $M=19$, which we found to provide impeccable agreement with genuine calculations from Eq. \ref{eq:demodulation} across the entire range of $\beta$.  (The factor $s_{3,0}$ ensures that $s_n=0$ at $\beta=0$.)  The poles $P_\nu$ so obtained are plotted as black data points surrounded by pink disks whose sizes (radii) are used to show (relatively) the respective residue amplitudes $|R_\nu|$.  To enable facile prediction of future nano-spectroscopy measurements from bulk media (with approximately local reflectance), Appendix \ref{app:barycentric} reports the parameters resulting from our barycentric rational approximations of $s_{2,3,4}(\beta)$ from the standard probe geometry (Fig. \ref{fig:2}b) for several choices of probe tapping amplitude.  Fig. \ref{fig4}f shows that all poles of consequence reside at $|P_\nu|<100$ largely in a continuum overlapping with ``$d$-trajectories" of underlying eigenvalues shown in Fig. \ref{fig4}b-c; the smallest of these arrayed near $|P_\nu|=1.2$ remain somewhat distinct in correspondence with their ``parent" modes at $\nu=0,1,\ldots$.  The lowest two of these reside close enough to the trajectory of $\beta(\omega)$ for SiC to quantitatively account for the $\delta \phi_3=2\times 2\pi$ phase increment (approximate double resonance) observable in both Fig. \ref{fig4}e and in previous nano-spectroscopies.\cite{McLeod2014} This prediction implies resonances associated to distinct nano-gap polaritons are experimentally addressable, as we explore in the next section.

\section{Position-resolved contrast from a scanning optical probe}

\subsection{Distance control of a nano-gap cavity polariton over SiC}\label{sec:probegapcavitypolariton}

In the previous section we reasoned that nano-gap polaritons are excited appreciably only for surface reflectivities entering the ``actively coupled" regime (\textit{cf.} Fig. \ref{fig:3}h; $|\beta-\rho_\nu|^{-1} \gtrsim 1$) of associated probe-cavity eigenmodes.  This condition is chiefly satisfied for strong surface optical phonons like that shown in Fig. \ref{fig5}a (inset) for the SiC surface where $\beta\gg1$, which is to be compared with cavity eigenvalues for experimentally relevant nanoscopy probe geometries. Fig. \ref{fig5}b resolves their strong $d$-dependence for the stout hyperboloid probe shown in Fig. \ref{fig:3}b (their $d$-dispersion resembles that for simpler analytic geometries only qualitatively; compare to \textit{Jiang et al.}\cite{Jiang2016}), showing that reducing $d$ simultaneously confines all eigenmodes but eases excitation of nano-gap polaritons.  On the other hand, that each $-\mathrm{Im}\,\rho_\nu$ drops to zero with gap size (nearly linearly for $d/a<1$) implies scattered fields from excited nano-gap polaritons simultaneously become increasingly undetectable. (Recall $-\mathrm{Im}\,\rho_\nu$ dictates an upper bound on the magnitude of the corresponding squared eigenmode brightness after integration over a large aperture of illumination/collection)  As predicted in Fig. \ref{fig5}c-d (for $a=40$nm, selected to emulate forthcoming experiments), the counterbalance of these effects makes evaluating scattered field amplitudes (per Eq. \ref{eq:scatteredamplitude}, $\sim \mathrm{Im}\,\rho_\nu/(\mathrm{Re}\,\rho_\nu-\beta)$) versus gap size $d$ chiefly a quantitative exercise, from which emerges rich structure in $\delta E_\mathrm{scat}$.  This reasoning is exact in the local limit (Fig. \ref{fig5}c), where near $\omega_{SO}$ each $|\rho_\nu(d)| \lesssim \mathrm{max}\{\beta\}$ contributes a distinct nano-gap polariton with measurable amplitude.  Analogous to that computed for the etalon (Eq. \ref{eq:etalonoscparams}), these \textit{configurational resonances} each inherit a linewidth $\gamma_\nu \approx \mathrm{Im}(\beta-\rho_\nu)/\mathrm{Re}\,\partial_\omega \beta$ wherever $\beta \approx \rho_\nu(d)$, which makes them increasingly distinct for $d/a<1$ where eigenmodes lose their radiative decay, albeit at the expense of their far-field visibility.  Curves labeled $d_{\nu=0}$ highlight the precise probe-sample gap of lowest order (and brightest) configurational resonance.  Fig. \ref{fig5}c predicts (via Eq. \ref{eq:genreflectanceq}) that strong nonlocal response realistic to the SiC surface optical phonon polariton blurs these idealities but does not eliminate them.

Details of the total scattered field predicted by Fig. \ref{fig5}d are difficult to measure directly except through purpose-built experiments\cite{wang_tomographic_2018,wang_scattering-type_2015}.  Clearly however, nanoscopies like s-SNOM that record tapping-demodulated scattering (Eq. \ref{eq:demodulation}) are sensitive to nano-gap polaritons whose excitation multiplied by brightness varies most rapidly with probe-sample gap $d$.  Comparing Fig. \ref{fig4}c-e confirms this hypothesis, where Fig. \ref{fig4}e-f apply Eq. \ref{eq:demodulation} to predict the associated ``time-integrated" contribution of nano-gap polaritons to the amplitude and phase of $s_3$ that might be measured versus ``minimum distance" of approach $d$ between probe and sample.  These simulations correspond to ``approach curves" experimentally addressible by s-SNOM, in which scattering from the tapping nanoscopy probe is recorded and demodulated while withdrawing the probe a distance $d$ from engagement with the surface.  Using a home-built s-SNOM illuminated by tunable monochromated infrared radiation from an ultrafast laser source (details in Appendix \ref{app:experimentaldetails})\cite{lukaskawcez_interfacial_2025,bragg_machine-learning-enabled_2025,steinle_ultra-stable_2016}, we evaluate these predictions with approach curves over the surface of 6H SiC using a metallic probe with apex curvature $a\approx 40$nm (calibrated by a method we detail in Sec. \ref{sec:calibration}).  In these experiments, approach curves were each acquired after sequentially tuning the laser emission energy from $\omega=926-950$ cm$^{-1}$ in steps of $1$ cm$^{-1}$ (the monochromated radiation linewidth is several cm$^{-1}$).  To remove the influence of variable illumination power or extrinsic chromatic sensitivity of detection, each approach curve was normalized by its maximum value. Fig. \ref{fig5}g shows these maxima occur where predicted by Fig. \ref{fig5}e, strictly for probe-sample gaps $d_{\nu=1}<d<d_{\nu=0}$ (between the first two predicted configurational resonances).

These findings confirm that nano-gap polaritons, while broadened somewhat by scattering demodulation and by nonlocal surface reflectivity, are both observable and reconfigurable by probe-sample gap $d$.  Furthermore, radiated fields from their superposition are ultimately responsible for the outcomes of scattering nanoscopy experiments, with enormous enhancement possible at energies wherever surface reflectivity counterbalances the deconfinement of a probe-cavity eigenmode ($\beta(\omega)\sim \rho_\nu$).  We emphasize in this non-perturbative scenario that optical nanoscopy does not provide record of surface optical excitations (polaritons), but rather of emergent probe-confined polaritons.

\subsection{Local scattering, generalized photonic density of states, and nano-imaging} \label{sec:imaging}

The defining feature of near-field microscopy is the probe's ability to nano-focus light to target regions of the analyte and so, by scanning the probe, to enable nano-resolved imaging.  As highlighted earlier, this ``focusing" feature is conceptually tied to recurring products $G_P G_E$ in the composite response function, which in our cavity eigenmode basis are described by the matrix $\bm{\rho}^{-1} \bm{\beta}$, with $\bm{\beta}$ the modal reflectance matrix of generalized reflection coefficients expressed earlier by Eqs. \ref{eq:genreflectanceq} -\ref{eq:genreflectancer}.  Our foregoing attention has focused on momentum-resolved reflectances associated with Fresnel coefficients of homogeneous, semi-infinite, or layered environments, where the nano-gap polaritons corresponding to $\rho_\nu$ are realizable excitations.  On the other hand, Eq. \ref{eq:genreflectancer} also soundly defines a matrix of ``nano-resolved" generalized reflectances with manifest dependence on the probe position, or equivalently on the ``origin" for the real-space eigenfields $\Phi_\nu(\bm{r})$ (\textit{e.g.} Fig. \ref{fig:2}a).  Consider the point dipole model, in which case the only relevant ``eigenfield" corresponds to the dipole field $\Phi_{dp}(\bm{r})$.  In this case Eq. \ref{eq:genreflectancer} provides a familiar expression for the field reflected back to the point dipole\cite{jing_phase-resolved_2023,jing_terahertz_2021,rizzo_nanometer-scale_2022,xu_deep_2021}, a quantity otherwise known as the ``photonic density of states"\cite{novotny2012principles} $\rho(\bm{r})$ in which $\bm{r}$ denotes the dipole coordinate.  For a dipolar emitter of susceptibility $\chi_e$ placed into an optical environment, $\chi_e \rho(\bm{r})$ comprises the leading term in a perturbative expansion for their composite response function $G_+$, and governs such phenomena as spontaneous emission, quenching, and strong-coupling to a cavity resonance (Sec. \ref{sec:strongcoupling}).  Of course, a physical near-field probe is not a dipole, although the eigencurrents $|j_\nu)$ so far discussed can be conceptually regarded as weighted distributions of dipoles over the probe surface.  To this extent, the probe position-resolved modal reflectances $\beta_{\mu\nu}$ comprise a generalization of photonic density of states to spatially extended emitters, of which an extended optical nanoscopy probe forms one example.  In a perturbative treatment of Eq. \ref{eq:scatteredamplitude}, radiative emission from our probe is modified through proximity to the sample environment by an amount $\delta E_\mathrm{scat}=\vec{B}^T \bm{\rho}^{-2} \bm{\beta} \vec{B}$, with $\vec{B}$ the column vector of eigenmode brightnesses.  By analogy to Purcell enhancement of radiative emission, we denote this the \textit{multi-modal Purcell effect}.  Clearly at a location of the probe where the modal reflectivity $\bm{\beta}(\bm{r}_\mathrm{probe})$ is sufficiently large, the non-perturbative regime applies, whereby we obtain the
\begin{equation} \label{eq:rdependentscattering}
    \text{\textit{Locally scattered field:}}\quad
    E_\mathrm{scat}(\bm{r}_\mathrm{probe})=\vec{B}^T \left[\bm{\rho}-\bm{\beta}(\bm{r}_\mathrm{probe})\right]^{-1} \vec{B}.
\end{equation}
\noindent Eq. \ref{eq:rdependentscattering} highlights the manifest ease of generalizing the probe-scattered field in our eigenfield framework to the case of variable probe position within an optical environment where the local modal reflectivity $\bm{\beta}$ mediates inhomogeneous coupling.

\begin{figure}[tbp]
\centering
\includegraphics[width=0.9\textwidth]{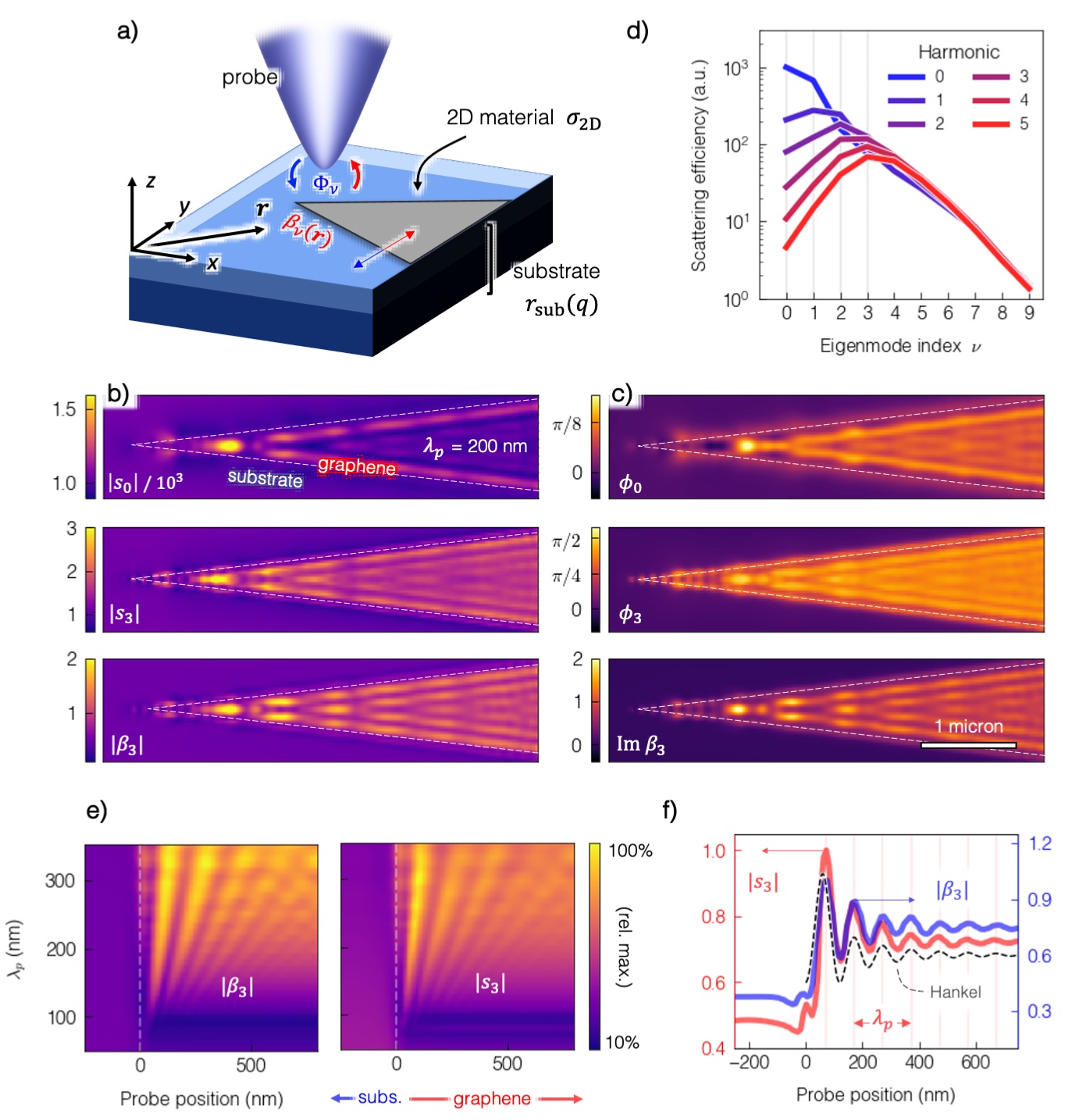}
\caption[Prediction of image contrasts from eigenmode reflectivities.]{\label{fig6} {\bf Prediction of image contrasts from eigenmode reflectivities.} a) Schematic configuration for predicting nanoscopy contrasts from a 2D material (\textit{e.g.} a plasmonic graphene nano-taper). b-c) Amplitude and phase of the zeroth and third-order demodulated scattered field contrasts predicted by Eqs. \ref{eq:Rmatrix}-\ref{eq:genreflectanceimaging}; $\lambda_p=2\pi/(\kappa q_p)$ denotes the screened polasmon wavelength (details in main text).  The amplitude and imaginary part of the local modal reflectivity $\beta_{\nu=3}$ (bottom panels) provide a semi-quantitative description of $s_3$ contrasts.  d) Distributions describe the relative contribution of distinct eigenmodes (associated with average probe-sample distance $d_0=a$) to the demodulated scattering amplitude at harmonics $0-5$. e) Position-resolved modal reflectivity and demodulated scattering predicted across the line shown in panel a) while varying the screened plasmon wavelength.  Edge-reflected plasmon (non-local) response combined with quasi-uniform (local) response dictate the imaging contrast.  f) Selections from e) at $\lambda_p=200$nm compare favorably to a Hankel function $H_0(2\pi \cdot 2x/\lambda_p)$ at probe-edge distance $x$, which semi-quantitatively accounts for cylindrical emanation of plasmons.}
\end{figure}

Here we show Eq. \ref{eq:rdependentscattering} is a concrete route to quantitatively predict imaging results from near-field microscopy of heterogeneous media, especially when we evaluate position-resolved $\bm{\beta}$ by the following \textit{spectral method}.  Shown schematically in Fig. \ref{fig6}a, we focus on the characteristic case of an inhomogeneous 2D material (with conductivity $\sigma_{2D}$) residing on a semi-infinite (conceivably layered) substrate.  Nano-imaging of mono- to few-layer graphene and other thin van der Waals crystals\cite{chen_optical_2012,Fei2012,rizzo_charge-transfer_2020,jing_phase-resolved_2023,jing_terahertz_2021}, and even conducting thin films, are all subsumed in this category.  For such applications we can compactly combine the laws of Gauss and Ohm with the composite response function of a 2D medium and substrate (Eq. \ref{eq:gcomposite2}) to rapidly compute the quasi-electrostatic optical response of such inhomogeneous material systems, as described in detail by Appendix \ref{app:eigenprobeimaging}.  Briefly, for a given location of the probe, a reflected eigenfield $\tilde{\Phi}_\nu(\bm{r})$ can be computed by projecting $\Phi_\nu(\bm{r})$ into a basis of $L^2$-orthonormal functions defined along the sample surface ($z=0$) for which the sample reflectance is easily represented by a matrix $\bm{R}$ describing:
\begin{gather} \label{eq:Rmatrix}
    \text{\textit{Inhomogeneous reflectivity:}}\quad
    \bm{R}=\frac{\beta_\mathrm{subs}-\bm{V} \bm{L}/(2\pi\kappa q_p)}{1-\bm{V}\bm{L}/(2\pi\kappa q_p)}, \\
    \text{where} \quad q_p \equiv i\omega/(2\pi \sigma_{2D}) \quad
    \text{and} \quad \kappa\equiv (\varepsilon_\mathrm{subs}+1)/2 \nonumber \\
    \text{so that} \quad \bm{\beta}(\bm{r}_\mathrm{probe}) = (2\pi)^{-1}\bm{\Phi}(\bm{r}_\mathrm{probe})^T \bm Q \bm R \bm{\Phi}(\bm{r}_\mathrm{probe}). \label{eq:genreflectanceimaging}
\end{gather}
\noindent Considering a homogeneous substrate for simplicity, here $\beta_\mathrm{subs}$ and $\varepsilon_\mathrm{subs}$ denote the substrate reflectivity and permittivity, respectively.  Maintaining bold fonts for matrix-valued operators, $\bm L$ denotes the Laplace operator ($\nabla^2$) acting over the spatial region of the 2D material with plasmon wavevector $q_p$, and $\bm V$ denotes the Coulomb operator (spatial convolution by kernel $1/r$).  Both are projected into matrices by operating on our basis functions in the plane $z=0$.  It is convenient to select a basis where $\bm L$ is diagonal -- namely, eigenmodes of the 2D Helmholtz (wave) equation along the 2D material surface.  In this basis $\vec{\Phi}_\nu(\bm{r}_\mathrm{probe})$ expresses a single eigenfield from the probe at $(\bm{r}_\mathrm{probe})$ as a vector, $-\bm R \vec{\Phi}_\nu$ is such a vector describing the reflected field, and the modal reflectivity is predicted (Eqs. \ref{eq:genreflectancer} and \ref{eq:genreflectancetransform}) by $\beta_{\mu\nu}(\bm{r}_\mathrm{probe})=(2\pi)^{-1}\vec{\Phi}_{\mu}(\bm{r}_\mathrm{probe})^T \bm Q \bm R \vec{\Phi}_{\nu}(\bm{r}_\mathrm{probe})$.  Here $\bm Q$ is the matrix projection of the operator $-\partial_z$, which in a basis of plane (or cylindrical) waves is simply a diagonal matrix of in-plane momenta $q$ that recovers Eq. \ref{eq:genreflectanceq}.  Assembling the column vectors $\vec{\Phi}_\nu$ into a matrix $\bm{\Phi}$ supplied the entire modal reflectance matrix as Eq. \ref{eq:genreflectanceimaging}.  This approach straightforwardly extends to the case where multiple 2D materials (overlapping or otherwise) are supported on a substrate described by complex Fresnel coefficient $r_\mathrm{subs}(q)$.  When all conducting media are homogeneous, plane (or cylindrical) waves form a suitable function basis for Eq. \ref{eq:Rmatrix} and $\bm V \bm L=2\pi \bm Q$.  In this basis the reflectance reduces to a Fresnel coefficient, and elements of $\vec{\Phi}_\nu$ (up to normalization) simply evaluate the Hankel transforms of the eigenfields, again recovering Eq. \ref{eq:genreflectanceq}.  (Note that when eigenmodes are appropriately normalized, $(2 \pi)^{-1} {\bm \Phi}^T {\bm Q} {\bm \Phi}$ is the identity matrix.)

We now demonstrate application of Eq. \ref{eq:Rmatrix} to predict imaging results from a near-field probe (of ``standard" geometry) rastered over a triangular nano-taper of monolayer graphene with (screened) plasmon wavelength $\lambda_p=200$ nm supported on a dielectric substrate ($\beta_\mathrm{subs}=0.5$).  We select this case to replicate in theory several experimentally reported features of plasmon propagation near graphene edges and constrictions, including boundary reflection, plasmonic nano-focusing, and edge-propagating plasmons.\cite{Fei2012,chen_optical_2012,gerber_phase-resolved_2014,rizzo_charge-transfer_2020,nikitin_real-space_2016,Ni2018,woessner_near-field_2016}  First, we select a probe gap $d_0=a$ at which to define our eigenmode basis (Fig. \ref{fig1}e) and compute $\beta_{\mu\nu}$ over a grid of probe positions in a chosen field of view over the conducting nano-taper (Fig. \ref{fig6}a).  We then apply Eq. \ref{eq:rdependentscattering} to compute the probe-scattered field $E_\mathrm{scat}(\bm{r}_\mathrm{probe})$.  To replicate the experimental process of harmonic demodulation in principle requires identical such computation for eigenfields defined at all probe-sample distances $z$ spanning a tapping cycle of the probe, which would be burdensome. 
Fortunately, here we can again apply Eq. \ref{eq:genreflectancetransform} to recover $E_\mathrm{scat}(d,\bm{r}_\mathrm{probe})$ at all $d$, with explicit calculation of $\bm \beta(\bm{r}_\mathrm{probe})$ at only a single probe-sample gap $d_0$.  Fig. \ref{fig6}b presents the result, where $s_n$ denotes the demodulated field at harmonic $n$ whereas $\beta_\nu$ denotes the modal reflectance $\beta_{\nu\nu}(d_0)$, and  Fig. \ref{fig6}c analogously presents the associated complex phase.  Whereas the zeroth-harmonic (time-averaged) scattered field $s_0$ is several orders of magnitude larger than $s_3$, as expected, it also provides a blurry view of the graphene flake.  On the other hand, $s_3$ showcases several features evocative of experimental demonstrations \cite{Fei2012,chen_optical_2012,rizzo_charge-transfer_2020}, including i) plasmon fringe patterns near graphene edges with characteristic periodicity $\lambda_p/2$, ii) nano-focusing of the plasmon field at the tapered end, and iii) so-called ``edge plasmons" propagating with wavelength less than $\lambda_p$ manifesting as bright-dark dots close to the graphene apex.

While this example demonstrates the achievement of rapid \textit{ab initio} calculation of complex near-field images using our eigenfield formalism, it also brings conceptual insights.  Despite the overall formal complexity of this computation, the evident resemblance \textit{e.g.} of $s_3$ to the foundational reflectance quantity $\beta_3$ is noteworthy, and originates in part from the harmonic content of the eigenfields themselves.  We anticipate the resemblance of $s_n$ to $\beta_{\nu=n}$ for $\nu \le 3$ to be a general result, as explored in Fig. \ref{fig6}d.  To first order in small scalar $\beta_\nu$, the $d$-dependent scattered field from eigenmode $\nu$ is perturbatively proportional to $\beta_\nu B_\nu(d)^2/\rho_\nu(d)^2$.  Fig. \ref{fig6}d presents the demodulated amplitude (absolute value) of this quantity versus eigenmode index $nu$ and harmonic $n$, showing that radiated fields contributing to $s_{n}$ are dominated by eigenfields near $\nu=n$.  At higher harmonics $n>3$, higher order eigenmodes fail to sustain the trend, since the tapping amplitude $A$ selected for simulation is already comparable to the probe apex radius $a$, so that even higher order eigenmodes are modulated only at lower harmonics.   Therefore, one can generally understand nanoscopy contrasts from scattering demodulated at tapping harmonic $n$ by considering the generalized reflectance associated with eigenmode $\nu=n$, provided that $|\beta_\nu| \lesssim |\rho_\nu|$, beyond which we expect no simple qualitative correspondence can persist in the actively coupled regime .

Finally, Fig. \ref{fig6}e-f examine line-cuts across graphene edge along the colored arrows shown in Fig. \ref{fig6}a, presenting how the plasmon reflection profile in $s_n$ evolves with plasmon wavelength.  The transition from dark contrast to bright within the graphene (relative to the reflective substrate, at left) with increasing $\lambda_p$ owes to the fact that the plasmon momentum (\textit{cf}. Fig. \ref{fig:2}d) ``sweeps" to lower $q$ where the angular spectrum of all eigenfields (\textit{cf}. Fig. \ref{fig:2}b, bounded by the probe radius $a$) is appreciable, leading to an enhanced local reflectance from the graphene surface.  Fig. \ref{fig6}f presents representative linecuts from $\lambda_p=200$ nm, revealing that despite all formal complexities, $\beta_3$ and even the demodulated scattering signal $s_3$ are still well described by a simple Hankel function with wavevector $2q_p$, as anticipated for cylindrical plasmons reflected from a hard wall.\cite{woessner_near-field_2016}  That this simple description does not account for the ``edge plasmon" fringe visible precisely at the graphene edge is its chief failure.   Notably, these profile are \textit{not} well described by a decaying exponential, which grossly underestimates amplitude of the first primary fringe of reflected polaritons relative to those that follow.

\section{Probe-cavity eigenmodes for quantitative metrology through optical nano-spectroscopy}

\subsection{Probe calibration through cavity eigenvalues} \label{sec:calibration}

Returning to one of the goals of our \textit{EigenProbe} formalism, realistic description of the probe response $\hat{S}_P$ is a precondition for inferring $\hat{S}_E$ through measurement of their their coupled response (Eq. \ref{eq:gcomposite2}).  By compactly describing this coupling, probe-cavity eigenmodes enable quantitative interpretation of nanoscopy experiments, as we will further show.  But as we have seen, the eigenmodes are also reasonably sensitive to probe geometry (Fig. \ref{fig:3}), which is not likely known with certainty, especially when experiments deploy ``off the shelf" probes for measurement that may be worn or otherwise deviate from nominal geometric specification.  Of what use then is the precision of our formalism without an accurate description of the probe?  The stout hyperboloidal probe described by Fig. \ref{fig:3}b has overall dimension and half-angle $\theta$ resembling\cite{McLeod2014} typical probes utilized for s-SNOM experiments at infrared energies (\textit{e.g.} \textit{PtSi-FM} or \textit{ARROW-NCPt} by \textit{NanoWorld}).  The utility of this probe geometry for infrared nanoscopies can be appreciated by first recalling that $-\mathrm{Im}\,\rho_\nu$ describe ``visibility" of eigenmodes participating in probe-sample interactions, which Fig. \ref{fig:3}e shows reach appreciate and roughly $\omega$-constant values for $\omega>10$ THz. Second, $\mathrm{Re}\,\rho_\nu$ describe eigenmodes' excitability, which Fig. \ref{fig:3}g shows to be likewise roughly $\omega$-constant, unlike for the more slender probe (Fig. \ref{fig:3}a).  Though inessential to our formalism, our calculations have for simplicity adopted an axisymmetric approximation for the probe's geometry, excitation, and scattering, admittedly unlike the directional illumination and pyramidal shape of most realistic probes.  However, far more problematic for the quantitative accuracy of our formalism is uncertainty in the probe radius relative i) to the probe-sample gap $d$ ($\bar{d}\equiv d/a$), ii) to the tapping amplitude $A$ ($\bar{A}\equiv A/a$) used in an s-SNOM experiment, or iii) to momentum scales ($q\cdot a$) essential to surfaces that host polaritons or other nonlocal optical response. For nanoscopy through demodulated scattering, here we propose, validate, and hereafter leverage a simple method to calibrate $\bar{A}$ that both restores the accuracy of our formalism and allows its quantitative ``inversion" for trustworthy inferences about \textit{e.g.} optical constants ($\hat{S}_E$) of surfaces under study.  Calibration like that proposed here is thus a necessary ``bridge" between the \textit{EigenProbe} formalism and quantitative nano-optical metrology, particularly through near-field spectroscopy.

\begin{figure}
  \begin{minipage}[t]{0.3\textwidth}
  \mbox{}\\[-\baselineskip]
  \vspace{-10pt}
    \caption[Standardized scattering contrast as probe calibrator]{{\bf Standardized scattering contrast as probe calibrator.} a) Predicted probe scattering demodulated at harmonic $n$ over silicon (relative to remotely measured gold) is predicted to depend moderately on s-SNOM tapping amplitude $A$ relative to probe radius $a$ as a power law when $A<a$.  b) The self-referenced signal $\xi\equiv s_3/s_2$ depends more strongly on tapping amplitude; data points were recorded with a probe exposed to successive degrees of wear, which degrades the probe radius (decreases $A/a$) in a fashion quantifiable from predictions of $\xi$. Demodulated scattering signals from silicon relative to that from gold when both are measured (c) ``near" (within 5 microns) and (e) ``far" (apart by more than 50 microns); spectra dependent moderately on measurement frequency even when reflectivities are constant and strongly on distance between materials.  d,f) The respective self-normalized spectra $\xi_\mathrm{Si}/\xi_\mathrm{Au}$ show no dependence on these complications, but only on material contrast.}
    \label{fig:7}
  \end{minipage}
  \hfill
  \begin{minipage}[t]{0.67\textwidth}
  \mbox{}\\[-\baselineskip]
    \includegraphics[width=\textwidth]{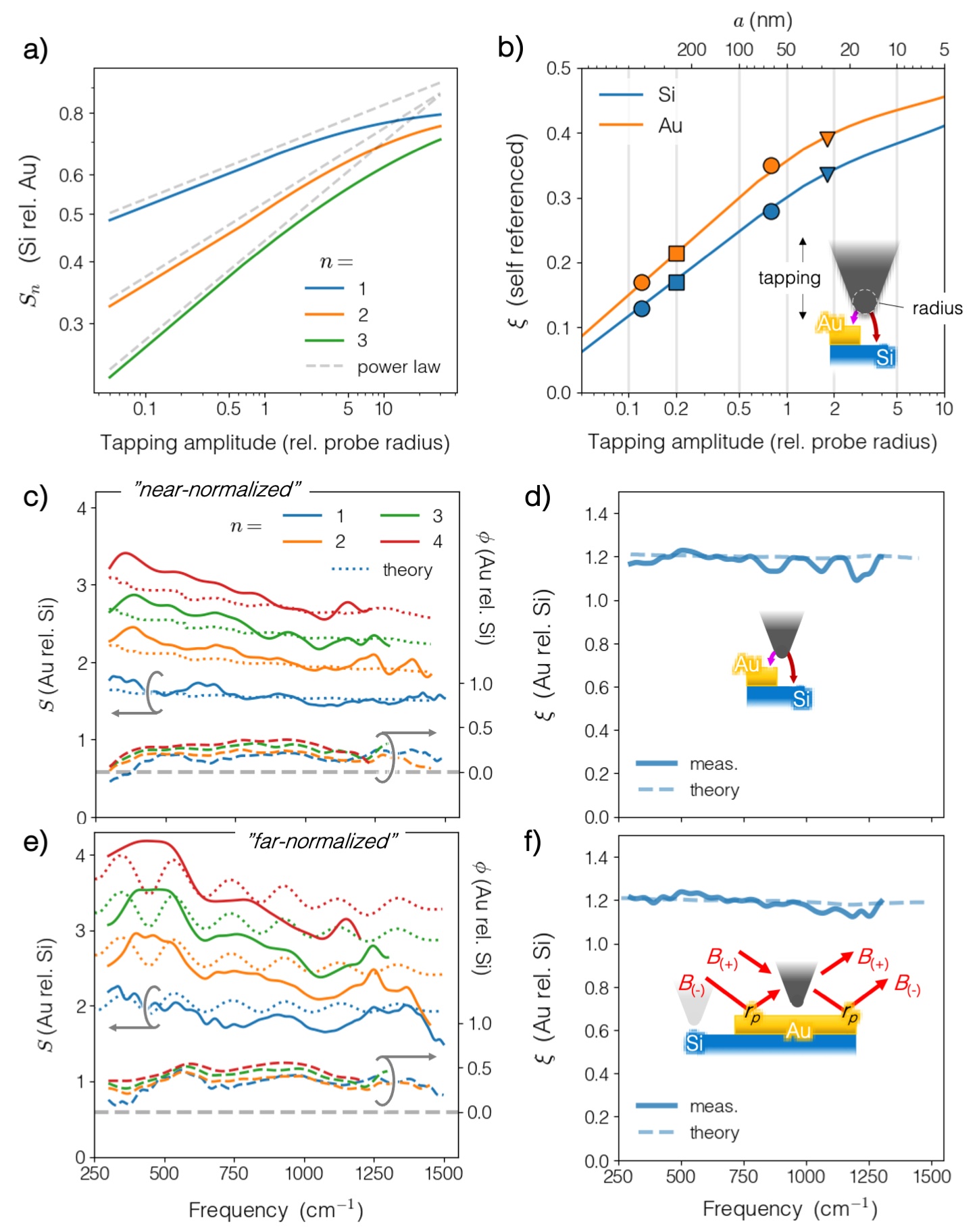}
  \end{minipage}
\end{figure}

Our model of demodulated probe scattering (Eq. \ref{eq:demodulation}) predicts how $s_n$ depends quantitatively on both the tapping amplitude relative ($\bar{A}$) to probe radius as well as the (local) reflectivity $\beta$, such as from silicon ($\beta_\mathrm{Si}\approx 0.83$ at infrared energies) and gold ($\beta_\mathrm{Au} \approx 1$) surfaces.  For reasons already mentioned, s-SNOM experiments can best report with certainty only carefully normalized quantities, like $S_n(\bar{A}) \equiv |s_n(\beta_\mathrm{Si})/s_n(\beta_\mathrm{Au})|$.  Fig. \ref{fig:7}a predicts how this quantity at harmonics $n=1$ to $3$ depends monotonically on $\bar{A}$, showing how (on dual log scale) a universal power-law dependence for $\bar{A}<1$ that could be predicted from a spheroidal description of the metallic probe relevant in this limit \cite{Jiang2016} gives way to geometry-specific dependence for $\bar{A}>1$.  Weak dependence on $\bar{A}$ in the latter regime (most relevant to experiments that aim to maximize observability of $s_n$) implies that optical contrast $S_n$ is model-insensitive with respect to probe radius but simultaneously of limited value for ``probe calibration".  For physically significant reasons \cite{mester_high-fidelity_2022} we discuss in Sec. \ref{sec:farfieldfactor}, such ``remote-normalized" contrasts $S_n$ are also susceptible to systematic error ($>10\%$) intolerable for precise calibration.  We therefore consider the recently advocated\cite{McLeod2021,chen_near-field_2022,mester_high-fidelity_2022} self-normalized quantity $\xi(\bar{A},\beta)\equiv s_3/s_2$ as a metric first for probe calibration, and later (Sec. \ref{sec:farfieldfactor}) for optical metrology. Fig. \ref{fig:7}b predicts how this quantity depends significantly on both surface reflectivity (for $\beta_\mathrm{Si,Au}$) and $\bar{A}$, which we can understand through the greatest possible simplification of Eq. \ref{eq:scatteredamplitude}.  For $\beta\lesssim 1$ and $\rho_\nu - 1 \propto \bar{d}$, the term in a perturbation series of the probe-scattered field at first-order in $\beta$ is $E_\mathrm{scat}^{(1)}\propto - \mathrm{Im}\left(\partial_{\bar{d}} \rho\right) \cdot \bar{d}/(1+\partial_{\bar{d}} \rho \cdot \bar{d}) \cdot \beta $.  The ratio of this quantity demodulated at harmonics $n+1$ and $n$ is analytic in $\bar{A}$ is approximately $\xi \sim \frac{1}{2} \partial_{\bar{d}}\rho \bar{A}$ when the quantity is small. Inspecting Fig. \ref{fig5}b shows $\partial_{\bar{d}}\sim 1$, which implies that although this approximation for $\xi$ is poor, its appreciable monotonic growth with $\bar{A}$ predicted (without approximation) by Fig. \ref{fig:7}b is expected; the rate of increase and sensitivity to $\bar{A}$ ``rounds off" only for $\bar{A} \gtrsim 1$.  (Although not shown here, we predict $\xi(\beta)$ evolves with $\beta$ similarly to the demodulated signal shown in Fig. \ref{fig4}f, although this similarity is only qualitative.)

We conclude that the ``self-normalized" signal $\xi$ may be useful not only for semiquantitative inferrence of local dielectric response from materials \cite{McLeod2021,chen_near-field_2022,mester_high-fidelity_2022} as we later discuss in Sec. \ref{sec:farfieldfactor}, but also for calibration of probe radius, provided the tapping amplitude $A$ is known \textit{a priori}.  We propose that, after the atomic force microscope (AFM) underlying most s-SNOM systems directly reports the tapping amplitude, comparing $\xi$ measured over (``perturbatively") reflective surface like $\beta_\mathrm{Si,Au}$ to values predicted by \ref{fig:7}b allows inferring whcih $\bar{A}$ (thus $a$) ``best describes" the probe via Eq. \ref{eq:demodulation}.  We have tested this proposal through s-SNOM experiments conducted at the interface of a gold layer on silicon (details in Appendix \ref{app:experimentaldetails}):  With a tapping amplitude of 50nm, a ``fresh" (unused) probe (\textit{PtSi-FM}, with geometry and eigenmodes qualitatively described by Figs. \ref{fig:3}b and \ref{fig5}b respectively) was used to measure $\xi$ over gold and silicon.  These values are recorded by reverse triangle data points in Fig. \ref{fig:7}b, indicating the experiment is best described by $A/a\approx 1.8$ and $a\approx 25$ nm; this calibration result is in reasonable agreement with the manufacturer specification of $10\text{ nm}<a<30\text{ nm}$ and consistent with former electron microscopies\cite{McLeod2014} of such probes before use in experiment.  Next, we purposefully exposed the probe to mechanical wear by executing scans across the sharp gold-silicon interface in contact-mode feedback with a probe-sample force of 500 nN.  This combination of probe and excessive force was selected to intentionally degrade the probe radius while preserving ``internal" conductivity of the probe apex, since the probe's PtSi conductive ceramic material is far thicker than the metallic coating of probes more conventionally used in s-SNOM experiments (whose apex may be rendered non-conductive by wear).  The remaining data points in Fig. \ref{fig:7}b were each recorded after such sequential contact mode scans each executed over about 12 hrs; each pair of recorded values $\xi_\mathrm{Si,Au}$ best matches predictions only for a specific value of probe radius, which evolves in steps from the initial ``sharp" value to larger than $400$ nm.

Clearly at the end of this process of tip wear, a theoretical model for probe scattering assuming a tapping amplitude $A>a$ would be completely unsuitable to describe measurements by this probe.  Crucially, since probe wear is inevitable in most practical experiments, the calibration method we propose here can both improve the precision of theoretical interpretations and reclassify blunt probes as useful (\textit{viz.} a ``bad" probe is merely one without reliable theoretical description).  While this method is far from the final word on s-SNOM probe calibration, measurement of ``standardized" materials (gold, silicon) with well known energy-independent (in the infrared) reflectivities makes our approach robust and easy.  Notably, for bespoke probe geometries like the slender ``antenna resonant" probe shown in Fig. \ref{fig:3}a that are better suited to unique experiments like s-SNOM at THz energies\cite{jing_phase-resolved_2023,jing_terahertz_2021,mastel_terahertz_2017}, the ``calibration curves" shown in Fig. \ref{fig:7}b will be different, but are easily calculable with the \textit{EigenProbe} formalism.  To better tailor theoretical models towards their application, future work should address simultaneous calibration of apex radius with other salient but likely uncertain geometric features of the probe (like half-angle $\theta$, only approximately known for commercial probes).

\subsection{Implicit energy dependence of the probe response} \label{sec:energydependence}

We have shown how knowledge of eigenvalues $\rho_\nu(d/a)$ allows inferring the probe radius from measurement of ``standard" materials like gold or silicon, so long as the overall probe geometry is known sufficiently to fix their functional form.  (Hereafter, we present measurements acquired exclusively with probes possessing the geometry in Fig. \ref{fig:3}b, tapping at 80 nm amplitude, with probe radii variously calibrated by the method above.)  On the other hand, brightness of eigenmodes is an electrodynamic and thus $\omega$-dependent feature that imparts dependence on its confinement by Kramers-Kronig relations, as perceptible in Fig. \ref{fig:3}g.  Using these standards with manifestly energy-independent reflectivity, we set out to test whether predicted energy dependence of eigenmodes might impact s-SNOM sensitivity to material contrasts.  Observing Fig. \ref{fig:3}e, we expect that with decreasing energy, $\beta_\mathrm{Au}\approx 1$ approaches $\rho_{\nu=0}$ signifying increasingly facile excitation of a nano-gap polariton as compared to when the probe is placed over $\beta_\mathrm{Si}$.  Conversely, eigenmodes overall deconfine with increasing energy, whence eigenmodes' excitation and thus probe scattering should grow increasingly linear in $\beta$.   Fig. \ref{fig:7}c presents measurements of the amplitude $S_n$ and phase $\phi_n$ of $s_{n,\mathrm{Au}}/s_{n,\mathrm{Si}}$ measured from gold ``near-normalized" to silicon using ultra-broadband synchrotron-based infrared nano-spectrosocpy (\textit{SINS}), which allows addressing $250\text{ cm}^{-1}<\omega< 1500\text{ cm}^{-1}$ with a single acquisition by nano-FTIR (detailed in Appendix \ref{app:experimentaldetails}).  By comparison, Fig. \ref{fig:7}e presents the corresponding ``far-normalized" s-SNOM spectra.  The schematics inset to Fig. \ref{fig:7}e,f, respectively, detail the difference between these acquisitions:  in the former case, spectra are acquired within 5 microns of the gold-silicon interface, whereas in the latter, spectra are purposefully acquired more than 50 microns distant from the interface (for simplicity, all spectra for this comparison were selected from a single 100 micron-long nano-FTIR linescan across the interface).

We seek first to rationalize the near-normalized spectra by demodulating Eqs. \ref{eq:demodulation}-\ref{eq:genreflectancetransform}, while recalling this describes strictly one portion of the probe-sample response function -- namely, that emphasized in pink by Fig. \ref{fig1}a.  Dashed curves in Fig. \ref{fig:7}c utilize $\rho_\nu(\omega)$ updated at each energy according to Fig. \ref{fig:3}e with \textit{EigenProbe encoding} $\bm{\Phi}(d)$ approximated by its representative evaluated once at both $\omega=1000$ cm$^{-1}$ and $d_0=a$, using a calibrated probe radius by $a\approx 30$ nm.  Predicted contrast between gold and silicon increases almost 30\% while decreasing energy by two octaves.  The excellent semi-quantitative agreement with measurements is remarkable given our rather approximate axisymmetric description of the probe geometry, whose large-scale features are solely responsible for the electrodynamic characteristics of $\rho_\nu(\omega)$.  We conclude that the electrodynamic response of micron-scale probes used for optical nanoscopy has \textit{non-neglible impact} on relative scattering collected from different dielectric environments, and \textit{should be considered} in quantitative treatments of remote-normalized nano-spectroscopies which measure over an octave or more in energy, including SINS.  In practice, a linear approximation of $\rho_\nu$ versus $\omega$ seems sufficient, especially for the most important low-order eigenmodes.  We observe however this ``implicit" energy dependence of $S_n(\omega)$ is quite similar in all harmonics and motivates examining the self-normalized spectrum $\xi(\omega)$.  Empirically, we find that the relative self-normalized signal $\xi_\mathrm{Au}/\xi_\mathrm{Si}\approx 1.23$ is completely insensitive to these variations.  As shown by Fig. \ref{fig:7}d, it is constant with energy to within 5\% both in theory and experiment (excluding variations attributed to noise), while still providing a quantitatively predictable (albeit more intricate) measure of optical contrast between $\beta_\mathrm{Au,Si}$.  (Not shown, individual spectra $\xi_\mathrm{Au}(\omega)$ incompletely remove the apparent energy dependence.)

\begin{figure}
  \begin{minipage}[t]{0.27\textwidth}
  \mbox{}\\[-\baselineskip]
  \vspace{-10pt}
    \caption[Nano-spectroscopy of 6H SiC with near-, far-, and self-normalization]{{\bf Nano-spectroscopy with near-, far-, and self-normalization.} a) Demodulated scattering over SiC normalized to that over gold when both spectra are recorded within a few microns of a SiC-gold interface.  b) Absolute value and imaginary part of the relative self-normalized SiC spectrum $\xi_\mathrm{SiC}/\xi_\mathrm{Au}$ (ratio of $n=3,2$ spectra in (a)).  c) Spectra acquired as in (a), but from measurements 50 microns distant from the SiC-gold interface, and d) the corresponding relative self-normalized SiC spectrum. }
    \label{fig:8}
  \end{minipage}
  \hfill
  \begin{minipage}[t]{0.7\textwidth}
  \mbox{}\\[-\baselineskip]
    \includegraphics[width=\textwidth]{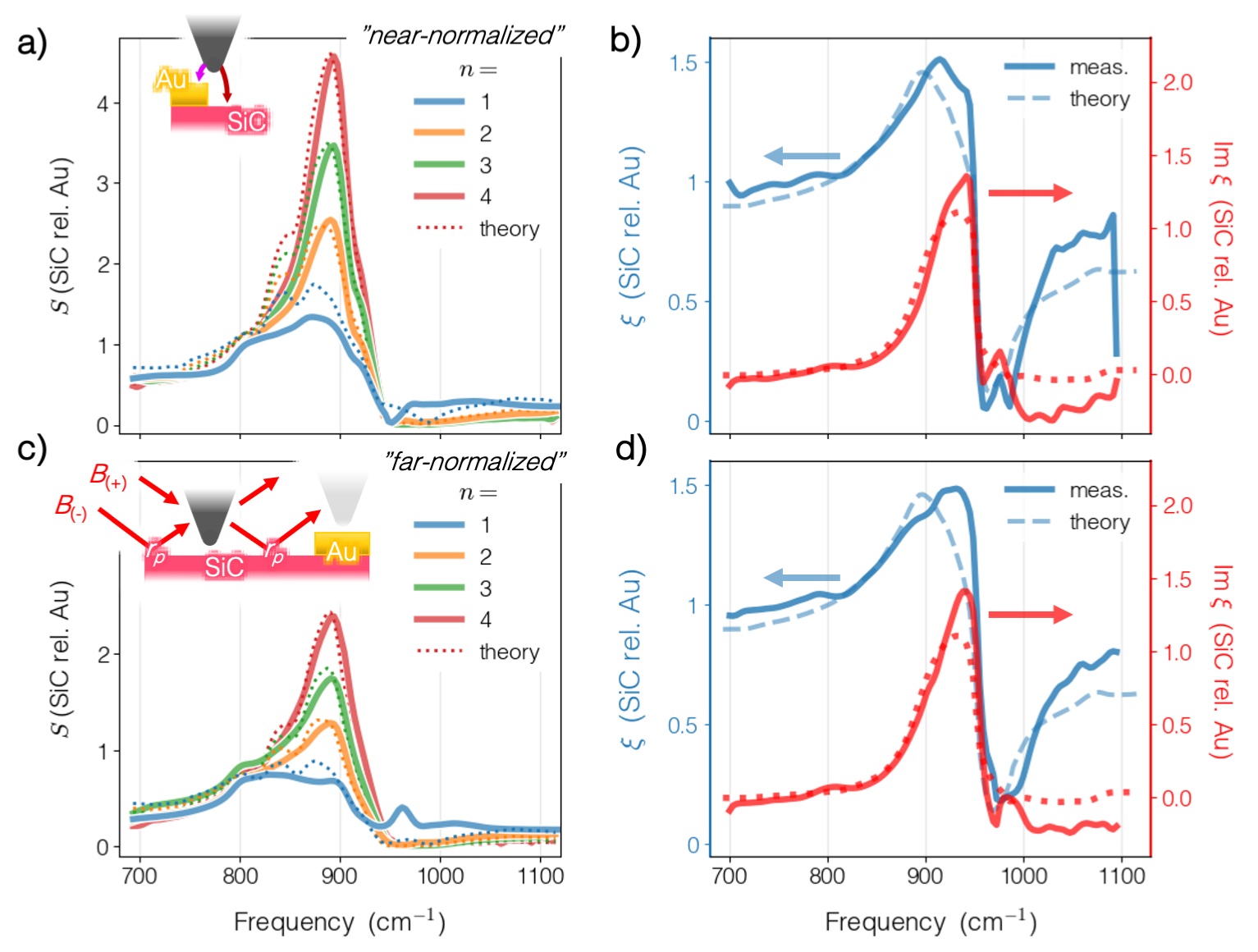}
  \end{minipage}
\end{figure}

\subsection{The impacts of far-field surface scattering} \label{sec:farfieldfactor}

Eigenmode confinement $\rho_\nu$ is not the only impactful feature of the probe-sample response that is manifestly dependent on energy.  Eigenmode currents $|j_\nu)$ likewise evolve with $\omega$, and may impact scattering contrasts when their $\omega$-dependent brightness affects measurement.  We now turn our attention to the far-normalized spectra which, as shown by Fig. \ref{fig:7}e, show contrasts even greater than near-normalized spectra and notably stronger variations with energy reminiscent of interference oscillations.  As recently reported,\cite{mester_high-fidelity_2022} far-normalized s-SNOM scattering contrasts are notoriously impacted by optical diffraction and other ``far-field responses" of the sample surface unaccounted for in $\hat{S}_{PS}\equiv(1-\hat{S}_P \hat{S}_s)^{-1} \hat{S}_P$ that so far has been the target of our \textit{EigenProbe} formalism. As emphasized in Fig. \ref{fig1}a, the entire probe-sample response function (Eq. \ref{eq:gcomposite2}) includes a leading (and trailing) factor of $(1+\hat{S}_E)$ that we expect to diffract or reflect illuminating radiation to (and scattered radiation from) the probe.  These effects have been phenomenologically included in models of s-SNOM contrast from planar media with an overall factor of $(1+r_p(\theta,\omega))^2$, where the Fresnel coefficient describes the surface's reflection of incoming/outgoing radiation from/to an aperture placed at angle $\theta$.  The \textit{EigenProbe} formalism straightforwardly includes this effect by generalizing each eigenmode's
\begin{gather}\label{eq:effectivebrightness}
    \text{\textit{Effective brightness:}}\quad B_{\nu,\mathrm{eff}} = \int_A d\Omega \Big( B_{(+)} + r_p(\theta) B_{(-)}  \Big) \\
    \text{where} \quad  B_{(\pm)} \equiv B_{\nu}(\theta)\; \text{and} \; B_{\nu}(\pi/2-\theta). \nonumber
\end{gather}
\noindent Here $B_\nu$ and $r_p$ depend implicitly on $d$ and $\omega$, respectively.  The latter factor in Eq. \ref{eq:effectivebrightness} equals $(j_\nu|\hat{S}_E|E_\mathrm{ext})$ and accounts for radiation to/from the aperture ``redirected"  by the nearby surface into angle $\pi/2-\theta$.  Relevant to top-illuminated s-SNOM, ``radiation patterns" for lowest order eigenmodes resembling Fig. \ref{fig:3}d are dominated by $\theta<\pi/2$.  While ``direct" $(+)$ is thus more significant than ``indirect" $(-)$ radiation for eigenmode excitation, their interference (as sketched in the inset to Fig. \ref{fig:7}f) depends on details of $B_\nu(\omega)$.  Even when $r_p$ is $\omega$-independent (as for gold and silicon), this interference generally may oscillate with energy, especially when the pattern in Fig. \ref{fig:3}d modulates across the antenna resonances shown in Fig. \ref{fig:3}e.  Furthermore, we expect interference and overall impact from $B_{(-)}$ to be negligible over silicon ($r_p<0.3$) when compared to gold, leaving its impact on spectra in Fig. \ref{fig:7}e regrettably robust against remote normalization.  By comparison, the fact that ``identical" effective brightness will impact both analyte and reference spectra in near-normalized nanoscopy leads us to expect that detailed ``angular distributions" considered for calculating $B_\nu$ (like Eq. \ref{eq:effectivebrightness}) will remain quantitatively unimportant for predictions.

Testing these hypotheses, dashed curves in Fig. \ref{fig:7}e present predictions that combine Eqs. \ref{eq:demodulation}-\ref{eq:genreflectancetransform} with \ref{eq:effectivebrightness} to rigorously predict the full $\omega$-dependence of far-normalized demodulated scattering.  (We note that the complication of Eq. \ref{eq:effectivebrightness} adds negligible computational expense to our predictions, but demands pre-computation and caching of $B_\nu(\theta,\omega)$ in addition $\rho_\nu(\omega)$ for $\nu \lesssim 20$.)  In qualitative agreement with experiment i) additional oscillations versus energy do arise from variable interference between direct and indirect illumination and ii) apparent material contrast increases owing to ``extra" indirect illumination over the medium with larger far-field reflectance (gold).  (Notably, we have found that feature ii) at least can be captured with qualitative accuracy by replacing the ``effective brightness" (Eq. \ref{eq:effectivebrightness}) with the ``direct brightness" multiplied by $(1+0.3 i\cdot r_p(\theta))^2$, where the complex factor $0.3i$ reasonably approximates the ``typical" ratio of indirect to direct brightness but is highly dependent on probe geometry.)  These features apparently show sensitivity to meso-scopic geometries of the experiment and probe far beyond what can be described with numeric certainty, thus \textit{far-normalized nanoscopy provides an imperfect quantitative metrology.}  Nevertheless, Fig. \ref{fig:7}f shows that self-normalized spectra acquired even across great distances (taken as ratio between spectra shown in Fig. \ref{fig:7}e) can be combined to form a consistent self-normalized spectroscopy through $\xi_\mathrm{Au}/\xi_\mathrm{Si}$.  This metric therefore remains quantitatively independent of both $\omega$ and probe position (\textit{viz.} identical to $\xi$ in Fig. \ref{fig:7}e) relative to mesoscopic features of the sample surface that effect far-field scattering.

For the token case of a gold-silicon interface, we have now established that i) near-normalized nanoscopy, where possible, provides a quantitative measure of material optical response, ii) that far-normalization can be understood only semi-quantitatively, and that iii) self-normalized scattering contrasts $\xi$ are robust to either experimental geometry.  We proceed to test these statements where the target of experiment is to distinguish spectroscopic contrasts in $\beta(\omega)$ against all aforementioned complications.  To predict realistic nano-spectroscopy measurements of the strong optical phonon mode of 6H SiC (Fig. \ref{fig4}a-c) provides a simultaneous test of the \textit{EigenProbe} formalism in the strongly non-perturbative regime and in presence of the practical complications discussed above.  To this end, we acquired demodulated probe scattering spectra from a specimen of single-crystal 6H SiC (cut from a commercial wafer) by nano-FTIR using the previously mentioned (Sec. \ref{sec:probegapcavitypolariton}) home-built s-SNOM with tunable illumination from an ultrafast laser system (details in Appendix \ref{app:experimentaldetails}).  The spectra comprising Fig. \ref{fig:8} were selected from a linescan spanning 100 microns across a sharp SiC-gold interface, in the same fashion as produced spectra in Fig. \ref{fig:7}c-f.  Namely, Fig. \ref{fig:8}a,c present near- and far-normalized demodulated scattering spectra, and Fig. \ref{fig:8}b,d present the associated self-normalized spectra $\xi_\mathrm{SiC}(\omega)/\xi_\mathrm{Au}$ obtained as the ratio of harmonic $n=3,2$ spectra in the former panels.  Dashed curves present \textit{EigenProbe} predictions using optical constants we  discuss later in Sec. \ref{sec:inversion}, the full energy-dependent eigenvalues $\rho_\nu(\omega)$ applied for Fig. \ref{fig:7}, a calibrated probe radius of $a\approx 30$ nm, and eigenmode brightness with and without inclusion of indirect illumination $B_{(-)}$ (Eq. \ref{eq:effectivebrightness}) for far- and near-normalized predictions, respectively.  Notably, since these calculations also employ the full non-local reflectivity $r_p(\omega,q)$ of the SiC surface optical phonon\cite{McLeod2014,tong_temperature-dependent_2018}, these present the most complex \textit{EigenProbe} calculations introduced so far.

We observe satisfying quantitative agreement between experiment and predictions, particularly to the extent that one set of SiC optical constants well describes both near- and far-normalized spectra.  Harmonic $n=1$ presents an exception which is difficult to predict quantitatively, owing likely to unaccounted tapping-induced modulation (interference) within the illumination factor $(j_\nu|1+\hat{S}_E|E_\mathrm{ext})$ that was previously proposed \cite{larson_detection_2024} to scale with $k_z A$, with $k_z$ the longitudinal momentum typical of the illumination field.  (Contamination of demodulated scattering at harmonic $n$ from such ``additive background" should scale as $(k_z A)^n$, justifying use of higher harmonics and lower tapping amplitudes for s-SNOM measurement at ever increasing energy \cite{hu_imaging_2017}.)  The few ``bumps" appearing in Fig. \ref{fig:8}a theoretical spectra are traceable to corresponding features in $\rho_\nu(\omega)$ as observable in Fig. \ref{fig:3}e. Lastly, much like the case for gold-silicon contrast, self-normalized spectra $\xi(\omega)$ shown in Fig. \ref{fig:7}b,d remain quantitatively independent of their acquisition location relative to the interface, providing an optical metrology of SiC apparently insensitive to dimensions beyond the nano-scale, as desired for quantitative nanoscopy.  Clearly the SiC optical phonon lineshape imprints upon $\xi(\omega)$ in a somewhat intuitive fashion, and as earlier, normalization to the self-normalized spectrum from gold further immunizes the spectroscopy against implicit energy dependence of the probe response.

\subsection{High-throughput extraction of causal optical constants from probe-calibrated spectra.} \label{sec:inversion}

\begin{figure}[tbp]
\centering
\includegraphics[width=\textwidth]{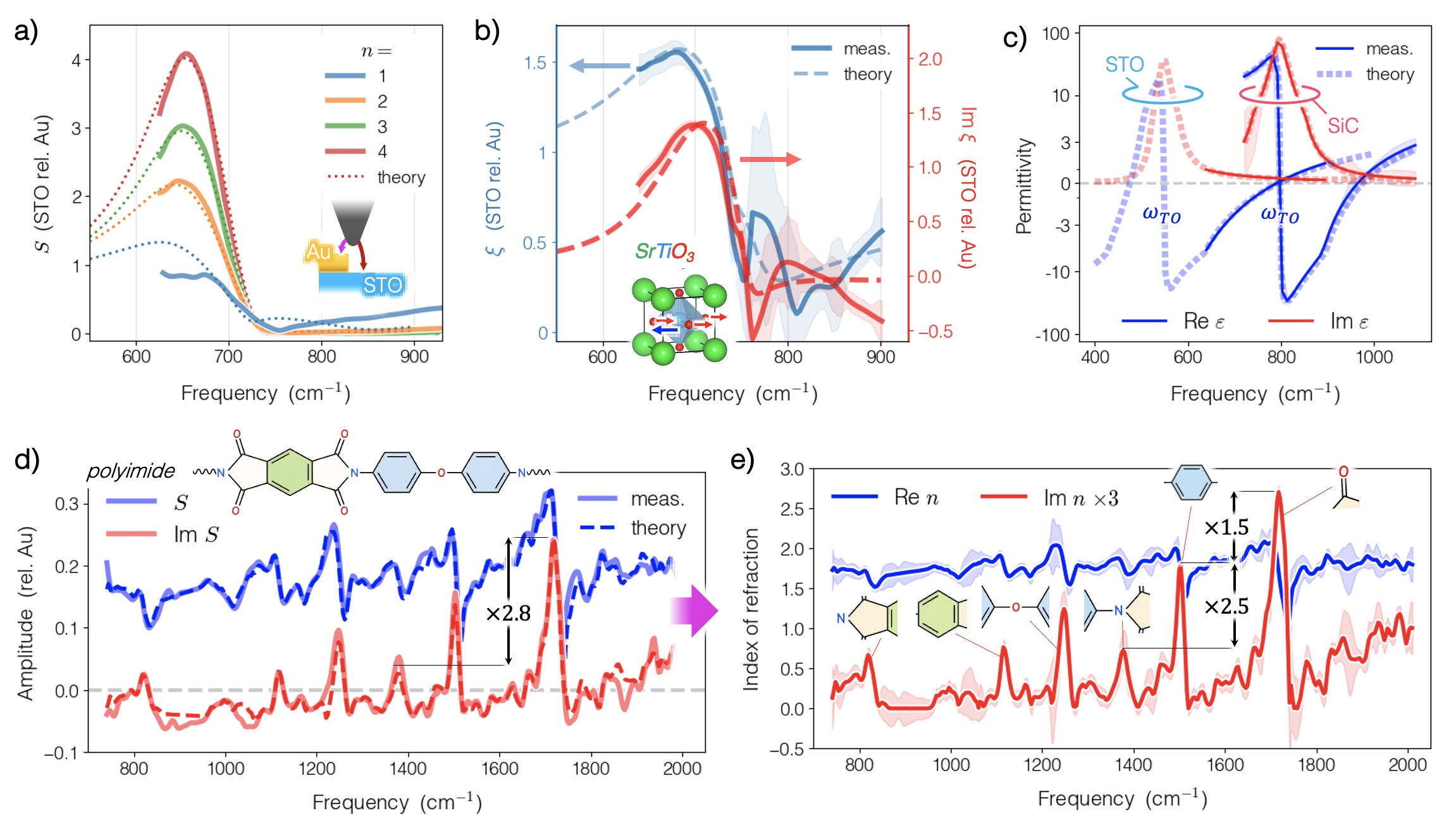}
\caption[Eigenmode inversion - from scattered contrast to optical constants]{\label{fig9} {\bf Eigenmode inversion - from scattered contrast to optical constants.} a) Nano-spectroscopy of the highest energy surface optical phonon of single crystal SrTiO$_3$; spectra of the probe-scattered field demodulated at harmonics $n$, near-normalized to those recorded over a gold surface, compare quantitatively to \textit{EigenProbe} predictions.  b) Self-normalized spectra present $\xi\equiv \xi_\mathrm{STO}/\xi_{Au}$; variance in $\xi(\omega)$ across the sample surface indicates uncertainty of this metric (shading) which is large where $S_{2,3}(\omega)$ over STO are small.  c) Optical permittivities associated with STO and SiC transverse optical (TO) phonons determined by \textit{EigenProbe} inversion of spectra presented in (a) and Fig. \ref{fig:8}a, respectively; measurements compare well to optical constants from literature (dashed curves), albeit with revised specimen-specific phonon linewidths.  d) Probe scattering spectra $S(\omega)$ demodulated at $n=3$ recorded over a bulk polyimide surface, near-normalized by spectra over gold; inset: molecular structure of a single monomer. Peaks in $\mathrm{Im}S$ ``fingerprint" absorption bands associated to molecular vibrations of structural elements labeled in (e).  Theoretical spectra (dashed curves) result from \textit{EigenProbe inversion} through a multi-oscillator description of the  polyimide permittivity.  e) Complex index of refraction inferred by the fit, where peak heights relative to \text{e.g.} the C-C vibration at $1495$ cm$^{-1}$ provide a quantitative measure of extinction and concentration for labeled structural elements.  Notably, strength of the imide vibration C-N at $1370$ cm$^{-1}$ reports degree of imidization.\cite{pryde_ftir_1993} }
\end{figure}

While measurements and predictions of SiC nanoscopy presented in Fig. \ref{fig:8}a resemble those in \textit{McLeod et al.}\cite{McLeod2014} (differing mainly in the SiC phonon linewidth, unique to the particular sample), we find that \textit{EigenProbe} calculations are more than one order of magnitude faster than ones at comparable accuracy by the \textit{Lightning Rod Model}, thanks to the strictly low-rank scattering matrix inversion (Eq. \ref{eq:demodulation}) needed for the eigenmode description.  Here we explore harnessing this tremendous speed-up for quasi-real-time ``inversion" from nanoscopy measurements to underlying optical constants of materials under study, even in the non-perturbative regime.  To supply another nanoscopy target for this exercise, Fig. \ref{fig9}a,b presents near-normalized and self-normalized nano-spectroscopies acquired over single-crystal SrTiO$_3$ (STO) acquired with the same probe and methodology as for SiC described earlier.  Dashed curves show \textit{EigenProbe} predictions employing a set of optical constants deduced by the following method, which is in essence an efficient nonlinear least-squares method.

The optical permittivity $\varepsilon(\omega)$ of insulators like STO and SiC with singular strong optical phonons in the infrared are reasonably described by a single harmonic (Lorentz) oscillator (\textit{e.g.} Eq. \ref{eq:resonance}), which enforces causal response with only four free parameters $\varepsilon_0$, $\omega_\mathrm{TO}$, $\gamma$, and $f_p$, whose set we collectively denote $\mathfrak{p}$.  These respectively denote the high-frequency permittivity, transverse optical phonon frequency, linewidth (scattering rate), and oscillator strength (the latter follows from the longitudinal optical phonon energy $\omega_\mathrm{LO}$ as $(\omega_\mathrm{LO}/\omega_\mathrm{TO})^2=f_p+1$).  We first select an initial guess for these parameters through which an automated program constructs a ``material model" that predicts the (full nonlocal) Fresnel coefficient $r_p(\omega,q)$.  Using precomputed eigenmodes properties ($|j_\nu)$, $\rho_\nu$, and $B_\nu$) and their \textit{EigenProbe encoding} (Eq. \ref{eq:genreflectancetransform}), the program next predicts demodulated scattering spectra $S_n(\omega)$ by Eq. \ref{eq:demodulation} relative to gold.  For available harmonics $n=2,3,4$, the program then computes a ``cost function" to be minimized $\mathcal{C}(\omega,\mathfrak{p}) \sim \sum_n w_n |S_{n,\mathrm{pred.}}(\omega)-S_{n,\mathrm{meas.}}(\mathfrak{p},\omega)|^\epsilon$, where $\epsilon>1$ is an exponent, and $w_n$ are weights selected proportional to the ``certainty" of each harmonic spectrum (roughly, inversely proportional to their noise level that increases with $n$).  Next, the program computes the Jacobian of the cost function at $\mathfrak{p}$ relative to all elements $p$ of the set by $\partial_{\varepsilon_\alpha} \mathcal{C}(\omega) \partial_p \varepsilon_\alpha(\omega)$ at every $\omega$, where $\alpha=1,2$ distinguishes real and imaginary parts of the permittivity (we estimate $\partial_\varepsilon \mathcal{C}(\omega)$ by finite differences).  Finally, this program is looped within a nonlinear least-square minimizer like the Levenberg-Marquardt algorithm \cite{virtanen_scipy_2020}  to minimize $\mathcal{C}$ over $\mathfrak{p}$, deducing the permittivity of ``best fit" with high precision (and accuracy limited only by assumptions of the model, including calibration of probe radius).  This method permits weighting over $\omega$ to reflect certainty of the measured spectrum at energies of higher \textit{e.g.} detection efficiency or probe irradiance from the light source.

Solid curves in Fig. \ref{fig9}c presents outcomes of this inversion method applied to the collective harmonic spectra for SiC and STO shown in Figs. \ref{fig:8}a and \ref{fig9}a, respectively.  The theoretical predictions already presented in Figs. \ref{fig:8}-\ref{fig9} employ these optical constants of ``best fit".  Thanks to computational efficiency of the \textit{EigenProbe} and its rapid convergence with the number of eigenmodes and probe-sample gap points ($\nu \lesssim 20$ and $k \lesssim 30$, respectively) used for quadrature in Eq. \ref{eq:demodulation}, inversions for SiC and STO presented here take less than one minute each when their respective measured spectra are first down-sampled (without loss of accuracy) to less than 50 frequency points across the measured range.  If ``inverted" optical constants are obtained with precision and speed, are they accurate?  Fig. \ref{fig9}c compares permittivities of best fit (solid curves) with models selected from literature \cite{tong_temperature-dependent_2018,servoin_infrared_1980} (dashed curves).  Notably, fits for both materials are strongly sensitive to the phonon linewidths, which were ascertained as $\gamma_\mathrm{SiC}\approx25$ cm$^{-1}$ and $\gamma_\mathrm{STO}\approx20$ cm$^{-1}$.  The agreement evidenced in Fig. \ref{fig9}c has required modifying literature material models only to use precisely the fitted phonon scattering rates (expected to vary with purity among samples).  The large scattering rate of the SiC sample is notably consistent with a (nitrogen-doped) carrier density of $~2\times10^{18}$ cm$^{-1}$ as we might expect from the deep green color of the selected specimen. \cite{lely_optical_1958,xiong_characterizations_2022}  Uncertainties on the fitted permittivities are indicated in the figure by colored shading, which across most energies are smaller than the width of dashed curves.  These were estimated by $\delta \varepsilon \approx \sum_n (\partial_\varepsilon S_n)^{-1} \delta S_n$, where $\delta S_n$ combines a 5\% uncertainty in values of measured spectra to the ``unexplained difference" between measured and ultimately predicted spectra.  While our nanoscopy of STO extends across $\omega_\mathrm{SO}$ but not as low as $\omega_\mathrm{TO}$, we can still infer agreement with literature constants across the available range, which already allows identifying oscillator parameters of the phonon with precision.

Our successful extraction of local optical constants from nanoscopy of polar crystals implies that, whereas the non-perturbative regime of probe-sample coupling is both theoretically challenging and essential to these measurements, the regime likewise renders spectroscopy across $\omega_\mathrm{SO}$ largely sufficient to infer properties of the transverse and longitudinal optical phonons.  We now proceed to evaluate capacity of our \textit{EigenProbe inversion} scheme to interpret nanoscopy of more complex materials, like polymers.  Fig. \ref{fig9}d presents infrared nano-spectroscopy (demodulated scattering $S_3$ relative to gold from a notably sharp calibrated probe with $a\approx 15$ nm) of kapton (PMDA-ODA) polymer acquired by our tabletop nano-FTIR method, accompanied by the (inset) schematic structure of a monomeric element of kapton.  Molecular vibrations of polymers rarely produce large reflectivity, and unlike the case of SiC or STO, we expect a local ($\beta$) description of nanoscopy contrast will suffice here, which enables using our purely algebraic expression for scattering (Eq. \ref{eq:scatteredamplitude}) as the ``engine" for \textit{EigenProbe inversion}.  Our resulting quasi-instantaneous prediction of $S_n$ from $\varepsilon(\omega)$ fortunately allows us to include manhy more oscillators in the description of this polymer permittivity.  Taking inspiration from now-popular toolkits for Kramers-Kronig (KK) constrained variational analysis of optical spectra \cite{kuzmenko_kramerskronig_2005}, we add to our set $\mathfrak{p}$ of variational parameters not only more than 10 ``free" Lorentz oscillators, but more than 50 ``narrow oscillators" fixed at equi-spaced energies that serve as KK-constrained local basis elements to account for fine details of measured spectra.  Fig. \ref{fig9}e shows the outcome of this inversion by presenting the complex index of refraction $n=\sqrt{\varepsilon}$ inferred for the kapton sample; shaded zones estimate uncertainties by the method already discussed.

The widely reported (average) kapton index of refraction $\mathrm{Re}\,n\approx 1.8$ is observable from this inversion, together with noteworthy molecular resonances arising from (sketched) local moieties of the monomer, including imide C-N ($\approx 1370$ cm$^{-1}$), benzene ring C-C ($\approx 1495$ cm$^{-1}$), and carbonyl C=O ($\approx 1705$ cm$^{-1}$) stretch modes. \cite{pryde_ftir_1993}  That the imaginary part of the demodulated probe scattering spectrum from kapton directly resembles its inferred absorption spectrum ($\mathrm{Im}\,n(\omega)$) adds support to adopted ``best practice" for nano-FTIR of polymers: comparing $\mathrm{Im}\,S_{n\ge2}(\omega)$ to ``libraries" of conventional FTIR transmission spectra may suffice to chemically identify specimens\cite{huth_nano-ftir_2012,govyadinov_quantitative_2013}, \textit{but strictly in the perturbative regime of probe-sample coupling.}  Going beyond this best practice, is then there added value in quantitative nano-metrology of $n(\omega)$ delivered by \textit{EigenProbe inversion}?  Consider the ratio of absorption band peak heights in $\mathrm{Im}\,n$ for imide, benzene ring, and carbonyl vibrations, related by the inferred factors indicated in Fig. \ref{fig9}e, which is a direct reporter of relative concentration and oscillator strength of underlying structural moieties.  On the other hand, conventional absorption spectroscopies resolve the same quantities through the Beer-Lambert Law only in thin samples over diffraction-limited length scales.  In polyimides like kapton, the degree of imidization is both integral to mechanical properties and chemical stability, and detectable by comparing the amide peak height with a reference mode like the benzene stretch \cite{pryde_ftir_1993}.  We first confirm the validity of this approach by comparing our inferred strength of carbonyl relative to benzene stretch modes ($\approx \times1.5$) to that identified by reference transmission spectra ($\times1.56$) \cite{pryde_ftir_1993}.  Thus, we can regard the $\approx \div 2.5$ relative strength of the $1370$ cm$^{-1}$ band a quantitative measure from which we infer only partial imidization of this specimen has likely resulted from cure temperatures of formation below $200^\circ$ C \cite{pryde_ftir_1993}.  On the other hand, such quantitative absorption metrology is ambiguated in the raw spectrum, firstly by an offset in ``baseline" of $\mathrm{Im}\,S$ that our model finds to be genuine, and secondly by a ``stretch" of relative peak heights ($\times 2.8 \ne 2.5\times1.5$) that owes to nonlinearity of the probe-scattered field in $\beta$, which persists even to the perturbative regime of probe-sample coupling.

We have shown that \textit{EigenProbe inversion} empowers quantitative nanoscopy of molecular absorption beyond what is possible through ``standard practice".  Still, we envision that faster methods are possible in this regime to infer complex refractive index.  A series expansion of Eq. \ref{eq:PSScatteringPerturbative} supplies the scattered field in powers $n$ of $\beta$ with coefficients equal to $\sum_\nu B_\nu(d)^2/(\rho_\nu(d))^{n+1}$.  These functions are straightforward to demodulate once, and this approach connects $S_n$ to $\beta$ through a fixed power series.  Whereas KK-constrained fitting of $\beta$ would still demands nonlinear optimization, quasi-instantaneous evaluation of the cost function should enable genuine on-line quantitative metrology of polymers through the \textit{EigenProbe} formalism.  Future work should explore performance and limitations of this approach applied to polymer nanoscopies.  We conclude this section by anticipating widespread application of the methods presented here for analysis of optical nano-spectroscopies, for which purpose the authors make these publicly available through the \textit{EigenProbe} software package \cite{mcleod_eigenprobe_2026}.

\section{Strong coupling and probe-polariton hybridization}
\subsection{Mode softening of a nano-gap polariton through strong coupling with a resonant antenna probe} \label{sec:STO}

\begin{figure}[tbp]
\centering
\includegraphics[width=\textwidth]{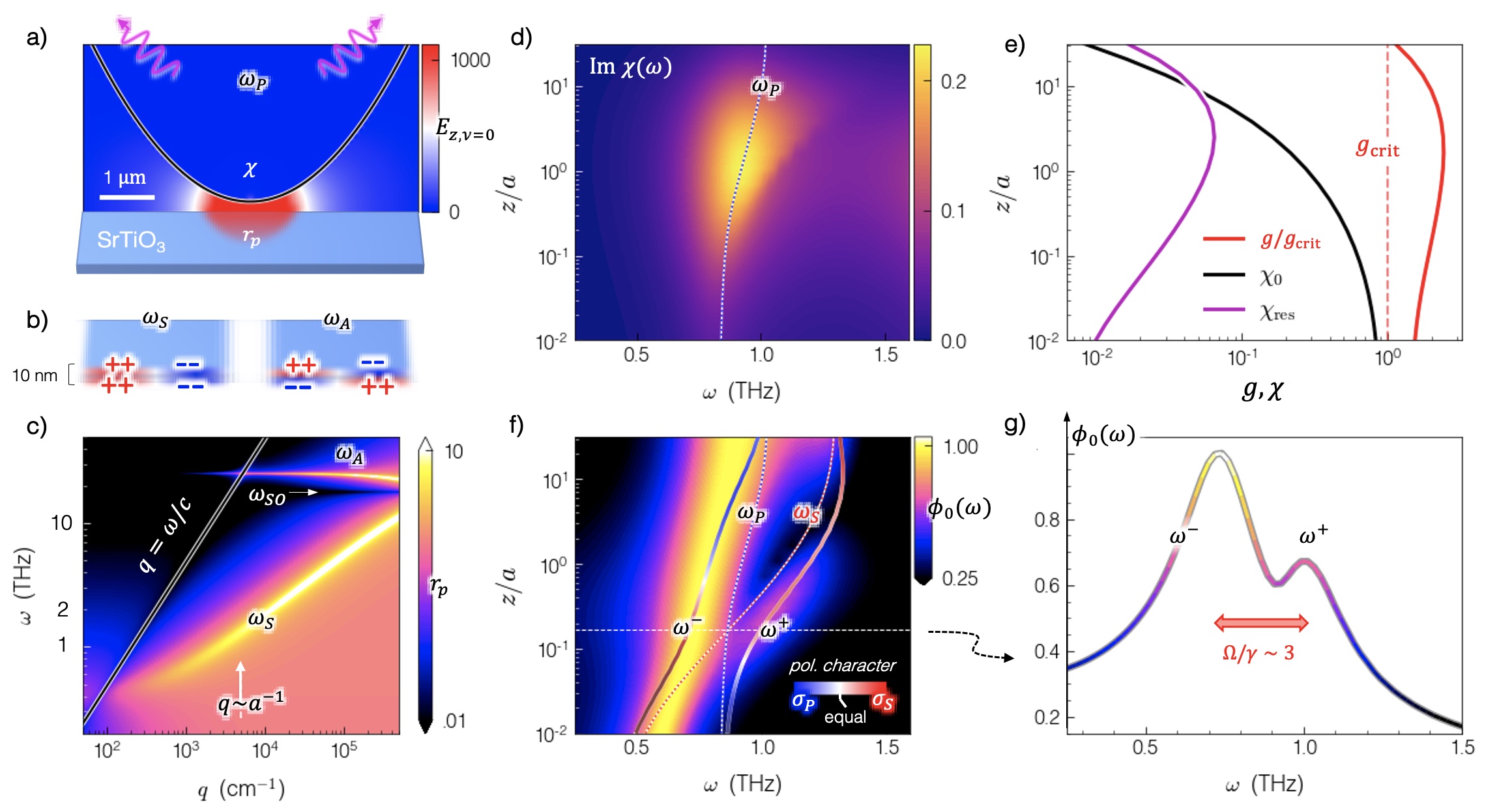}
\caption[Nano-gap polariton strong coupling with an antenna probe]{\label{fig10}  {\bf Nano-gap strong coupling with an antenna probe.} a) Experimental configuration of an antenna probe apex described by susceptibility $\chi$ over a thin SrTiO$_3$ (STO) membrane; shown is the lowest order eigenmode $z-$polarized field. b) The soft mode-resonant membrane hosts symmetric $\omega_S$ and antisymmetric $\omega_A$ polaritons; shown is the transient charge density. c) Dispersion of the two modes $|r_p(\omega,q)|$, highlighting the momentum $a^{-1}$ of greatest probe-sample coupling.  d) Imaginary part of the lowest order eigenmode susceptibility quantifies the probe's antenna resonance.  e) Predicted non-resonant susceptibility $\chi_0$ and resonant oscillator strength $\chi_\mathrm{res}$ along with the predicted probe-sample coupling strength $g$ relative to its critical value $g_\mathrm{crit}$.  f) Coupling depends on the probe-sample distance $d/a$ relative to probe radius; dashed curves track evolution of ``non-resonantly screened" probe antenna resonance $\omega_P$ and symmetric STO polariton $\omega_S$, while solid curves show the strongly coupled nano-gap polaritons $\omega_\pm$ with ``avoided crossing" near $d/a\approx0.2$; $\omega_\pm$ are colored according to dominant character of the polariton.  Background colormap predicts the phase of the probe-scattered field that might be spectroscopically recorded versus gap distance, with g) an exemplary spectrum at the configuration of strongest coupling; $\Omega\gamma$ quantifies the coupling strength relative to characteristic linewidth.}
\end{figure}

Strong coupling between resonant subsystems can enable manipulation of the combined system response without any modification to its constituent elements, making this phenomenon an appealing tool for engineering modified states of matter\cite{schlawin_cavity_2022}.  The goal of this section is to quantitatively predict realizable experiments of strong coupling between a nanoscopy probe and fundamental excitations of a quantum material like the ``quantum paraelectric" SrTiO$_3$, the result of which could realize proposed enhancement of unconventional superconductivity\cite{keren_cavity-altered_2026}.  First, we revisit our formal description of strong-coupling:  As discussed in Secs. \ref{sec:configurationalresonance}-\ref{sec:strongcoupling} and there demonstrated for the simplest models of an optical cavity (the etalon and point-dipole model), the dual-system response function Eq. \ref{eq:gcomposite2} in the non-perturbative regime can accomplish strong-coupling either between its dual elements, or with an added optical emitter, when modal quantities exceed quantifiable thresholds (Appendix \ref{app:fabryperotandpointdipole} and Eqs. \ref{eq:coupledpoles}-\ref{eq:coupledpoles2}).  We now explore how the \textit{EigenProbe} formalism provides an answer to the quantitative question of electromagnetic strong coupling, in particular, between a nanoscopy probe and a targeted material surface.  Here we harness the fact that quantitative description of the point dipole model (Eq. \ref{eq:pointdipolemodel}) and the full multi-modal response of the probe-sample system (Eq. \ref{eq:PSScatteringlocal}) are structurally identical, with $\chi_\nu \equiv \rho_\nu(d,\omega)^{-1}$ playing the role of distance $d$- (and possibly energy $\omega$-)dependent probe susceptibility to eigenmode $|E_\nu)$.  Whereas the point dipole model (Eq. \ref{eq:pointdipolemodel}) conventionally describes the non-resonant response of a small polarizable \textit{e.g.} metallic sphere, there is no reason in practice why the susceptibility $\chi(\omega)$ could not display resonance, and in this case might be harnessed to mediate strong coupling to a fundamental excitation (\textit{e.g.} surface polariton) of a material described by resonant $\beta(\omega)$.  For concreteness, we consider when $\chi$ describes the lowest order eigenmode $|E_{\nu=0})$ of a slender probe like that in Fig. \ref{fig:3}c for which antenna resonances imprint strongly upon the susceptibility.

If we approximate $\chi$ and $\beta$ by the harmonic form given in Eq. \ref{eq:resonance} with oscillator strengths $f_\alpha$, resonant energies $\omega_\alpha$, linewidths $\gamma_\alpha$ (with $\alpha \equiv P,S$), and non-resonant parts $\chi_0,\,\beta_0$, respectively, then normal modes of the coupled system are ``nano-gap polaritons" dressed by the resonant probe and occur at energies $\omega_\pm$ (\textit{cf.} Eq. \ref{eq:coupledpoles}):

\begin{gather} \label{eq:coupledpolesPS}
    \left(\bar{\omega}_P^2-i \gamma_P \omega_\pm-\omega_\pm^2 \right)
    \left(\bar{\omega}_S^2-i \gamma_S \omega_\pm-\omega_\pm^2 \right)-g^2=0 
    \\
    \text{with} \quad \bar{\omega}_P\equiv \sqrt{1-\mathcal{N} f_P \beta_0}\, \omega_S, \quad \bar{\omega}_S\equiv \sqrt{1-\mathcal{N} f_S \chi_0}\, \omega_S, \nonumber \\
    \text{and} \quad 
    g^2 \equiv \mathcal{N}f_P f_S \cdot \omega_P^2\omega_S^2 \quad \text{with} \quad \mathcal{N}\equiv \frac{1}{1-\chi_0\beta_0}. \nonumber
\end{gather}

\noindent Here $\bar{\omega}_\alpha$ represent the probe or surface polariton resonance frequencies nonresonantly screened by the presence of the other system.  Eq. \ref{eq:coupledpolesPS} describes a system of coupled oscillators under coupling strength $g$ whereby $\omega_\pm$ will show hybridization (deviation from $\bar{\omega}_\alpha$) when $g^2 \gg g_\mathrm{crit}^2 \equiv (\gamma_P-\gamma_S)^2/4 \cdot \bar{\omega}_{PS}^2$, with $\bar{\omega}_{PS}$ the average of $\bar{\omega}_{P}$ and $\bar{\omega}_{P}$ \cite{dolfo_damping_2018}.   When $\delta \omega^2 \equiv (\bar{\omega}_P-\bar{\omega}_S)^2$ is small, this condition signifies the onset of ``avoided crossing" and strong-coupling characterized by resonant energy exchange between the probe and surface at Rabi frequency $\omega_+-\omega_- \equiv \Omega=\sqrt{g^2-g^2_\mathrm{crit}}/\bar{\omega}_{PS}$.  As defined earlier (Eq. \ref{eq:strongcouplingsimple}), \textit{strong coupling} is here established\cite{flick_strong_2018,simpkins_control_2023} when squared mode-splitting exceeds the typical squared resonance linewidth $\langle \gamma^2 \rangle$: 
\begin{equation} \label{eq:strongcouplingconditionPS}
    \Omega^2> \langle \gamma^2 \rangle
    \quad \text{or equivalently} \quad
    g/g_\mathrm{crit} > 1 + \langle \gamma^2 \rangle/\bar{\omega}_{PS}^2.
\end{equation}
\noindent We now emphasize that the coupling strength depends manifestly on the probe-sample gap through geometric structure of the eigenmode and its susceptibility $\chi(d)$ (\textit{e.g.} Fig. \ref{fig5}b), and $\mathcal{N}$ provides opportunity for \textit{configurational enhancement} of the strong coupling, a vestige of the \textit{configurational resonance} or nano-gap polariton already discussed in Sec. \ref{sec:probegapcavitypolariton} where the probe was non-resonant.  Lastly, Appendix \ref{app:strongcoupling} describes how the strongly coupled nano-gap polariton normal modes at $\omega_\pm$ can be described through two amplitudes $\sigma_{P,S}$ as a coherent superposition of a fundamental (\textit{e.g.} antenna) mode on the probe and a surface polariton.\cite{dolfo_damping_2018}

To demonstrate how strong coupling may arise in practice between an antenna-resonant nanoscopy probe and a confined surface polariton, we consider the experimentally realizable case of a long slender THz-resonant probe (a version of that in \ref{fig:3}a lengthened ten-fold to $L\approx 200$ microns, similar to that applied for resonantly enhanced THz s-SNOM \cite{mastel_terahertz_2017}) coupled to a remarkable low-energy excitation of the quantum paraelectric SrTiO$_3$ (STO, as explored earlier with nano-spectroscopy at higher infrared energies).  The lowest energy optical phonon\cite{peng_room-temperature_2017} in bulk crystals of STO softens incompletely to zero with decreasing temperature,\cite{sirenko_soft-mode_2000} signifying an incipient ferroelectric phase ``frustrated" by quantum fluctuations, and most conventionally triggered only through epitaxial, elastic, or plastic strain.\cite{xu_strain-induced_2020,hameed_enhanced_2022,wang_multiferroicity_2024}  This soft mode has been implicated in STO's unconventionally robust superconductivity compared with its low carrier densities \cite{gastiasoro_superconductivity_2020}  and has been target of recent efforts to condense ferroelectricity from the ``quantum paraelectric" of unstrained crystals, whether through direct high-field THz excitation of the soft mode into an anharmonic regime \cite{li_terahertz_2019} or by triggering a metastable ``optically strained" lattice \cite{nova_metastable_2019}.  So-called ``cavity-quantum electrodynamic" modification of the STO soft mode has been proposed as a means to enhance superconductivity \textit{even without} external drive, mediated rather through molding the spectrum of vacuum fluctuations of the electromagnetic field near STO, as \textit{e.g.} when a layer is placed into resonance within an etalon cavity\cite{latini_ferroelectric_2021}.  Here we consider replacing the etalon by a nano-gap cavity formed with a THz-resonant nanoscopy probe, with analogous impact on the spectrum of emergent electromagnetic modes, which are no more than poles of the coupled response function $\hat{S}_{PS}$ that our eigenmode description fully quantifies (Eq. \ref{eq:PSScattering}). 

Shown schematically by Fig. \ref{fig10}a, our proposal couples the lowest energy antenna resonance and lowest order prove-cavity eigenmode of a long metallic probe to the THz-range surface optical phonon of an unstrained thin ($\sim10$ nm) film or single-crystal membrane of STO.\cite{varshney_hybrid_2024,xu_highly_2024,lukaskawcez_interfacial_2025}.  This STO geometry is selected to deliver a polariton associated with the established soft mode ($\omega_{TO}\approx 2.5$ at room temperature\cite{matsumoto_measurement_2011}) into THz-scale energies by ``splitting" the surface optical phonon into symmetric $\omega_S(q)$ and antisymmetric $\omega_A(q)$ ``membrane polaritons" that disperse with in-plane momentum $q$ positively and negatively, respectively, above and below $\omega_\mathrm{SO}$, as recently described by \textit{Lukaskawcez et al.}\cite{lukaskawcez_interfacial_2025} and observed directly with s-SNOM for the highest energy STO optical phonon\cite{lukaskawcez_interfacial_2025,xu_highly_2024}.  Fig. \ref{fig10}b-c present the charge modes and dispersions (surface reflectance $|r_p(q,\omega)|$) associated with these polaritons, emphasizing the intersection of $\omega_S(q)$ with excitation momenta $a^{-1}$ characteristic of probe-cavity eigenmodes.  We select large a probe radius $a\approx 500$ nm whose great disparity with the STO thickness is intended to reduce the values of $\omega_S(q)$ important to our problem to the sub-THz regime where ``polariton softening" might be most impactful.  Predicted by solving Eq. \ref{eq:generalizedeigenvalue} over a span of relevant probe-surfaces distances $d$ and energies, Fig. \ref{fig10}d shows the (imaginary part of the) probe susceptibility $\chi \equiv \rho_{\nu=0}(d,\omega)^{-1}$ with strong antenna resonance at $\omega_P$ fixed chiefly by the probe length and only weakly dispersive with $d$.

Can this configuration achieve strong probe-sample coupling to soften the lowest energy surface phonon of thin STO?  The answer follows simply from quantifying the constituent resonances that interact according to Eq. \ref{eq:coupledpolesPS}: at each probe-sample distance $d$, we estimate the eigenmode oscillator strength $f_P(d)$ from Fig. \ref{fig10}d and its non-resonant background value $\chi_0$ as reported in Fig. \ref{fig10}e by magenta and black curves, respectively.  The antenna resonance linewidth $\gamma_P\approx 0.15$ THz is quite similar to the polariton linewidth $\gamma_S\approx 0.1$ THz\cite{matsumoto_measurement_2011} and largely sets the threshold constant to achieve probe-polariton strong coupling.  A resonant description of the effective surface reflectivity follows from the single-mode approximation $\beta(d,\omega)\approx \beta_{\nu=0}$ computed with $d-$dependent eigenmodes of the probe via Eq. \ref{eq:genreflectanceq}.  The resulting effective reflectance function is physically understood at each probe-sample gap $d$ by the quantitative overlap between the surface polariton dispersion (Fig. \ref{fig10}c) and the eigenmode scalar potential, and peaks with minimal variation near $\omega_S\approx 1$ THz, as expected (the remaining parameters $f_P\approx 0.5$ and $\beta_0\approx 0$ so that $\mathcal{N}\approx 1$, both nearly constant with $d$, are likewise inferred from this numerical evaluation).  Fig. \ref{fig10}f predicts (dashed curbes) how ``nonresonantly screened resonances" $\bar{\omega}_{PS}$ disperse with probe-sample gap according to Eq. \ref{eq:coupledpolesPS}, showing that even without the antenna resonance, the STO polariton is increasingly screened to lower energies, roughly in proportion to $1/\mathrm{Re}\rho_\nu(d)$, as the metal probe approaches and the eigenmode increasingly confines.

Fortunately, there exists a \textit{configurational condition} $d/a\approx 0.2$ where screened polariton and antenna resonances cross $\delta \omega=0$, and here lies opportunity for strong coupling.  Fig. \ref{fig10}e shows that $g(d)/g_\mathrm{crit}\approx 2$ at this probe-sample gap (predicting a Rabi frequency $\Omega\approx 0.4$ THz) which robustly satisfies the strong-coupling condition (Eq. \ref{eq:strongcouplingconditionPS}), even with our realistic description of both STO and a practicable nano-probe.  Fig. \ref{fig10}f shows (solid curves) how coupled nano-gap polaritons at $\omega_\pm$ disperse with the probe-sample gap.  Each polariton curve is colored according to its superposed character, whether antenna mode-like (blue, $\sigma_P\approx 1$) or surface polariton-like (red, $\sigma_S\approx 1$);  where $\delta \omega\approx 0$, the superposition is nearly equal ($\sigma_P/\sigma_S\approx 1$).  When strong coupling is established, how can it be experimentally confirmed?  To address this question, Fig. \ref{fig10}f renders the phase of the $\nu=0$ eigenmode excitation amplitude (colormap) predicted by Eq. \ref{eq:PSScattering} for this case, which for sSNOM nano-spectroscopy is regarded a phenomenogical measure of local absorption.  The result presents a clear experimental observable for THz spectroscopy\cite{koch_terahertz_2023}  in the near-field, which might directly resolve the Rabi splitting shown in Fig. \ref{fig10}g and, we predict, its strong dependence on probe-sample gap.  Might such an experiment perturb or stabilize ferroelectricity in unstrained thin STO?  Reliable prediction demands a microscopic description of STO similar to recent theoretical studies,\cite{latini_ferroelectric_2021} but is certainly beyond our present scope.  However, the \textit{EigenProbe} formalism clearly supplies needed tools to predict ``strong coupling nanoscopy" of quantum materials like SrTiO$_3$.  The parameter space for further optimization of our particular proposal is expansive, and experimentally addressible by \textit{e.g.} manipulating geometric and antenna characteristics of the probe, or nano-structuring the STO membrane to enhance its modal reflectivity.

\section{Discussion} \label{sec:discussion}

We have shown that a simple formalism of coupled response functions provides a precise and compact route to understand optical coupling between ``pointed probes" and their environment, which highlights optical nanoscopy as an unconventional ``active microscopy" capable of both forming a reconfigurable nano-gap optical cavity\cite{keller_configurational_1993,schlawin_cavity_2022,keren_cavity-altered_2026}  and triggering strongly coupling to target optical systems. \cite{simpkins_control_2023,flick_strong_2018,park_tip-enhanced_2019,dolfo_damping_2018}  Our \textit{EigenProbe} formalism expands upon former work by \textit{Jiang et al.} \cite{Jiang2016} and harnesses a basis of probe-cavity eigenmodes to describe and predict how observables including in-gap electric field, polarization, and scattering of the probe, depend on optical properties of the environment (such as optical constants of a material surface) as well as multi-scale features of the probe, including its geometry, overall length, and near-field proximity to the target of study.  While Appendix \ref{app:eigenmodeproperties} provides in-depth description of how physical features of our formalism emerge from eigenmode solutions to Eq. \ref{eq:generalizedeigenvalue}, our following results are most crucial to optical scattering nanoscopies, including s-SNOM:  Eqs. \ref{eq:PSScattering}-\ref{eq:PSScatteringPerturbative}, Eqs. \ref{eq:genreflectanceq}-\ref{eq:genreflectancer}, Eq. \ref{eq:scatteredamplitude}, and Eq. \ref{eq:demodulation}.  As prescribed by the \textit{EigenProbe imaging} approach, generalizing our formalism to the regime of inhomogeneous environments allows describing near-field imaging experiments.  We have shown that Eqs.  \ref{eq:rdependentscattering}-\ref{eq:genreflectanceimaging} enable rapid and accurate numerical prediction of optical scattering contrasts even in seminal s-SNOM imaging experiments \cite{Fei2012,chen_optical_2012,rizzo_charge-transfer_2020,Ni2018,gerber_phase-resolved_2014} where non-local surface responses like graphene plasmons both complicate and enrich the interpretation of images.  These findings expand upon and refine previous approaches to simulate nanoscopy imaging experiments\cite{xu_deep_2021,rizzo_nanometer-scale_2022,chen_rapid_2022,jing_phase-resolved_2023,jing_terahertz_2021} that have largely approximated s-SNOM scattering as a measure the local density photonic density of states (PDOS).  By comparison, \textit{EigenProbe} predicts how a multi-modal ``generalized PDOS" (Eq. \ref{eq:genreflectancer} governs local probe-sample interactions, with additional and experimentally realizable emergence of \textit{configurational resonances}, or \textit{nano-gap polaritons}, that arise uniquely from non-perturbative probe-sample coupling.  Future work should harness our formalism to i) quantify how probe geometry and other experimental parameters might be tuned to refine performance of nanoscopic imaging, especially towards achieving ultimate spatial resolution and sensitivity approaching the single-molecule level,\cite{mastel_terahertz_2017,xu_pushing_2012} and ii) predict measurables by nanoscopies beyond s-SNOM, such as optical forces detected by PIFM \cite{shcherbakov_photo-induced_2025} and field-enhanced tunneling currents recorded by THz-STM.\cite{cocker_ultrafast_2013}

Our formalism provides a pathway towards \textit{real-time} quantitative interpretation and \textit{extraction} (inversion) of \textit{e.g.} nano-resolved optical constants from materials through optical nanoscopy experiments, of particular importance for spectroscopic s-SNOM, nano-FTIR, and SINS.  Empowered by a simple calibration scheme we propose to elucidate essential properties of the probe, we find that \textit{EigenProbe} predictions of spectroscopic nanoscopy compare quantitatively with newly presented experimental findings by table-top and synchrotron-based s-SNOM, and with it have rationalized some cautionary features that arise in practical experiments, including manifest energy dependence of the probe response (as might be anticipated from an antenna) and the impact of surface far-field scattering in remote-normalzied nanoscopy.  As we have demonstrated even for strongly non-perturbative nanoscopies of infrared phonons in SiC and SrTiO$_3$, our predictions rigorously support widespread adoption of so-called \textit{self-normalized} nanoscopy \cite{McLeod2021,chen_near-field_2022,mester_high-fidelity_2022} as a bullwark against these challenges and a robust route to quantitative optical nano-metrology.  We have shown how the compact eigenmode description of probe-sample scattering enables rapid Kramers-Kronig-constrained\cite{kuzmenko_kramerskronig_2005} extraction of optical constants from nano-spectroscopies of media spanning the strongly resonant (lattice phonons; \textit{e.g.} ionic crystals) to weakly resonant (molecular vibrations; \textit{e.g.} polymers) regimes, and with a combination of accuracy and speed far outstripping previous approaches to ``nano-scale ellipsometry"\cite{McLeod2014,govyadinov_quantitative_2013,govyadinov_recovery_2014,hillenbrand_complex_2000} (nano-metrology of optical constants).  We envision that future work should extend these capabilities to the self-normalized nanoscopies advocated above, and harness further speed-ups available to inversion in the perturbative regime characteristic of polymer nanoscopies, and foreseeably attained through physics-informed machine learning models built upon \textit{EigenProbe} training data.\cite{chen_validity_2021,xu_deep_2021}  Finally, we have shown that distilling quantitative details predicted by the \textit{EigenProbe} formalism into an eigenmode description of strong coupling between probe and sample predicts new routes to cavity-control of quantum materials in the nanoscale regime.  Our particular proposal harnesses coupling with the antenna-resonance of probes deployed for THz s-SNOM to manipulate \cite{schlawin_cavity_2022,keren_cavity-altered_2026,flick_strong_2018,simpkins_control_2023} nano-gap polaritons associated with the ``soft polar mode" of SrTiO$_3$\cite{sirenko_soft-mode_2000,peng_room-temperature_2017}.  When quantified through \textit{EigenProbe} modeling, this and similar experiments are likely to carve new paths towards \textit{cavity quantum materials} that can arise solely through passive control of optical environments mediated by pointed probes.

A software implementation of our formalism, enabled by a compact purpose-built boundary element method (Appendix \ref{app:sommerfeld})\cite{gibson_method_2008} and object-oriented interface to expressions presented in this work, is made publicly available for use by nanoscopy researchers and metrologists alike under an open-source license.\cite{mcleod_eigenprobe_2026}

\begin{acknowledgments}
The author acknowledges instructive discussions with Dr. M. Zhang, and software support from Dr. M. Berkowitz and Mrs. C. F. B. Lo and W. Z.-C. Zheng.  This research used resources of the Advanced Light Source, which is a DOE Office of Science User Facility under contract no. DE-AC02-05CH11231.
\end{acknowledgments}

\appendix

\section{Experimental methods} \label{app:experimentaldetails}

Table-top s-SNOM experiments are enabled by coupling a custom-configured scattering-type scanning near-field optical microscope (s-SNOM) based on a commercial atomic force microscope (NX-10 AFM, Park Systems) to a tunable ultrafast infrared laser.  The light source combines a 40 MHz pulsed oscillator and optical parametric amplifier (Primus and Alpha, respectively, Stuttgart Instruments) supplying near-infrared radiation from an optical parametric amplifier to an integrated difference frequency generation stage producing infrared pulses of spectral bandwidth 50 cm$^{-1}$ (350 fs pulses) centered at energies tunable from 600-2000 cm$^{-1}$.\cite{steinle_ultra-stable_2016}  Radiation is monochromated (Stuttgart Instruments) to 5 cm$^{-1}$ bandwidth for imaging experiments. This radiation is focused by nano-positioned parabolic mirror onto a sharp atomic force microscope (AFM) probe (\textit{PtSi-FM} distributed by \textit{NanoAndMore USA}) whose tip radius was calibrated according to the method described in the main text. Probe-scattered light was collected by a nano-positioned (positioners by \textit{Micronics USA}) parabolic mirror and directed into a nitrogen-cooled mercury cadmium telluride photodetector (\textit{Judson Teledyne}). The photovoltage is demodulated (HF2LI, \textit{Zurich Instruments}) at harmonics of the probe tapping frequency to suppress background-scattered radiation unrelated to the probe-sample near-field interaction.  Simultaneous recording of amplitude and phase of the probe-scatted light (phasor) is enabled by an asymmetric Michelson interferometer with reference mirror alternatively phase-modulated with a piezoelectric actuator for nano-scale imaging (time-resolved pseudo-heterodyne detection \cite{ocelic_pseudoheterodyne_2006}) or scanned by a voice-coil stage (scanDelay 50, \textit{APE}) for nano-scale Fourier transform infrared (nano-FTIR) spectroscopy\cite{huth_nano-ftir_2012} recorded at 3 cm$^{-1}$ resolution.  ``Line-scan" nano-spectroscopy was enabled by positioning the AFM probe along a pre-defined path of pixels, typically from a membrane edge to its interior, and at each position recording a probe-scattered spectrum by nano-FTIR.  Owing to the limited energy span at a single tuning of our light source, to record a broadband spectrum line-scan, the line-scan path is re-traced several times at incremental tunings of our laser (\textit{e.g.} 5 tunings).  The resulting complex-valued spectra are simply co-added to yield a broadband measurement of nano-reflectance and absorption. This approach leverages the ultra-high long-term positional stability of our AFM system ($<50$ nm positional drift per hour) and spectral stability of our ultrafast light-source.\cite{steinle_ultra-stable_2016}

Far- and near-normalized demodulated probe-scattering spectra were collected from commercially obtained samples of 6H-SiC, SrTiO$_3$, and Kapton with a $100$ nm-thick surface layer of gold transferred to select regions of their surface.  This gold layer was produced by sputtering gold onto a commercial silicon wafer, peeling gold from the silicon surface with a layer of thermal release tape (\textit{Nitto Denko Corporation}), placing the tape-with-gold onto the target surface upon a hot plate, heating to 150$^\circ$ C, and removing the tape.  This method allowed reliable transfer of 100 micron-scale conformally adhered gold ``islands" to the studied surfaces, each with sharply defined gold edges.  Near- and far-normalized spectra were acquired by measuring within 5 microns of such gold interfaces or 50 microns away from them, respectively.

Synchrotron-based infrared nano-spectroscopy (SINS) measurements were performed with a commercial scattering-type near-field microscope (Neaspec GmbH, Germany) located at beamline 2.4, Advanced Light Source (ALS), Berkeley, USA. ALS provides broadband IR radiation of high illuminance \cite{BECHTEL2020100493}. The AFM was operated in tapping mode with 80nm oscillation amplitude, using \textit{ARROW-NCPt} probes (by \textit{NanoWorld}) distributed free of PDMS or other polymer contamination, which was important for the sensitive ultra-broadband measurements conducted here. Detectivity down to $200$ cm$^{-1}$ was enabled using a custom-built liquid helium-cooled HgCdTe detector at the beamline.\cite{wehmeier_ultrabroadband_2023}

\section{Reciprocity relations, Poynting's theorem, and far-field radiation} \label{app:reciprocity}

While results of this section are presented in any standard text for classical electromagnetism\cite{jackson1999classical,novotny2012principles}, we state them here for ease of reference.  We begin by stating Maxwell's equations for the time-harmonic electric field $\bm{E}_n$, driven in a space defined by permittivity $\varepsilon(\omega,\bm{r})$ by a current density $\bm{j}_n$ at frequency $\omega$, as an inhomogeneous ordinary differential equation with implied boundary conditions of outgoing radiation at infinite distance.  This is the

\begin{equation}\label{eq:jinhomog}
    \text{\textit{Maxwell problem:}}\quad
    \frac{4\pi i \omega}{c^2} \bm{j}_n = -\left( \frac{\omega^2}{c^2}\varepsilon(\omega,\bm{r})-\nabla \times \nabla \times \right) \bm{E}_n.
\end{equation}

\noindent Below we suppress the $\omega$- and $\bm{r}$-dependence of the permittivity.  We thus obtain reciprocity relations for the field $\bm{E}_n$ with a secondary field $\bm{E}_m$ through their product in a volume $\Omega$:

\begin{align}
    \frac{4\pi i \omega}{c^2} \int_\Omega dV\, \bm{j}_n \cdot \bm{E}_m
    =& -\int_\Omega dV\, \bm{E}_m \cdot \left( \frac{\omega^2}{c^2}\varepsilon-\nabla \times \nabla \times\right) \bm{E}_n \\
    =& -\frac{\omega^2}{c^2} \int_\Omega dV\,\varepsilon \bm{E}_n \cdot \bm{E}_m - \int_{\partial\Omega} d\bm{A}\cdot (\bm{E}_m \times \nabla \times \bm{E}_n) \\
        & \qquad \qquad +\int_\Omega dV \left(\nabla \times \bm{E}_n\right) \cdot \left(\nabla \times  \bm{E}_m\right) \\
    = & -\frac{\omega^2}{c^2} \int_\Omega dV \left(\varepsilon \bm{E}_n \cdot \bm{E}_m + \bm{B}_n \cdot \bm{B}_m \right)
        - \frac{i\omega}{c} \int_{\partial\Omega} d\bm{A} \cdot (\bm{E}_m \times \bm{B}_n)
\end{align}

\noindent Here we have applied Eq. \ref{eq:jinhomog} together with Faraday's law of induction $\nabla \times \bm{E}_n=i\omega/c \bm{B}_n$, and integration by parts, yielding our first reciprocity relation:
\begin{equation}\label{eq:reciprocity}
    -\frac{1}{2} \int_\Omega dV \,\bm{j}_n \cdot \bm{E}_m
        = -\frac{i \omega}{8\pi} \int_\Omega dV \left( \varepsilon\bm{E}_n \cdot \bm{E}_m + \bm{B}_n \cdot \bm{B}_m \right) + \frac{c}{8\pi} \int_{\partial\Omega} d\bm{A} \cdot (\bm{E}_m \times \bm{B}_n)
\end{equation}

\noindent Repeating the preceding steps with the complex-conjugated form of Eq. \ref{eq:jinhomog}, conjugated Faraday's law, and the conjugated trio $\bm{j}_n^*$, $\bm{E}^*_n$, $\bm{B}^*_n$, then yields a second reciprocity relation:
\begin{equation}\label{eq:reciprocityconj}
    -\frac{1}{2} \int_\Omega dV \,\bm{j}_n^* \cdot \bm{E}_m
        = \frac{i \omega}{8\pi} \int_\Omega dV \left(\varepsilon^* \bm{E}_n^* \cdot \bm{E}_m - \bm{B}_n^* \cdot \bm{B}_m \right) + \frac{c}{8\pi} \int_{\partial\Omega} d\bm{A} \cdot (\bm{E}_m \times \bm{B}_n^*)
\end{equation}

\noindent If we take $m=n$, then we should understand $\mathbf{j}_n$ as the ``source current" for fields $\mathbf{E}_{n},\,\mathbf{B}_{n}$ appearing in the equality.  Taking the real part Eq. \ref{eq:reciprocityconj}, we can associate terms with contributions to the total time derivative $\partial_t \mathcal{U}$($=0$ in steady state) of electromagnetic energy $\mathcal{U}$ in volume $\Omega$.  From left to right, respectively, terms identify in the most typical case with energy $(\partial_t \mathcal{U})_\mathrm{curr}>0$ added injected into the volume by source currents, the energy absorbed $-(\partial_t\mathcal{U})_\mathrm{abs}>0$ by media (typically with $\mathrm{Im}\,\varepsilon>0$), and the rate of energy loss $-(\partial_t \mathcal{U})_\mathrm{rad}>0$ due to radiation through the surface $\partial \Omega$ (\textit{viz.} the Poynting vector).  (Scenarios are possible where any terms in Eq. \ref{eq:reciprocityconj} are positive or negative, although equality maintains energy balance.)   Importantly, applying the ``bra-ket" notation introduced in the main text, Eq. \ref{eq:reciprocityconj} indicates the total energy radiated by a current $|j_n)$ (into its ``scattered field" $|E_n)$) is calculated simply (taking $\Omega$ to enclose the current and exclude its environment) by
\begin{equation} \label{eq:braketradiation}
    -\frac{1}{2}(j_n^*|E_n) = -(\partial_t \mathcal{U}_n)_\mathrm{rad}.
\end{equation}

\noindent Another instructive scenario is where $\varepsilon$ is taken as real and all conduction currents induced in media are accounted in the left-most volume integral of Eq. \ref{eq:reciprocityconj} through $\bm{j}=\bm{j}_{\mathrm{ext}} + \sigma \bm{E}$, with $\bm{j}_{\mathrm{ext}}$ external (driving) currents and $\sigma$ the complex conductivity of material media.  In this case the volume integral over $\varepsilon$ is pure imaginary and the remaining real parts of Eq. \ref{eq:reciprocityconj} express the energy balance of Poynting's theorem, with $\mathrm{Re}\,\sigma$ mediating Ohmic loss:
\begin{equation} \label{eq:ohmicloss}
-\frac{1}{2} \,\mathrm{Re}\hspace{-3pt} \int_\Omega\, dV \bm{j}^*_{\mathrm{ext}} \bm{E} = \frac{1}{2}\int_\Omega dV\, \mathrm{Re}\,\sigma |\bm{E}|^2 + \frac{c}{4\pi}\,\mathrm{Re}\hspace{-3pt} \int_{\partial\Omega} \hspace{-5pt} d\bm{A} \cdot \left( \bm{E} \times \bm{B}^*\right).
\end{equation}

Next, interchanging indices $m$ and $n$ in Eqs. \ref{eq:reciprocity} and \ref{eq:reciprocityconj} provides two equally valid reciprocity relations; subtracting these from the original relations after complex-conjugating Eq. \ref{eq:reciprocityconj} results in:

\begin{gather}
    \int_\Omega dV \left( \bm{j}_n \cdot \bm{E}_m - \bm{j}_m \cdot \bm{E}_n\right)
        = \frac{c}{4\pi} \int_{\partial \Omega} d\bm{A} \cdot \left(\bm{E}_n \times \bm{B}_m-\bm{E}_m \times \bm{B}_n\right) \label{eq:reciprocity2} \\
    \int_\Omega dV \left( \bm{j}^*_n \cdot \bm{E}_m - \bm{j}^*_m \cdot \bm{E}_n\right)
        = \frac{c}{4\pi} \int_{\partial \Omega} d\bm{A} \cdot \left(\bm{E}^*_n \times \bm{B}_m-\bm{E}^*_m \times \bm{B}_n\right). \label{eq:reciprocityconj2}
\end{gather}

\noindent Clearly, when $\bm{j}_{n,m}$ are bounded and $\partial\Omega$ is taken to infinity where $\bm{E}_{n,m}=\bm{B}_{n,m}$, the right hand side of Eq. \ref{eq:reciprocity2} vanishes, and reciprocity can be written as $(j_n|E_m)=(j_m|E_n)$, which we use throughout the main text.  A particular application of Eq. \ref{eq:reciprocity2} enables calculating the radiation pattern from a source distribution $\bm{j}_P$, as from an optical nanoscopy probe serving as an antenna.  Firstly, we highlight a result from geometric optics stating that the far-field $\bm{E}_\infty(\bm{r}_0)$ measured at location $\bm{r}_0$ distant (in comparison to to the light wavelength) from its source current distribution at the origin is determined solely by the Fourier (\textit{viz.} plane wave) component $\tilde{\bm{E}}(\bm{k}_0)$ of the field directed with wave-vector $\bm{k}_0 \parallel \bm{r}_0$ according to \cite{novotny2012principles}:

\begin{equation} \label{eq:geometricoptics}
    \tilde{\bm{E}}(\bm{k}_0) = \frac{i r_0 e^{-ik r_0}}{2\pi k} \bm{E}_\infty(\bm{r}_0)
\end{equation}

\noindent This expression applies when the direction of $\bm{k}_0$ is expressed with polar angles $\varphi,\theta$ whereby the radiated power emitted into a solid angle $\Delta \Omega$ is proportional to $k^2 \int_{\Delta \Omega} d\varphi d\theta \sin\theta \,|\tilde{\bm{E}}(\varphi,\theta)|^2$ and the total field collected (and perhaps later focused) by an aperture over the same solid angle is proportional to $k\int_{\Delta \Omega} d\varphi d\theta \sin\theta\,\tilde{\bm{E}}(\varphi,\theta)$.  In either case, $\bm{E}_\infty(\bm{r}_0)$ can be evaluated with Eq. \ref{eq:reciprocity2} taking $\bm{j}_n=\bm{j}_P$ and $\bm{j}_m=i\omega\,\hat{\bm{p}}\, \delta(\bm{r}_0)$, whereby
\begin{equation} \label{eq:radiationreciprocity}
    i\omega \,\hat{\bm{p}} \cdot \bm{E}_\infty(\bm{r}_0) =  \int \hspace{-3pt} dV \,\bm{j}_P \cdot \bm{E}_m.
\end{equation}

\noindent Here we have taken the surface integral of Eq. \ref{eq:reciprocity2} to such distances from the current distributions that it vanishes due to the equality of electric and magnetic far-fields: $\bm{B}_{m,n}=\bm{E}_{m,n}$.  Meanwhile $\hat{\bm{p}}$ denotes the orientation of our choosing for a distant point dipole.  Field polarizations are constrained to $\hat{\bm{\theta}}$ when $\bm{j}_P$ is axisymmetric about the $z$-axis, and accordingly in this case we select $\hat{\bm{p}}=\hat{\bm{\theta}}$.  Then for $r_0$ a great distance from evaluation points $\bm{r}$ near the origin, $\bm{E}_m$ matches the usual expressions for far-fields radiated by the point dipole:
\begin{equation} \label{eq:dipolefield}
    \bm{E}_m \approx k^2 \frac{e^{i \bm{k}_0\cdot (\bm{r}_0-\bm{r})}}{r_0} \left(\sin\theta \hat{\bm{z}} - \cos\theta \hat{\bm{\rho}}\right) \equiv \frac{k^2}{r_0} \bm{E}_{\mathrm{PW},\theta}(\bm{r})
\end{equation}

\noindent Here $\bm{E}_{\mathrm{PW},\theta}$ denotes the corresponding plane wave of unit amplitude with wave vector $-\bm{k}_0$, and the vector in parentheses is simply $\hat{\bm{\theta}}$ transported to the origin.  Summarizing Eqs. \ref{eq:geometricoptics}, \ref{eq:radiationreciprocity}, and \ref{eq:dipolefield}, the average field radiated into an aperture with solid angle $\Delta \Omega$ can be determined by:
\begin{equation} \label{eq:brightness}
    \langle \bm{E}_{\infty} \rangle_{\Delta \Omega} =  \int_{\Delta \Omega} \hspace{-3pt} \frac{d\varphi d\theta}{\Delta \Omega} \sin\theta  \left(-\hat{\bm{\theta}} \frac{i\omega}{r_0 c^2} \int \hspace{-3pt}  dV\, \bm{j}_P \cdot \bm{E}_\mathrm{PW,\theta} \right)
\end{equation}

\noindent The $\theta$-dependent term in parentheses has units of electric field and can be denoted the ``brightness" $B_{\bm{j}_P}(\theta)$ of the current distribution and is apparently determined by its spatial overlap with a plane wave propagating \textit{from} the observation point at $\bm{r}_0$.  Results in the main text utilize Eq. \ref{eq:brightness} to predict the brightness $B_\nu(\theta)\propto (j_\nu|E_{\mathrm{PW},\theta})$ of a nanoscopy probe's eigenmode currents $|j_\nu)$.  Unsurprisingly, Eq. \ref{eq:brightness} is formally equivalent to the volume integral of (the far-field component of) the Green dyadic function in vacuum with $\bm{j}_P$, but the present form makes explicit the reciprocity between emission and excitation from a corresponding plane wave.

\section{Composite response functions} \label{app:compositeresponsefunction}

The response function $\hat{S}_+$ for a composite of $N$ interacting sub-systems described by response functions $\hat{S}_n$ can be deduced from the following $N$ self-consistency relations:
\begin{equation} \label{eq:gcomponents}
    \hat{S}_n \bm{E}_\mathrm{ext} = \bm{E}_n-\hat{S}_n \sum_{m\ne n}^N \bm{E}_m
\end{equation}

\noindent Informally, Eq. \ref{eq:gcomponents} states that field $\bm{E}_n$ is scattered by $\hat{S}_n$ which acts on all ``external" fields including $\bm{E}_\mathrm{ext}$ together with $\bm{E}_{m\ne n}$. We can systematize Eqs. \ref{eq:gcomponents} as a linear system defined in matrix form, \textit{e.g.} with rank $N=2$ for the probe-analyte system of active microscopy.  Denoting by $\hat{S}_{P,E}=\hat{G}_{P,E}-1$ the two ``bare" scattering operators associated with a ``probe" and ``environment", respectively (and connected to their corresponding Green functions $\hat{G}_{P,E}$ as described in Sec. \ref{sec:coupledresponsefunction}), $\hat{S}^+ \equiv \hat{S}_P^+ + \hat{S}_E^+$ emerges as a sum of ``dressed" scattering operators for which $\bm{E}_{P,E}\equiv \hat{S}_{P,E}^+ \bm{E}_\mathrm{ext}$ and $\bm{E}_\mathrm{scat}=\bm{E}_P+\bm{E}_\mathrm{E}$:
\begin{align}
    \bm{E}_\mathrm{scat}&=
    \begin{pmatrix} \hat{I} \\
                    \hat{I} \end{pmatrix}^T
    \begin{pmatrix} \hat{I} & -\hat{S}_P \\
                    -\hat{S}_E & \hat{I} \end{pmatrix}^{-1}
    \begin{pmatrix} \hat{S}_P \\
                    \hat{S}_E \end{pmatrix} \bm{E}_\mathrm{ext}
                    \label{eq:gcompositematrices} \\
    &=  \left( \frac{\hat{S}_P (1+\hat{S}_E)}
                    {1-\hat{S}_P \hat{S}_E}
            +  \frac{\hat{S}_E (1+\hat{S}_P)}
                    {1-\hat{S}_E \hat{S}_P}
         \right)\, \bm{E}_\mathrm{ext}  \label{eq:gcomposite} \\
    \mathrm{where}\quad &(1-\hat{S}_P \hat{S}_E)\bm{E}_P =\hat{S}_P(1+\hat{S}_E)\bm{E}_\mathrm{ext} \label{eq:SelfConsistentEp}\\
    \mathrm{and} \quad &(1-\hat{S}_E \hat{S}_P)\bm{E}_E =\hat{S}_E(1+\hat{S}_P)\bm{E}_\mathrm{ext} \label{eq:SelfConsistentEe}
\end{align}

\noindent Eqs. \ref{eq:SelfConsistentEp} and \ref{eq:SelfConsistentEe} define the denominators of Eq. \ref{eq:gcomposite} in the sense of inverse operators \textit{pre-multiplying} the numerators.  The composite response function evidently combines the elementary response functions in a nontrivial way.  While this abstract approach clearly generalizes beyond $N=2$, it does so at the expense of rapidly proliferating terms, since we must carefully preserve orderings where operators like $\hat{S}_{P,S}$ need not commute.  However, we can bring focus to $\hat{S}_P$ by noting that $(1-\hat{S}_E \hat{S}_P)^{-1} \hat{S}_E=\hat{S}_E (1-\hat{S}_P \hat{S}_S)^{-1}$ which allows rewriting Eq. \ref{eq:gcomposite} as the

\begin{equation} \label{eq:gcomposite2app}
    \text{\textit{Composite response:}}\quad
    \hat{S}^+ = \hat{S}_E+\left(\hat{S}_E+1\right) \frac{\hat{S}_P}{1-\hat{S}_P \hat{S}_E} \left(1+\hat{S}_E\right).
\end{equation}

\noindent Eq. \ref{eq:gcomposite2app} is the central result of this appendix.

\section{Strong coupling in the composite response function} \label{app:strongcoupling}

Here we present the conditions by which the interacting response functions $\hat{S}_A$ and $\hat{S}_B$, each presenting a resonant character in energy $\omega$, endow their composite response function  $\hat{S}^+$ with entirely new resonant characteristics.  In the scenario denoted "strong coupling", the combined response function $\hat{S}^+$ is described by new poles that did not originally comprise $\hat{S}_{A,B}$.  For a most pedagogical case, consider where $\hat{S}_{A,B}$ both act in a single ``scattering channel", which might describe, \textit{e.g.} 1) the amplitude of specular reflection for a particular plane wave (specified \textit{e.g.} by in-plane wavenumber $q$) from a pair ($A$ and $B$) of stacked and strongly interacting planar media, or 2) the electric field polarized along the polarizable axis of a point dipole ($A$) over a dielectric surface ($B$), both discussed in the main text Sec. \ref{sec:strongcoupling}.  The probe eigenmode description presented in the main text is also a modal expansion for the dressed probe response in such ``independent" scattering channels.  In such cases, employing the notation of Eq. \ref{eq:Gdef}, $\hat{S}_\alpha\approx |E_\alpha) g_\alpha(\omega) (j_\alpha|$ (with $\alpha = A,B$ denoting either of the interacting subsystems) and $g_\alpha(\alpha)$ describe scalar scattering susceptibilities which in the simplest nontrivial case might be described by a non-resonant ``background" plus a resonant response with Lorentzian lineshape at energy $\omega_\alpha$:

\begin{equation} \label{eq:resonance}
    g_\alpha(\omega) \approx \chi^0_\alpha+f_\alpha/\delta_\alpha(\omega) \quad \text{and} \quad \delta_\alpha(\omega)^{-1}\equiv \frac{\omega_\alpha^2}{\omega_\alpha^2-i \gamma_\alpha \omega-\omega^2} \quad \text{with} \quad \alpha\equiv A,B.
\end{equation}

\noindent Here $\chi^0_\alpha$ denotes an impulsive ($\omega$-independent) contribution to the susceptibility, while $f_\alpha$ describes the oscillator strength at resonance with quality factor $Q_\alpha\equiv\omega_\alpha/\gamma_\alpha$ (which differs by $2$ from the conventional definition, for our notational ease), and $\delta_\alpha$ quantifies ``distance from resonance".  Inserting Eq. \ref{eq:resonance} into the dressed probe response function (\textit{viz}. Eq. \ref{eq:gcomposite2}) reveals a spectrum of dressed resonance frequencies $\omega_\pm$ satisfying:

\begin{gather} \label{eq:coupledpoles}
    \left(\bar{\omega}_A^2-i \gamma_A \omega_\pm-\omega_\pm^2 \right)
    \left(\bar{\omega}_B^2-i \gamma_B \omega_\pm-\omega_\pm^2 \right)-g^2=0 
    \\
    \text{with} \quad \bar{\omega}_A\equiv \sqrt{1-\mathcal{N} f_A \chi_B}\, \omega_A, \quad \bar{\omega}_B\equiv \sqrt{1-\mathcal{N} f_B \chi_A}\, \omega_B, \\
    \text{and} \quad 
    g^2 \equiv \mathcal{N}f_A f_B \cdot \omega_A^2\omega_B^2 \quad \text{with} \quad \mathcal{N}\equiv \frac{1}{1-\chi_A\chi_B} \label{eq:coupledpoles2}
\end{gather}

\noindent Here $\bar{\omega}_{A,B}$ represent the resonance frequencies of each of the two subsystems screened by the nonresonant response of the other, and we should define $\bar{\omega}_{AB}\equiv (\bar{\omega}_A+\bar{\omega}_B)/2$ as their average.  Eq. \ref{eq:coupledpoles} represents a system of coupled oscillators under coupling strength $g$ whereby $\omega_\pm$ will show strong hybridization (deviation from $\bar{\omega}_{A,B}$) when $g^2 > g_\mathrm{crit}^2 \equiv (\gamma_A-\gamma_B)^2 \bar{\omega}_{AB}^2/4$.\cite{dolfo_damping_2018}  This condition signifies the onset of strong coupling characterized by resonant energy exchange between the subsystems $A$ and $B$ at frequency $\Delta \omega = \omega_+-\omega_-$ when $\Delta \omega > (\gamma_A+\gamma_B)/2$.  The occurrence of finite Rabi frequency $\Delta \omega\equiv \Omega=\sqrt{g^2-g^2_\mathrm{crit}}/\bar{\omega}_{AB}$ even when $\bar{\omega}_A \approx \bar{\omega}_B$ signifies an ``avoided crossing" that in optical contexts is denoted electromagnetically-induced transparency (EIT).\cite{fleischhauer_electromagnetically_2005}

The relevant oscillator degrees of freedom internally responsible for Eq. \ref{eq:coupledpoles} are identifiable from the linear system of interacting subsystems $A,B$ whose normal modes comprise $\omega_\pm$.  Consider an external ``drive" field $|E_\mathrm{ext})$ together with the fields $|\mathcal{E}_{A,B})$ generated by the subsystems' response functions.  Together these are:
\begin{eqnarray}
     |E_\mathrm{tot})&=&|E_\mathrm{ext})+|\mathcal{E}_{A})+|\mathcal{E}_{B}) \label{eq:coupledsystemsAandB}\\
    \text{with}\quad |\mathcal{E}_A) &=& |E_A) g_A(\omega) (j_A| \bigg(|E_\mathrm{ext}+|\mathcal{E}_{B})\bigg) \label{eq:coupledsystemsAandB1} \\
    \text{and}\quad |\mathcal{E}_B) &=& |E_B) g_B(\omega) (j_B| \bigg(|E_\mathrm{ext}+|\mathcal{E}_{A})\bigg) \label{eq:coupledsystemsAandB2}
\end{eqnarray}
\noindent The scalar products $\sigma_A\equiv(j_A|\mathcal{E}_B)$ and $\sigma_B\equiv (j_B|\mathcal{E}_A)$ are sufficient to reconstruct the total field $|E_\mathrm{tot})$ from Eq. \ref{eq:coupledsystemsAandB} only once determined from the linear system defined by Eqs. \ref{eq:coupledsystemsAandB1}-\ref{eq:coupledsystemsAandB2}:

\begin{equation}
\left( \begin{matrix}
    1 & -(j_B|E_A) g_A(\omega) \\
    -(j_A|E_B) g_B(\omega) & 1
\end{matrix} \right)
\left( \begin{matrix}
    \sigma_A \\
    \sigma_B
\end{matrix} \right)
    = \left( \begin{matrix} 
    (j_B|E_A) g_A(\omega) (j_B|E_\mathrm{ext}) \\
     (j_A|E_B) g_B(\omega) (j_A|E_\mathrm{ext})
     \end{matrix} \right)
\end{equation}

\noindent Normal modes correspond to the case $|E_\mathrm{ext})=|0)$ and, after subsuming the inner products $(j_B|E_A)$ and $(j_A|E_B)$ into numeric definitions of $\chi_A^0,\,f_A$ and $\chi_B^0,\,f_B$ (\textit{c.f.} Eq. \ref{eq:resonance}), respectively, the linear system becomes:

\begin{equation} \label{eq:coupledsystem}
\left( \begin{matrix}
    1 & -\left(\chi_A^0+f_A/\delta_A(\omega)\right) \\
    -\left(\chi_B^0+f_A/\delta_B(\omega)\right) & 1
\end{matrix} \right)
\left( \begin{matrix}
    \sigma_A \\
    \sigma_B
\end{matrix} \right)
    = 0.
\end{equation}

\noindent A solution requires vanishing determinant of the matrix at left, leading immediately to Eq. \ref{eq:coupledpoles}.  We interpret $\sigma_{A,B}$ as self-consistent ``polarizations" of subsystems $A$ and $B$ that self-sustain indefinitely only at frequencies $\omega=\omega_\pm$ and which ``source" the self-consistent field (\textit{i.e.} cavity resonance) given by $|E)=|\mathcal{E}_A)+|\mathcal{E}_B)$.

Provided that $\bar{\omega}_A\sim\bar{\omega}_B$, see immediately that the conditions for observing the mode-splitting characteristic of strong-coupling of a probe with its environment (or generically within any pair of interacting systems) requires that the Rabi splitting exceed the mode-split linewidth \cite{bylinkin_-chip_2024,dolfo_damping_2018,simpkins_control_2023}, which demands sufficient oscillator strength:
\begin{gather}
    \sqrt{g^2/\bar{\omega}_{AB}^2-\frac{1}{4} \big|\gamma_A-\gamma_B\big|^2}>\left(\gamma_A+\gamma_B\right)/2 \nonumber \\
    \text{so that} \quad f_A f_B \gtrsim \frac{1}{2} \left(Q_A^{-2}+Q_B^{-2}\right) \left(1-\chi_A\chi_B \right).
    \label{eq:criteria2}
\end{gather}
\noindent Eq. \ref{eq:criteria2} holds when $f_A f_B \ll 1$; otherwise, $g^2$ scales sub-linearly in the product of oscillator strengths, which makes strong coupling more difficult to attain when simultaneously $\chi_{A,B}$ are nonzero and quality factors are small. In the case that $\chi_A\chi_B \ll 1$, the right-hand side is evidently the reciprocal ``average" quality factor squared, $Q_{AB}^{-2}$.  Thus, the condition $f_A f_B\, Q_{AB}^2=\mathrm{max}_{\,\omega}\, G_A G_B >1$ coincides with the non-perturbative regime defined by those energies $\omega$ where $|G_P G_E|>1$.  Eqs. \ref{eq:coupledpoles}, \ref{eq:coupledpoles2}, \ref{eq:criteria2}, and the accompanying expression for $\Omega$ are the central results of this appendix.

\section{Conditions for strong coupling of an emitter to a Fabry-P\'{e}rot or probe-surface optical cavity.} \label{app:fabryperotandpointdipole}

Sec. \ref{app:strongcoupling} identified conditions for strong coupling between two subsystems.  Here we demonstrate application of those conditions to the two practical examples shown in Fig. \ref{fig1}b-c and discussed in Sec. \ref{sec:configurationalresonance}:  1) the two-mirror Fabry-P\'{e}rot cavity (``FP"), and 2) a polarizable point-dipole over a resonantly reflective surface (``PD", corresponding \textit{e.g.} the point-dipole model of probe-sample near-field interaction), each coupled to a point-like optical emitter.  We model the latter by its own resonant response function $G_e$ of the form shown in Eq. \ref{eq:resonance}, denoted by $\alpha = e$, with oscillator strength $f_e$, emission energy $\omega_e$, and narrow linewidth $\gamma_e \sim 0$, with $\chi_e=0$ for simplicity.  This demonstration requires to first identify resonant response functions $G_\mathrm{FP}$ and $G_\mathrm{PD}$ for 1) and 2), respectively (as composited from their constituent elements by Eq. \ref{eq:gcomposite}), then to identify their important parameters (the poles, their linewidths, and oscillator strengths), and finally to identify the conditions where ``optical cavities" 1) and 2) enable manipulation of the emitter.  The latter conditions coincide with those for (ideally controllably) ``strong coupling" and thus also emergent characteristics from the emitter-cavity response function $G_+$ that composites $G_\mathrm{FP}$ and $G_\mathrm{PD}$ (Eq. \ref{eq:gcomposite}).  We will consider an emitter with resonant response 

We begin with 1), the Fabry-P\'{e}rot cavity, comprising two parallel mirrors with reflectivities $r_{1,2}$ separated by a distance $d$ as shown in Fig. \ref{fig1}b.  To describe the response function $\hat{S}_\mathrm{FP}$ with a scalar, we consider only the amplitude of only a single plane wave ``mode" $|E_\mathrm{pw})$ with transverse (in-plane) momentum $q$ and longitudinal momentum $k_z=\sqrt{(\omega/c)^2-q^2}$.  Application of Eq. \ref{eq:gcomposite} composites the reflectance of both mirrors and the wave retardation between them to obtain a familiar result for this scalar response function of the cavity,\cite{hecht_optics_2017} which we denote $g_\mathrm{FP}(\omega)$, as defined by $\hat{S}_+\equiv \hat{S}_\mathrm{FP}=|E_\mathrm{pw})g_\mathrm{FP}(\omega)(j_\mathrm{pw}|$, Here $(j_\mathrm{pw}|E_\mathrm{pw})=1$ defines a modal current localized to one of the two mirrors.

\begin{equation} \label{eq:etalonresponse}
    g_\mathrm{FP} = \frac{\epsilon}{1-r_1 r_2 e^{2ik_z d}} \equiv \frac{\epsilon}{\delta (\omega)} \approx \frac{-\epsilon/\partial_\omega \delta}{-\delta_0/\partial_\omega \delta + (\omega_0-\omega)} .
\end{equation}

\noindent Here scalar $\epsilon\equiv e^{i\phi}$ through phase $\phi$ depends only on the longitudinal position where the response field is evaluated within the cavity, and scalar $\delta$ describes ``distance" from resonance (mathematically, proximity to the nearest pole).  The approximation expands $\delta\approx \delta_0 + \partial_\omega \delta \cdot (\omega_0-\omega)$ about (one such) $\omega_0$ where $|\delta|$ is minimized.  Eq. \ref{eq:etalonresponse} has the form of Eq. \ref{eq:resonance} when we identify:

\begin{equation} \label{eq:etalonoscparams}
    \gamma_\mathrm{FP} = -i\frac{\delta_0}{\partial_\omega \delta} \quad \text{and}\quad f_\mathrm{FP}=-\frac{\epsilon}{\omega_0 \partial_\omega \delta}.
\end{equation}

Expressions in Eq. \ref{eq:etalonoscparams} are quite general to a ``single pole" (and "single channel") cavity response function, whence the condition of cavity strong coupling to the emitter follows Eq. \ref{eq:criteria2} as:

\begin{equation}  \label{eq:etalonstrongcoupling1}
    f_e f_{FP} > \frac{1}{2\omega_0^2} \gamma_{FP}^2, \quad \text{implying}\quad -i\epsilon f_e > -i \frac{\delta_0}{\omega_0 \partial_\omega \delta}=Q^{-1}.
\end{equation}

\noindent Here quantities on either side of the inequalities are assumed real-valued, and $Q\equiv \omega_0/\gamma_\mathrm{FP}$ denotes the cavity finesse.  Resonance at $\omega_0$ demands $k_z(\omega_0)=\pi n/d$ for integer $n$, whence $\omega_0=\pi n c/d$ when $q=0$ (for simplicity).  Near resonance then we have $\partial_\omega \delta = -2i d R/c$ where $R\equiv r_1 r_2 = (1-\delta_0)$. The cavity finesse is then $\omega_0/\gamma_\mathrm{FP}=2\pi R/(1-R)$ and the resonance oscillator strength is $f_\mathrm{FP}=-i\epsilon/(2 \pi R)$. The Fabry-P\'{e}rot cavity is unique in the sense that oscillator strength and linewidth are mutually determined by $R$.  After placing the emitter at a position where $\epsilon=+i$ and tuning the cavity to $\omega_0 \approx \omega_e$, cavity strong coupling to the emitter following Eq. \ref{eq:etalonstrongcoupling1} then demands a ``proximity to resonance" smaller than:

\begin{equation} \label{eq:etalonstrongcoupling2}
    \delta_0=1-R < \frac{2\pi f_e}{1+2\pi f_e}.
\end{equation}

\noindent For an emitter with weak oscillator strength (\textit{e.g.} supposing when polarized it emits weakly into the $q$-mode considered here), strong coupling simply demands ever smaller $\delta_0$ and ever larger finesse $Q>f_e^{-1}$.

We next consider 2), the cavity formed by a (non-resonantly) polarizable point dipole (``pd") over a resonantly reflective surface as shown in Fig. \ref{fig1}c, which grossly approximates a near-field probe over surface with resonant reflectivity $\beta(\omega)$.  To render $\hat{S}_\mathrm{pd}$ a scalar response, we consider only a single ``mode", a dipolar field $|E_\mathrm{pd})$ generated by a $z$-oriented polarization of a point-like sphere with polarizability $\alpha_0$ and radius $a<<d$, where $d$ is the height of the sphere above the surface.  The importance part of Eq. \ref{eq:gcomposite2app} is the quotient $\hat{S}_P/(1-\hat{S}_P\hat{S}_E) \equiv |E_\mathrm{pd}) \alpha_\mathrm{pd}(\omega) (j_\mathrm{pd}|$, and $(j_\mathrm{pd}|E)$ essentially measures the amplitude of the $z$-polarized component of a field $|E)$ at the sphere's center, with $(j_\mathrm{pd}|E_\mathrm{pd})\equiv 1$.  The well-known result\cite{Raschke2003,keilmann_near-field_2004,hillenbrand_complex_2000} is:

\begin{equation} \label{eq:pointdipolemodel}
    \alpha_\mathrm{pd}(\omega)=\frac{\alpha_0}{1-\beta(\omega) \chi_0(2d)} \quad \text{with} \quad \chi_0(2d)\equiv \alpha_0\cdot 2/\left(2d\right)^3.
\end{equation}

\noindent Here $\chi_0(2d)\equiv(j_\mathrm{pd}|E_\mathrm{pd}^M)$, where $|E_\mathrm{pd}^M)$ denotes the field from an equivalent ``mirror dipole" positioned distance $d$ below the surface.  Unlike in former literature, we utilize a Gaussian unit system for which a perfectly conducting sphere of radius $a$ has polarizability $\alpha_0=a^3$. As in the main text we consider the case where $\beta(\omega)\approx f_\beta \omega_\beta/(\omega_\beta-\omega-i\gamma_\beta)$ is resonant at energy $\omega_\beta$ with quality factor $Q_\beta=\omega_\beta/\gamma_\beta$.  As described in the main text, Eq. \ref{eq:pointdipolemodel} in this case produces a Fano resonant lineshape.

We can analyze the conditions for strong-coupling to a ($z-$oriented) emitter on the material surface a vertical distance $d$ below the polarizable point dipole (the ``probe") again using the approximation introduced in Eq. \ref{eq:etalonresponse} together with definitions Eq. \ref{eq:etalonoscparams} and the conditions of Eq. \ref{eq:etalonstrongcoupling1}.  To uniformize our treatment of the point dipole with that of the Fabry-P\'{e}rot cavity, we can rewrite Eq. \ref{eq:pointdipolemodel} as 

\begin{equation}
    (j_e|E_\mathrm{pd})^2 \alpha_\mathrm{pd}(\omega) = \frac{\epsilon}{\delta(\omega)} \quad \text{with} \quad \epsilon\equiv \chi_0(d)^2/\chi_0(2d) \quad \text{and} \quad \delta(\omega)\equiv\chi_0^{-1}(2d)-\beta(\omega).
\end{equation}

\noindent Here $|j_e)$ signifies the position (a distance $d_e$ from the point dipole) and orientation of the point-like emitter, and we subsume the dimensionless form factor $(j_e|E_\mathrm{pd})^2=\chi_0(d_e)^2$ into the oscillator strength according to the prescription of Eq. \ref{eq:coupledsystem}. We first identify the resonant frequency $\omega_\mathrm{pd}$ and linewidth $\gamma_\mathrm{dp}$ of the combined ``optical cavity" system that minimizes $\delta(\omega)$ and then identify its effective oscillator strength by Eq. \ref{eq:etalonoscparams} as follows:

\begin{equation} \label{eq:pdoscparams2}
    \omega_\mathrm{pd}=\left(1-f_\beta \chi_0(2d)\right) \omega_\beta,
    \quad \gamma_\mathrm{dp}=\gamma_\beta,
    \quad \text{and} \quad
    f_\mathrm{pd}=\chi_0(d_e)^2 \frac{f_\beta \chi_0(2d)}{1-f_\beta \chi_0(2d)}
\end{equation}

\noindent The Fano lineshape accompanying $f_\mathrm{pd}$ is discussed in the main text.  As with the Fabry-P\'{e}rot cavity, we next consider how the ``optical cavity" formed by Eq. \ref{eq:pdoscparams2} accentuates strong coupling to an emitter over the case where the emitter is placed directly into interaction over the surface with $\beta(\omega)$.  Assuming this cavity ($\omega_\mathrm{pd}$) is brought into resonance with the emitter, the enhancement is identifiable from Eq. \ref{eq:criteria2} as $\left(f_\mathrm{pd}/f_\mathrm{\beta}\right) \left(Q_\mathrm{pd}/Q_\beta\right)^2 = \chi_0(d)\chi_0(2d) \left(1-f_\beta \chi_0(2d)\right)$.  Although increasing $f_\beta$ leads to reduction of cavity  quality factor through $\omega_\mathrm{pd}$, the factor $\chi_0(d)\chi_0(2d)$ represents enhancement of the electric field within the gap between the polarizable dipole and surface.  In summary then, the strong coupling condition for the simplified ``point dipole probe-surface cavity" is:

\begin{equation} \label{eq:pdstrongcoupling}
    2 \chi_0(d_e)^2 \cdot \frac{f_\beta \chi_0(2d)}{1-f_\beta \chi_0(2d)}  \left(\frac{\omega_\mathrm{pd}}{\gamma_\beta}\right)^2 > f_e^{-1}.
\end{equation}

\noindent Here we have assumed that configurational resonance was leveraged to bring $\omega_\mathrm{pd}(d)$ (Eq. \ref{eq:pdoscparams2}) into resonance with the emitter.  Moreover, strong coupling clearly depends on maximizing oscillator strength of the probe-sample cavity by bringing $f_\beta \chi_0(2d)$ as close to unity as possible.  Comparison to Eq. \ref{eq:etalonstrongcoupling2} for the Fabry-P\'{e}rot cavity allows identifying the left hand side of Eq. \ref{eq:pdstrongcoupling} as an effective ``cavity finesse" for the ``probe-surface" configuration, when a single probe-cavity eigenmode is considered.  These equations are the central results of this appendix.  The realistic $d$-dependent value of $\chi_0$ is thus an essential parameter towards forming a practical probe-cavity capable of strong coupling to an emitter with oscillator strength $f_e$ on the material surface.

\section{The quasi-static probe-surface interaction and impedance matrices for a method of moments} \label{app:quasistaticsurface}

Sec. \ref{sec:eigenmodesbasis} develops a basis of eigenmodes $|j_\nu)$ that were carefully chosen to ``most closely" mutually diagonalize the response functions $\hat{S}_{P,S}$ of both a nanoscopy probe and surface, respectively.  The condition for this mutual diagonalization is that eigenmodes solve the generalized eigenvalue problem Eq. \ref{eq:generalizedeigenvalue}, which employs a quasi-electrostatic approximation for surface-scattering of fields from the probe, $\hat{\mathcal{E}}_S^{\mathrm{QS}}/i\omega \equiv \hat{\theta}_{\partial \Omega_P} \hat{M}_z \hat{\mathcal{G}}_0^\mathrm{\,QS}/i\omega $. (Operator definitions are given in Sec. \ref{sec:eigenmodesbasis}.) This particular choice of interaction is not \textit{ad hoc}. In addition to its simplicity ($\omega$-independence), here we show that the operator $\hat{O}\equiv \hat{\mathcal{E}}_S^{\mathrm{QS}}/i\omega$ produces a negative-definite Hermitian inner product equal to the quasi-electrostatic energy of interaction between a current distribution and its mirror image across $z=0$. Specifically, $(j^*|\hat{O}|j)<0$ for any $|j)$ distributed in the half-space $z>0$ which vanishes sufficiently rapidly at infinite distance:

\begin{equation} \label{eq:posdefquestion}
   (j^*|\hat{\mathcal{E}}_S^{\mathrm{QS}}/(i\omega)|j) =\lim_{\omega\rightarrow 0} \int \hspace{-3 pt} dV dV'\; \bm{j}(\bm{r})^H \,\overleftrightarrow{\bm{\mathcal{G}}_0}(\bm{r}|\bm{r}')\, \big( \hat{M}_z\, \bm{j}(\bm{r}')\big) / (i\omega).
\end{equation}

\noindent Here $\bm{j}^H$ is the vector Hermitian transpose of $\bm{j}$ and we have used the fact that $\hat{M}_z$ and  $\overleftrightarrow{\bm{\mathcal{G}}_0}$ are symmetric operators.  Meanwhile, $\overleftrightarrow{\bm{\mathcal{G}}_0}$ denotes the coordinate-space (dyadic) representation of the free-space Green function $\hat{\mathcal{G}}_0$:\cite{novotny_near-field_2006}
\begin{equation}
    \overleftrightarrow{\bm{\mathcal{G}}_0} \equiv \frac{i\omega}{c^2} \left( \overleftrightarrow{\bm{I}}+\frac{1}{k^2}\nabla \nabla^T\right) g_0(\bm{r}
    |\bm{r}') \quad 
    \text{with} \quad \left(\nabla^2 +k^2\right) g_0(\bm{r}|\bm{r}')=-4\pi \delta(\bm{r}-\bm{r}').
\end{equation}
\noindent Differentiation $\nabla$ is taken with respect to coordinate $\bm{r}$, $k\equiv \omega/c$, and $g_0$ is the Green function for the Helmholtz equation that tends to $1/|\bm{r}'-\bm{r}|$ in the limit of zero frequency.  In this limit the vector potential component of Eq. \ref{eq:posdefquestion} (from  $\overleftrightarrow{\bm{I}}$ in the Green dyadic) is irrelevant and, using $\nabla g_0 = -\nabla'g_0$, as well as integration by parts once in each of the integrals over $\bm{r}$ and $\bm{r}'$, we obtain:
\begin{equation}
     (j^*|\hat{\mathcal{E}}_S^{\mathrm{QS}}/(i\omega)|j) = - \lim_{\omega\rightarrow 0} \frac{1}{\omega^2} \int \hspace{-3pt} dV \left(\nabla \cdot \bm{j}(\bm{r})^*\right) g_0(\bm{r}|\bm{r}') \left(\nabla'\cdot \hat{M}_z \bm{j} (\bm{r}') \right).
\end{equation}
\noindent The charge density $\varrho$ associated to $\bm{j}$ satisfies periodic charge  $\nabla \cdot \bm{j} = +i\omega \varrho $, whence we obtain:
\begin{equation} \label{eq:negdefinite1}
    (j^*|\hat{\mathcal{E}}_S^{\mathrm{QS}}/(i\omega)|j) 
    = -\int dV dV' \frac{\varrho(\bm{r})^* \hat{M}_z \varrho(\bm{r')}}{|\bm{r}-\bm{r}'|}
    = -\int dV dV' \frac{\varrho(\bm{\rho},z)^* \varrho(\bm{\rho}',z')}{|\bm{\rho}-\bm{\rho}'|^2+(z+z')^2}
\end{equation}
\noindent which is simply the quasi-electrostatic interaction energy between a charge distribution $\varrho$ and its (sign- and orientation-inverted) mirror image $\hat{M}_z \varrho$ across the plane $z=0$.  The latter equality of Eq. \ref{eq:negdefinite1} separates vector positions $\bm{r}(')$ into in-plane radial vector $\bm{\rho}(')$ and $z(')$ coordinate components.  Integrals over radial coordinates amount to a convolution with the Coulomb kernel and can be replaced by a single integral over in-plane momentum $\bm{q}$.  The in-plane Fourier transforms of $\varrho(\bm{r})$ and $\varrho(\bm{r}')^*$ are $\varrho(\bm{q},z)$ and $\varrho(-\bm{q},z')^*$, respectively, and  the in-plane Fourier transform of our reflected Coulomb kernel is given by $2\pi e^{-|\bm{q}|(z+z')}/|\bm{q}|$.  Integrals over $z(')$ thus become separable:
\begin{equation} \label{eq:negdefinite2}
    (j^*|\hat{\mathcal{E}}_S^{\mathrm{QS}}/(i\omega)|j) = -\int d^2 \bm{q}\, \frac{2\pi}{|\bm{q}|} \left(\int_{z>0} \hspace{-6pt} dz\, e^{-|\bm{q}|z} \varrho(\bm{q},z)\right) \left(\int_{z'>0} \hspace{-6pt}  dz'\, e^{-|\bm{q}|z'} \varrho(\bm{q},z')^*\right).
\end{equation}
\noindent The integrals over $z,z'$ identify respectively with a function $F(\bm{q})$ and its complex conjugate.  Since the charge distribution $\varrho(\bm{r})$ is assumed to integrate to zero over space (charge neutral), $F(\bm{q})$ is zero at $\bm{q}=0$ so that Eq. \ref{eq:negdefinite2} converges.  The resulting integral $-2\pi \int d^2\bm{q} |F(\bm{q})|^2/|\bm{q}|$ is unconditionally negative, which establishes that $\hat{\mathcal{E}}_S^{\mathrm{QS}}/i\omega$ is a negative definite operator with respect to the Hermitian inner product.  We utilize this result in Appendix \ref{app:eigenmodeproperties} to demonstrate several properties of the eigenmodes.

We next describe how we solve Eqs. \ref{eq:PECboundarycondition} and \ref{eq:generalizedeigenvalue} in practice by use of a ``self"- and ``surface"-impedance matrices for an axisymmetric (with respect to the $z$-axis, for simplicity, in a cylindrical system of coordinates $z,r,\phi$) perfectly conducting nanoscopy probe.  As elaborated in $\textit{Gibson} \cite{gibson_method_2008}$, each element $\bm{Z}^P_{z,z'}$ of the probe ``self"-impedance matrix associates with the (dynamical) electric field emitted from  basis function $|j_z)$ (\textit{e.g.} a ``pulse" of current tangent to the probe surface at height $z$) overlapping with a complementary element at position $z'$:  $\bm{Z}^P_{z,z'}\equiv(j_{z}|\hat{\mathcal{G}}_0|j_{z'})=(j_{z}|\hat{\mathcal{E}}_P|j_{z'})$.  In practice the number of distinct $z$ (and $z'$) coordinates and associated current basis elements are discretized over the length of the probe; these basis elements can take a variety of functional forms but are each localized around $z$ (\textit{e.g.} as an annulus), are smoothly differentiable, and are polarized parallel to the probe surface with vanishing $\hat{\phi}$ component.  In this way, an arbitrary axisymmetric ``state" of current along the probe can be expressed as $|j_P)=\sum_z J^P_z |j_z)$ where $J^P_z$ are complex-valued coefficients forming a finite-length vector $\bm{J}^P$ that discretizes $|j_P)$.  In this way, the ``weak form" of integral equation Eq. \ref{eq:PECboundarycondition} for response of the perfectly conducting probe is:
\begin{equation} \label{eq:PECboundarycondition_weak}
    \bm{V}^\mathrm{ext}-\bm{Z}^P \bm{J}^P = 0 \quad \text{with} \quad \bm{V}^\mathrm{ext}_z \equiv (j_z|E_\mathrm{ext}).
\end{equation}
\noindent Similarly, the ``weak form" of Eq. \ref{eq:generalizedeigenvalue} is solved with a quasi-electrostatic surface impedance matrix $\bm{Z}^{S,\mathrm{QS}}$ by eigenvectors $\bm{J}^P_\nu$ as:
\begin{equation} \label{eq:generalizedeigenvalue_weak}
    \frac{1}{i\omega} \left(\bm{Z}^P-\rho_\nu \bm{Z}^{S,\mathrm{QS}}\right) \bm{J}^P_\nu = 0 \quad \text{with} \quad \bm{Z}^{S,\mathrm{QS}}_{z,z'}/(i\omega)\equiv (j_z| \hat{\theta}_{\partial \Omega_P} \hat{M}_z \hat{\mathcal{G}}_0^\mathrm{\,QS}/i\omega |j_{z'}).
\end{equation}
\noindent Note that $\bm{Z}^P$ and $\bm{Z}^{S,\mathrm{QS}}$ depend explicitly on the excitation energy $\omega$ and probe-sample gap distance $d$, respectively.  (When taken in ratio to $i\omega$ as written, both have well defined limits as $\omega \rightarrow 0$.)  We calculate impedance matrix elements by adaptive numerical quadrature as well described in \textit{Gibson} \cite{gibson_method_2008} by a ``method of moments" for a body of revolution.

\section{Properties of the probe-cavity eigenmodes} \label{app:eigenmodeproperties} 
This Appendix addresses the following targets: {\bf i)} We derive the expression for elements of the eigenmode scattering matrix which describe the response of a material surface at $z=0$ below the probe.  {\bf ii)} We relate the integrated energy flux $S_P$ radiated by an eigenmode excited on the probe to the imaginary part of its corresponding eigenvalue $\rho_\nu$.  {\bf iii)} We related the electromagnetic energy density associated with a single eigenmode excited on an isolated probe to the real part of its corresponding eigenvalue $\rho_\nu$.  {\bf iv)} We show that Ohmic losses on reasonably metallic probes are practically negligible compared to radiative ones and quantitatively unimportant for the predictions in this work.  {\bf v)} In the case of a local surface reflectivity, we quantify how $\beta$ controls the rate of Ohmic absorption by the surface associated to a single eigenmode $|E_\nu)$.  This determination is essential to predicting the ``lineshape" of optical nanoscopy experiments that utilize \textit{e.g.} spectroscopically resolved photothermal expansion as a measure of surface composition.

{\bf i)} Probe-cavity eigenmodes provide an orthogonal basis of fields in the plane of the sample to fully describe probe-sample interactions mediated by $p$-polarized fields.  Except when surface reflection is entirely local (\textit{i.e.} $r_p(q)\approx \beta$), reflection from the sample surface generally induces ``mixing" between eigenmodes described by a reflectivity matrix $\beta_{\mu\nu}$. Here $\hat{\mathcal{G}}_0$ and $\hat{S}_S$ will denote the free-space Green dyadic function and the scattering response function of the material surface below the probe, respectively.  Thus $\hat{S}_S \hat{\mathcal{G}}_0|j_\nu)$ supplies the surface-scattered field in response to eigenmode $|j_\nu)$.  We find four suitable expressions for $\beta_{\mu\nu}$ evaluated in the $z=0^+$ plane of the sample:
\begin{align}
    \text{\textit{Modal reflectivity:}}\quad
    \beta_{\mu\nu} &\equiv -(j_\mu|\hat{S}_S \hat{\mathcal{G}}_0|j_\nu) = \frac{i\omega }{4\pi} \int_{z=0^+} \hspace{-8pt} dA\, \sum_{\alpha=0}^3 \left( \mathcal{A}_\mu^\alpha \partial_z \tilde{\mathcal{A}}_\nu^\alpha - \partial_z \mathcal{A}_\mu^\alpha  \tilde{\mathcal{A}}_\nu^\alpha \right) \label{eq:genreflectance_app} \\
    &= i\omega \int_{0}^\infty dq \,q \cdot i k_z\cdot \left(-r_p(q)\right) \sum_{\alpha=0}^3 s_\alpha \, \mathcal{A}_\mu^\alpha(q)\, \mathcal{A}_\nu^\alpha(q) \label{eq:genreflectanceq_app2} \\
    &\approx \frac{i\omega}{4\pi} \int_{z=0^+} \hspace{-8pt} dA\,\left( \Phi_\mu(\bm{r}) \partial_z \tilde{\Phi}_\nu(\bm{r}) - \partial_z \Phi_\mu(\bm{r}) \tilde{\Phi}_\nu (\bm{r}) \right) \label{eq:genreflectancer_app} \\
    &\approx i\omega \int_{0}^\infty dq \,q^2\, \cdot r_p(q) \, \Phi_\mu(q)\, \Phi_\nu(q) \label{eq:genreflectanceq_app}
\end{align}

\noindent Here $\tilde{\mathcal{A}}_\nu^\alpha$ ($0\le \alpha \le 3$) is a component of the electromagnetic four-potential scattered by the sample surface, $k_z=\sqrt{\omega^2/c^2-q^2}$, and $\mathcal{A}_\nu^\alpha(q)$ is the angular spectrum representation (Hankel transform) of the potential $\mathcal{A}_\nu^\alpha(\bm{r})$ evaluated at $z=0$, with $\bm{r}$ the in-plane radial coordinate.  In Eq. \ref{eq:genreflectanceq_app2} we have also used that at $z=0$ each ``cylindrial wave" at wave-vector $q$ satisfies the Laplace equation so that \textit{e.g.} $\partial_z \mathcal{A}_\mu(q) = -i k_z \mathcal{A}_\mu(q)$, and signage is opposite for the reflected field.  The latter satisfies $\tilde{\mathcal{A}}^\alpha(q)=-s_\alpha r_p(q) \mathcal{A}^\alpha(q)$ with Fresnel coefficient $r_p$ for \textit{p}-polarized $\mathcal{A}^\alpha$, and the component-dependent sign $s_\alpha=(+1,+1,+1,-1)_\alpha$ enacts the mirror operation $\hat{M}_z$. Eq. \ref{eq:genreflectanceq_app} adopts the quasistatic approximation for the probe-sample interaction and replaces the four-potentials with the scalar potentials $\Phi_\mu,\tilde{\Phi}_\nu$, whereas Eq. \ref{eq:genreflectancer_app} is its real-space counterpart.  These expressions are used, respectively, to predict spectra and imaging results showcased in the main text. The scattering matrix $\bm{\beta}$ with elements $\beta_{\mu\nu}$ together with the diagonal matrix $\bm{\rho}$ with entries $\rho_\nu$ determine the excited eigenmode amplitudes through $(\bm{\beta}-\bm{\rho})^{-1}$.  Note that the prefactor $i\omega$ appearing above throughout is a by-product of the normalization condition $(j_\mu|\hat{M}_z\hat{\mathcal{G}}_0^\mathrm{QS}|j_\nu)=\delta_{\mu\nu}=\beta^\mathrm{QS}_{\mu\nu}$ that describes a hypothetical quasi-electrostatic (``QS") surface with unit reflectance.  Consider that the product $f_{\mu\nu}(q)\equiv q^2 \Phi_\mu(q)\Phi_\nu(q)$  in Eq. \ref{eq:genreflectanceq_app} represents a ``form factor" for inter-mode scattering.  To emphasize that only the orthogonality among $|j_\nu)$ is physically essential (\textit{viz.} $\int_0^\infty dq f_{\mu\ne\nu}(q)=0$), a more general statement of Eq. \ref{eq:genreflectanceq_app} allows computing modal reflectivity for eigenmodes with
\begin{gather}
    \text{\textit{Arbitrary normalization:}}\quad \beta_{\mu\nu}=\frac{\int_0^\infty dq \,f_{\mu\nu}(q)\, r_p(q)}{\sqrt{\mathcal{N}_\mu \mathcal{N}_\nu}} \\
    \text{where}\quad \mathcal{N}_\nu\equiv \int_0^\infty \hspace{-5pt} dq\, f^\mathrm{QS}_{\nu\nu}(q). \label{eq:genreflectance_norm}
\end{gather}
\noindent Above, $f^\mathrm{QS}_{\nu\nu}$ indicates that normalization factors $\mathcal{N}_{\mu,\nu}$ should be computed with scalar potentials $\Phi_{\mu,\nu}(q)$ generated in the quasi-electrostatic (``QS") approximation, since only in this approximation are $\mathcal{N}_{\mu,\nu}$ meaningful to the generalized eigenvalue problem Eq. \ref{eq:generalizedeigenvalue_weak}.  Appendix \ref{app:sommerfeld} details explicit methods to evaluate Eq. \ref{eq:genreflectanceq_app2} for arbitrary eigenmodes.

The properties (i) above follow from Eq. \ref{eq:genreflectance_app} which we prove as follows.  We first define $\bm{j}$ as the real-space current density describing eigenmode $|j_\mu)$ on the probe which is associated with a four-current $j^\alpha\equiv (c\varrho, j_{x}, j_y,j_z)_\alpha$, where $\varrho$ is the associated electric charge density.  We also define $|\tilde{E}_\nu)$ as the electric field generated by eigenmode $|j_\nu)$ reflected from the sample surface, which has real-space scalar and vector potential components $\tilde{\Phi}$ and $\tilde{\bm{A}}$, respectively (tilde quantities are surface-reflected) that can be combined into a reflected four-potential $\tilde{\mathcal{A}}^\alpha \equiv (\tilde{\Phi},\tilde{A}_x,\tilde{A}_y,\tilde{A}_z)_\alpha$.  (Distinction between co- and contravariant vectors is inessential here.)  With these definitions and working in the Lorenz gauge for the four-potential:
\begin{align}
    -(j_\mu|\hat{S}_S \hat{\mathcal{G}}_0|j_\nu) & = -(j_\mu|\tilde{E}_\nu) \\
    &=-i\omega\int_{z>0}\hspace{-5pt} dV\, \bm{j}\cdot \left(-\frac{\nabla \tilde{\Phi}_\nu}{i\omega}+\frac{\tilde{\bm{A}}_\nu}{c}\right) \\
    &\equiv -\frac{i\omega}{c} \int_{z>0}\hspace{-5pt} dV\,j^\alpha \tilde{\mathcal{A}}^\alpha. \label{eq:H7}
\end{align}
\noindent In Eq. \ref{eq:H7} we have used integration by parts, \textit{viz.} $\int dV \bm{j}\cdot -\nabla \tilde{\Phi} = \int dV\, \nabla \cdot \bm{j}\,\tilde{\Phi} = i\omega \int dV \varrho\, \tilde{\Phi}$, and Einstein summation convention is implied.  The four-current $j^\alpha$ is related to a four-potential component $\mathcal{A}^\alpha$ via the Helmholtz equation:  $\left(\nabla^2+k^2\right)\mathcal{A}^\alpha = -4\pi j^\alpha/c$.  Inserting this into Eq. \ref{eq:H7} and integrating by parts twice yields:

\begin{align}
    -(j_\mu|\tilde{E}_\nu)&= +\frac{i\omega}{4\pi c} \int_{z>0} \hspace{-5pt} dV\,\left(k^2+\nabla^2\right)\mathcal{A}^\alpha\,\tilde{\mathcal{A}}^\alpha \\
    & = \frac{i\omega}{4\pi c} \Big[k^2 \int_{z>0} \hspace{-5pt} dV \mathcal{A}^\alpha\,\tilde{\mathcal{A}}^\alpha + \int_{z=0^+} \hspace{-8pt} dA\, (-\hat{z})\cdot \left( \nabla \mathcal{A}^\alpha\,\tilde{\mathcal{A}}^\alpha - \mathcal{A}^\alpha\,\nabla \tilde{\mathcal{A}}^\alpha\right) \\\
    & \hspace{40pt} + \int_{z>0}\hspace{-5pt}dV\,\mathcal{A}^\alpha \nabla^2 \tilde{\mathcal{A}}^\alpha \Big].
\end{align}
\noindent The reflected field $\tilde{\mathcal{A}}^\alpha$ is source-free in the region $z>0$ so that here $\nabla^2\tilde{\mathcal{A}}^\alpha=-k^2 \tilde{\mathcal{A}}^\alpha$ whence the volume integrals cancel.  Restoring indices $\mu,\nu$ then yields Eq. \ref{eq:genreflectance_app} as desired.

{\bf ii)}  Eigenvalues $\rho_\nu$ associated to the eigenmodes describe not only their excitation in proximity to a surface, but also the flow of radiant energy from the probe.  For instance, let us consider the energy radiated by current distributions on the probe proportional to the first eigencurrent $|j_P)=|j_\nu)(j_\nu|E_\mathrm{ext})$, where the trailing matrix element describes the eigenmode excitation amplitude.  According to Poynting's theorem (Appendix \ref{app:reciprocity}) for a single eigenmode current then the \textit{integrated radiation flux $S_P$} is:
\begin{align} 
    S_P\equiv \frac{c}{8\pi} \int_{r\rightarrow \infty}\hspace{-10pt} d\bm{A} \cdot \bm{E}_P \times \bm{B}^*_P 
    &=-\frac{1}{2} \mathrm{Re} (j_P^*|E_P)=-\frac{1}{2} \mathrm{Re} (j_\nu^*|\hat{\mathcal{G}}_0|j_\nu) \cdot \big|(j_\nu|E_\mathrm{ext})\big|^2 \nonumber \\
    &=-\frac{u_\nu}{2}\,\mathrm{Im}\,\rho_\nu \cdot \big|(j_\nu|E_\mathrm{ext})\big|^2 \label{eq:radiance}
\end{align}
\noindent Here we have used the fact that $(j_\nu|\hat{\mathcal{G}}_0|j_\nu)\equiv(j_\nu|\rho_\nu \hat{\mathcal{E}}_S^{\mathrm{QS}}|j_\nu)$ and defined a dimensionless constant $u_\nu\equiv i\, (j_\nu^*|\tilde{E}_\nu^\mathrm{QS})>0$.  This is unconditionally positive owing to the fact that $\hat{\mathcal{E}}_S^{\mathrm{QS}}/i\omega$ was shown in Appendix \ref{app:quasistaticsurface} to provide a negative definite Hermitian inner product, equal to the interaction energy between a charge distribution and its sign-inverted mirror image. The energy factor $u_\nu$ appearing here is of order unity, and although it could be removed through suitable normalization of $|j_\nu)$, our choice of normalization for eigenmodes presented in the main text instead opts to avoid analogous factors in Eqs. \ref{eq:PSScattering} \& \ref{eq:PSScatteringlocal}; this choice though \textit{ad hoc} prioritizes simplicity in the probe-sample response function.  Eq. \ref{eq:radiance} implies that $\mathrm{Im} \rho_\nu<0$, which Eq. \ref{eq:generalizedeigenvalue} in fact guarantees owing to the structure of $\hat{\mathcal{E}}_P$.

The cumulative brightness of each eigenmodes is therefore also related to the imaginary elements of $\rho_\nu$ as follows. Consider an aperture of solid angle $A$ large enough to collect most scattered radiation from each excited eigenmode; in this case the aperture-averaged amplitude $\langle E_\mathrm{scat,\nu} \rangle_A$ associated with eigencurrent $|j_\nu)$ is bounded by:
\begin{align}
    |\langle E_\mathrm{scat,\nu} \rangle_A| & \propto  \left|\frac{\langle B_\nu(\theta) \rangle_A^2}{\rho_\nu-\beta} \right|
     \le \frac{-u_\nu \cdot \mathrm{Im}\, \rho_\nu}{|\rho_\nu-\beta|}  \label{eq:radiationfromrho}, \\
    \text{since}\quad |\langle B_\nu(\theta)\rangle_A^2| & \propto |\langle (j_\nu|E_\mathrm{PW,\theta})^2 \rangle_A| \nonumber \\
    &\le \langle |(j_\nu|E_\mathrm{PW,\theta})|^2 \rangle_A=\langle |\bm{E}_\nu(\theta)|^2 \rangle_A = -u_\nu \cdot \mathrm{Im}\,\rho_\nu. \nonumber
\end{align}

\noindent The ultimate proportionality invokes $\bm{B}_P\approx \bm{E}_P$ at distant aperture points and thus identifies the integrated Poynting flux from $|j_\nu)$ per unit eigenmode amplitude.  Eq. \ref{eq:radiationfromrho} assumes that illumination and detection of modes with complicated (\textit{e.g.} quadrupole and higher order) radiation patterns are subject to aperture-cancellation. Therefore Eq. \ref{eq:radiationfromrho} nevertheless establishes relative bounds on the ``detectability" of eigencurrents induced through probe-sample scattering.  Supposing that $|j_\nu)$ are ordered by increasing $|\rho_\nu|$, then while $|\rho_\nu| < |\beta|$ their contribution increases with $\nu$ up to a maximum where $\rho_\nu \sim \beta$, whereas eigencurrents for which $\rho_\nu \gg \beta$ will provide ever decreasing contributions due to aperture cancellation of their complex radiation patterns $B_\nu(\theta)$.  While rapidity of this cancellation may be known only through explicit calculation of $\langle B_\nu \rangle_A$, the main text shows that $B_\nu$ for increasing $\nu$ describe radiation from increasingly multipolar eigenmodes $|j_\nu)$, which supports the present argument.

{\bf iii)}  By an analogous calculation to Eq. \ref{eq:radiance}, for a single eigenfield we find the stored 
\begin{align} \label{eq:storedenergy}
    \text{\textit{Electromagnetic energy:}} \quad 
    \mathcal{U}_P&=-\mathrm{Im} \frac{1}{2\omega}(j^*_P|E_P) \nonumber \\
    &= +\frac{u_\nu}{2\omega}\,\mathrm{Re}\,\rho_\nu \cdot  \big|(j_\nu|E_\mathrm{ext})\big|^2.
\end{align}
\noindent Similarly, this result enables description of the confinement of eigenmode fields.  The energy of interaction between the probe and its ``mirror image" (\textit{e.g.} when the sample surface below presents the reflectivity $\beta=1$ of a perfect mirror) is proportional to $(j_P^*|E_S)\approx (j_P^*|\hat{\mathcal{E}}_S^\mathrm{QS}|j_P)$.  Using the fact that $|j_\nu)$ solve the generalized eigenvalue problem Eq. \ref{eq:generalizedeigenvalue}, the ratio of this energy scale and that given in Eq. \ref{eq:storedenergy} is precisely $1/\mathrm{Re}\,\rho_\nu$.  As discussed in the main text, this factor therefore provides an acceptable quantitative measure of the confinement of eigenmode energy into the probe-sample gap.  Furthermore, combining Eq. \ref{eq:storedenergy} with Eq. \ref{eq:radiance} implies that the ratio $S_P/\mathcal{U}_P=\omega \times -\mathrm{Im}\,\rho_\nu\,/ \mathrm{Re}\,\rho_\nu\equiv \omega \tan(-\varphi_\nu)$ describes the ``radiance" of the eigenfield analogous to the ``loss tangent" of dielectrics, except that here $-\varphi_\nu$ describes the complex angle of $\rho^*_\nu$ and only radiative loss of energy is considered from the (assumed perfectly conducting) probe considered. 

{\bf iv)}  We now consider whether the assumption of a perfect electrically conducting probe is quantitatively relevant to the case of a realistic nanoscopy experiment.  The answer can be evaluated with Eq. \ref{eq:ohmicloss} by calculating Ohmic loss associated with excited eigenmodes when the probe conductivity $\sigma$ is finite, and comparing the result to radiative losses that are assumed the only contribution to $\mathrm{Im}\,\rho_\nu$.  In the hypothetical case where only a single eigenmode is excited on the probe, Ohmic absorption associated with $|j_\nu)$ is proportional to $\mathrm{Re}(j_\nu^*|E)$, where the contributing part of $|E)=(1+\hat{S}_S)|E_\nu)$ is the total field ``transmitted" into the probe volume and $\hat{S}_S$ is the response function of the sample.  On the probe, this field differs from the ``incident" one $\hat{\theta}_{\partial \Omega_P} \hat{S}_S|E_\nu)\approx -\beta \mathcal{E}_S^{\mathrm{QS}}|j_\nu)$ by approximately the surface impedance $Z_P \equiv \sqrt{\frac{\omega}{4\pi \sigma}}\ll 1$ of the probe conductivity $\sigma$.  Approximating the material surface reflectivity in the quasistatic limit by constant $\beta$, we approximate the rate of
\begin{align}
    \text{\textit{Ohmic loss:}}\quad 
   +\frac{1}{2} Z_P \mathrm{Re}(j^*_\nu| \beta \mathcal{E}_S^{\mathrm{QS}}|j_\nu)
   &=\frac{1}{2} Z_P \mathrm{Re}\left(\frac{\beta}{\rho_\nu} (j^*_\nu|\mathcal{E}_P^{\mathrm{QS}}|j_\nu)\right) \nonumber \\
   & =\frac{1}{2} Z_P  \mathrm{Re}\left(-i \frac{\beta}{\rho_\nu} u_\nu\right) 
   = Z_P  \frac{u_\nu}{2}\mathrm{Im}\frac{\beta}{\rho_\nu} .
\end{align}
\noindent This result should be compared with Eq. \ref{eq:radiance} for unit eigenmode amplitude.  For realizable values of $\beta$ of order unity and a metallic probe conductivity with plasma frequency $\omega_p$ and scattering rate $\gamma$, Ohmic losses are greatest at frequencies $\omega<\gamma$ but still smaller than radiative ones by approximately $Z_P\sim \sqrt{ \frac{\omega\gamma}{\omega_p^2}}$. For metallic probes with $\gamma<\omega_p$ utilized at infrared frequencies $\omega \ll \omega_p$, we see that Ohmic losses are orders of magnitude less relevant to the ``actual" total eigenmode loss encoded in (the imaginary part of) $\rho_\nu$ than the radiative ones we have fully included in their calculation.

{\bf v)} Finally, we quantify the rate of optical absorption by a material surface described by local reflectivity $\beta(\omega)$ when excited by a single eigenmode $|E_\nu)$ from a nanoscopy probe.  This quantity is motivated by the suggested mechanism of so-called photothermal expansion nanoscopy, whereby a nanoprobe mechanically detects optical absorption by the proximate sample surface through thermal expansion, which exerts a (transient) force on the probe.\cite{shcherbakov_photo-induced_2025} We note that in a practical experiment, the amplitude of the eigenmode $|j_\nu)$ on the probe is given by $e_\nu$ in square brackets in Eq. \ref{eq:inversematrix}.  We omit this amplitude below and restore it only in the final result.  An elegant means of computing power dissipated by the surface reflectivity $\beta$ replaces physical currents within the medium ($z<0$) with equivalent ones on a ``mirror probe" (\textit{cf.} Fig. \ref{fig1}a) that, after inverting the $z$-axis, occupies now the same physical space previously occupied by the physical probe.  In this case, an eigencurrent on the physical probe $\hat{M}_z |j_\nu)$ associates with a current $|j_\mathrm{MP})=-\beta |j_\nu)$ on the mirror probe.  Here the operator $\hat{M}_z$ inverts the vector $z-$component and the position of $|j_\nu)$ across $z=0$.  Aside from their relative spatial inversion, currents on the physical probe differ in magnitude from those on the mirror probe by a factor $1/\beta$.   Eq. \ref{eq:reciprocityconj} now describes the rate of optical absorption as:
\begin{align}
    +\frac{1}{2}\mathrm{Re}(j_\mathrm{MP}^*|E_\mathrm{tot}) & = \frac{1}{2}\mathrm{Re}(j_\mathrm{MP}^*| \Big(|E_\mathrm{MP})+|E_P) \Big) \nonumber \\
    &=\frac{1}{2}\mathrm{Re}\left((j_\mathrm{MP}^*|\hat{\mathcal{E}}_P|j_\mathrm{MP})-\frac{1}{\beta} (j_\mathrm{MP}^*|\hat{\mathcal{E}}_S^\mathrm{QS}|j_\mathrm{MP})\right) \nonumber \\
    &=  \frac{1}{2} |\beta|^2 \mathrm{Re}\left((j_\nu^*|\rho_\nu \hat{\mathcal{E}}_S^\mathrm{QS}|j_\nu)-\frac{1}{\beta} (j_\nu^*|\hat{\mathcal{E}}_S^\mathrm{QS}|j_\nu)\right) \nonumber \\
    & =  \frac{u_\nu}{2} |\beta|^2 \left(-\mathrm{Im}\,\rho_\nu + \mathrm{Re}\frac{i}{\beta}\right) \nonumber \\
    & = \frac{u_\nu}{2}\left(\mathrm{Im}\,\beta- |\beta|^2 \mathrm{Im}\,\rho_\nu\right). \label{eq:sampleabsorption}
\end{align}
\noindent Above we have used the previously defined energy factor $u_\nu$, the previously defined inverting and non-inverting ``field generators" $\hat{\mathcal{E}}_P$ and $\hat{\mathcal{E}}_S^{\mathrm{QS}}$, and their relation to the eigencurrents $|j_\nu)$ given by Eq. \ref{eq:generalizedeigenvalue}.

Restoring the $\beta$-dependent excitation amplitude $e_\nu$ of eigenmodes, we arrive at an expression for probe-induced absorption by the material surface:
\begin{equation} \label{eq:photothermalabsorption}
    \left(\partial_t \mathcal{U}\right)_\mathrm{abs.} \approx \sum_\nu^\infty \frac{u_\nu}{2}\left(\mathrm{Im}\,\beta- |\beta|^2 \mathrm{Im}\,\rho_\nu\right) \left| \sum_\mu^\infty \left(\frac{1}{\bm{\rho}-\beta}\right)_{\nu\mu} (j_\mu|E_\mathrm{ext}) \right|^2.
\end{equation}
\noindent This result combines Eq. \ref{eq:sampleabsorption} with the $\hat{\mathcal{E}}_S^\mathrm{QS}$-orthogonality of excited eigenmodes $|j_\nu)$. Eq. \ref{eq:photothermalabsorption} presents a striking observation:  although absorption nanoscopy methods like photothermal expansion microscopy might be assumed to probe strictly the lossy part of the surface response $\mathrm{Im}\beta$, it is clear that photothermal spectroscopies remain subject to the ``Fano resonant" lineshape inherited from the eigenmode excitation amplitudes.  Furthermore, recalling that $\lim_{d\rightarrow 0} \mathrm{Im}\,\rho_\nu(d)=0$, the term proportional to $|\beta|^2$ may or may not be important, depending on the probe-sample proximity $d$, which is typically brought to zero (contact) in photothermal absorption nanoscopies.  (Although the coordinate $d=0$ is strictly ``singular" for the \textit{EigenProbe} expansion, predictions for $d\gtrsim0$ evolve smoothly to that limit and allow practical calculation with a finite number of modes $\nu \lesssim 20$.)  This result (Eq. \ref{eq:photothermalabsorption}) thus suggests that the spectral lineshape of photothermal nanoscopies around absorptive modes of a sample surface is \textit{not} a reliable indicator to distinguish between so-called ``photo-induced force" and photothermal expansion mechanisms for mechanical signals induced on a nano-probe.\cite{ocallahan_photoinduced_2018}  Future work should explore alternate means of distinguishing these physically distinct mechanisms of nano-optical contrast in forthcoming nanoscopy experiments.

\section{Sommerfeld integrals for the modal reflectivity} \label{app:sommerfeld}

Sec. \ref{sec:eigenmodeproperties} defines the \textit{modal reflectivity} between eigenmodes $j_{\mu,\nu}$ as $\beta_{\mu\nu}\equiv -(j_\mu|\hat{S}_S \hat{\mathcal{G}}_0|j_\nu)$, where $\hat{\mathcal{G}}_0$ is the free-space Green function and $\hat{S}_S$ is the scattering response of a (possibly inhomogeneous) optical surface beneath a nanoscopy probe.  Appendix \ref{app:eigenmodeproperties} supplies Eq. \ref{eq:genreflectanceq_app2} as the most general route to compute the modal reflectivity in the axisymmetric approximation even beyond the quasi-electrostatic approximation.  Here we detail explicit means to evaluate the \textit{Sommerfeld integrals} of Eq. \ref{eq:genreflectanceq_app2}, in practice via fixed precision numerical quadrature over $q$, using eigenmodes obtained by a method of moments as described in Appendix \ref{app:quasistaticsurface}.  We adopt cylindrical coordinates $\rho,\phi,z$, where the latter directs along the probe axis of rotational symmetry, and we exclude $\hat{\phi}$-polarized components of electric fields.  As described in Appendix \ref{app:quasistaticsurface}, eigenmodes $|j_\nu)$ are constructed from annular basis elements $|j_\mathfrak{z})$ at discrete positions $z=\mathfrak{z}$ along the probe.  For this discussion, we consider a coordinate system where the closest point on the probe to the surface is at $z=0$, and where the surface is a distance $d$ below. To utilize Eq. \ref{eq:genreflectanceq_app2}, it then suffices to generate the (Hankel transform of the) relevant elements of the electromagnetic four-potential, which are $\Phi(q|\mathfrak{z})$, $A_\rho(q|\mathfrak{z})$, and $A_z(q|\mathfrak{z})$ emitted from such an element at $z=\mathfrak{z}$ and evaluated at $z=-d$.  Here subscripts denote polarization, and $q|\mathfrak{z}$ signifies evaluation at in-plane momentum $q$ following emission from coordinate $z=\mathfrak{z}$.  Here we again utilize the Lorenz gauge for the four-potential.

In terms of $g_0$ defined in Appendix \ref{app:quasistaticsurface}, the scalar potential generated by axisymmetric charge density $\varrho$ and associated current density $\bm{j}$ is given by
\begin{align}
    \Phi(\bm{r}) &= \int dV' g_0(\bm{r}|\bm{r}') \varrho(\bm{r}') \\
     &= -(i\omega)^{-1} \int dV'\, \nabla' g_0(\bm{r}|\bm{r}')\cdot  \bm{j}(\bm{r}') \\
     & = -(i\omega)^{-1} \int dV' \left(   j_z \partial_{z'} + j_\rho \partial_{\rho'}\right) g_0(\bm{r}|\bm{r}')
\end{align}
\noindent where we have used charge continuity and integration by parts.  We next consider where $\bm{j}$ describes basis element $|j_{\mathfrak{z}})$ distributed uniformly around an annulus (``ring") near $z=\mathfrak{z}$ with radius $\mathcal{R}$.  Since $\bm{j}$ is isolated and locally tangent to the probe surface, a body of revolution entirely defined by the radial function $\mathcal{R}(z')$, then we can write $j_z \equiv \mathcal{C}_\mathfrak{z} J$ and $j_\rho \equiv \mathcal{S}_\mathfrak{z} J$, where local ``cosine" $\mathcal{C}_\mathfrak{z}\equiv 1/\sqrt{1+(\partial_{z'} \mathcal{R})^2}$ and ``sine" $\mathcal{S}_\mathfrak{z}\equiv \partial_{z'} \mathcal{R}/\sqrt{1+(\partial_{z'} \mathcal{R})^2}$ scalars describe directional flow of current on the annular element, and $J=1$ is \textit{ad hoc} amplitude for the basis element.  Thus:

\begin{equation}
    \Phi(\bm{r}|\mathfrak{z})=\int dz'\,  \frac{2\pi \mathcal{R}}{-i\omega} \int_0^{2\pi} \frac{d\phi'}{2\pi}\, 
    \left[\Big(\mathcal{C}_\mathfrak{z} \partial_{z'}+\mathcal{S}_\mathfrak{z} \partial_{\rho'}\Big)
    g_0\left(\bm{r}|\rho',\phi',z' \right)\right]_{\rho'=\mathcal{R},z'=\mathfrak{z}}
\end{equation}

\noindent Since the basis element is infinitesimal, the integrand is approximately constant on $z'$ and can be replaced with the $z'$-span $\Delta$ of the basis element. Next we decompose $g_0$ by the Weyl identity \cite{novotny2012principles} as:

\begin{equation} \label{eq:H4}
    g_0(\bm{r}|\rho',\phi',z')=\int dq\, \frac{i q}{k_z} \, e^{i k_z |z-z'|} \int_0^{2\pi} \frac{d\phi_q}{2\pi} \,e^{iq \left(\rho \cos \phi_q - \rho' \cos(\phi_q-\phi')\right)}
\end{equation}

\noindent Here $q$ is the magnitude of in-plane momentum associated with a plane wave propagating along $\bm{k}=k_z \hat{\bm{z}}+\bm{q}$, where $k_z\equiv \sqrt{k^2-q^2}$ and angle $\phi_q$ denotes the orientation of $\bm{q}$ relative to that of the observation point $\bm{r}$.  Since $\rho'$ will be held independent of azimuthal angle, we can evaluate angular integrals on $\phi'$ (Eq. \ref{eq:H4}) and $\phi_q$ straight away:

\begin{align}
    \int_0^{2\pi} \frac{d\phi'}{2\pi} \,g_0(\bm{r}|\rho',\phi',z')  & = 
    \Big[ \int_0^k \hspace{-5pt} dq\, \frac{i q}{k_z}\, J_0(q\rho) J_0(q \rho') e^{i k_z |z-z'|} \label{eq:scalarpotentialFF} \\
        &\qquad +\int_0^\infty \hspace{-5pt} d\kappa\, J_0(\sqrt{\kappa^2+k^2}\,\rho) J_0(\sqrt{\kappa^2+k^2}\,\rho') e^{-\kappa|z-z'|} \Big] \label{eq:scalarpotentialNF}
\end{align}

\noindent Here $\kappa\equiv\sqrt{q^2-k^2}$ is the propagation constant along $z$ associated with a rotationally symmetric evanescent ``Bessel beam" field, and $J_n$ denotes a Bessel function of the first kind of order $n$.  Eqs. \ref{eq:scalarpotentialNF} and \ref{eq:scalarpotentialFF} identify the near-field $q>\omega/c$ and far-field $q<\omega/c$ parts, respectively, of the scalar potential.  This separation of momentum scales allows convenient exclusion of the far-field part of the probe-sample interaction as desired when calculating the modal reflectivity.  Inserting into Eq. \ref{eq:H4}, we obtain for $z<\mathfrak{z}$:

\begin{equation}
    \Phi(\bm{r}|\mathfrak{z})=\int_0^\infty\hspace{-5pt} dq\,q J_0(q\rho) \left[ w_\mathfrak{z}  \frac{i}{k_z} \Big( \mathcal{C}_\mathfrak{z} \cdot i k_z J_0(q \mathcal{R}) - \mathcal{S}_\mathfrak{z} \cdot q  J_1(q \mathcal{R})\Big) e^{i k_z |z-\mathfrak{z}|} \right]
\end{equation}

\noindent Here we define a weight $w_\mathfrak{z}\equiv 2\pi \mathcal{R}\Delta/(i\omega) $  local to the basis function at $z=\mathfrak{z}$.  Note that $-ik_z\rightarrow+q$ in the quasi-electrostatic limit $\omega\rightarrow 0$.  Evaluating $z=-d$ in the plane of the surface, we identify the term in square brackets as the sought-after Hankel transform of the
\begin{eqnarray}
    &\text{\textit{Scalar potential from} $|j_\mathfrak{z})$:} \quad \Phi(q|\mathfrak{z}) & = w_\mathfrak{z}  \frac{i}{k_z} \Big( \mathcal{C}_\mathfrak{z} i k_z J_0(q \mathcal{R}) - \mathcal{S}_\mathfrak{z}q  J_1(q \mathcal{R})\Big) e^{i k_z (d+\mathfrak{z})} \\
    &\text{\textit{in the QS approximation:}} \quad 
    & \approx -w_\mathfrak{z} \Big( \mathcal{C}_\mathfrak{z} J_0(q \mathcal{R}) + \mathcal{S}_\mathfrak{z} J_1(q \mathcal{R})\Big) e^{i k_z (d+\mathfrak{z})}.
\end{eqnarray}

\noindent The vector potential components are similarly and more easily evaluated, yielding the  

\begin{eqnarray}
    \text{\textit{Vector $z$-potential from} $|j_\mathfrak{z})$}& \quad
  & A_z(q|\mathfrak{z}) = w_\mathfrak{z}  \mathcal{C}_\mathfrak{z} \cdot \frac{i k}{q} J_0(q \mathcal{R})  e^{i k_z (d+\mathfrak{z})}  \label{eq:H12} \\
    \text{and \textit{vector $\rho$-potential:}}&\quad
   & A_\rho(q|\mathfrak{z}) =w_\mathfrak{z}  \mathcal{S}_\mathfrak{z} \cdot \frac{i k}{q} J_1(q \mathcal{R})  e^{i k_z (d+\mathfrak{z})}. \label{eq:H13}
\end{eqnarray}

\noindent Note that appearances of $1/q$ above produce no physical divergence since the quantity of interest (Eq. \ref{eq:genreflectanceq_app}) is weighted by $q^2$.  Evaluation of $A_\rho(q|\mathfrak{z})$ involves an angular integral over the vector product of unit vectors $\hat{\rho}(\phi)\cdot\hat{\rho}'(\phi')$ which renders a radial dependence $J_1(q\rho)J_1(q\mathcal{R})$. Therefore $A_\rho$ vanishes at $\rho=0$, whereby Eq. \ref{eq:H13} is in fact the Hankel transform of $A_\rho$ of order $J_{n=1}$.  Since Eq. \ref{eq:genreflectanceq_app} holds for a Hankel transform of arbitrary order $n\ge0$ for each component $\mathcal{A}^\alpha$ of the electromagnetic four-potential individually, $n=1$ supplies a convenient choice for $A_\rho$.

Finally, the multimodal reflectance can now be computed from the basis contributions $\mathcal{A}^\alpha(q|\mathfrak{z})$ enumerated above when the Fresnel coefficient $r_p(q)$ of a surface is known.  When $i k_z\approx -q$, all potentials decay exponentially in the probe-surface separation $d$.  Consider a large matrix $\mathbf{B}$ whose elements comprise each of the following Sommerfeld integrals, which we evaluate \textit{e.g.} by Gauss-Legendre quadrature over discretely sampled nodes $q$ ranging from $0$ (or from $k=\omega/c$) to several times the inverse probe-sample gap $1/d$.  Below, $\mathcal{A}^{\alpha=0,1,2}$ denote $\Phi$, $A_\rho$, and $A_z$, respectively, and $s_\alpha=(+1,+1,-1)_\alpha$ implements the action of $\hat{M}_z$:
\begin{equation}
    \mathbf{B}_{\mathfrak{z}\mathfrak{z'}} \equiv \int_0^\infty \hspace{-5pt} dq \,q\cdot ik_z\,\left(-r_p(q)\right)\sum_{\alpha=0}^2 s_\alpha \mathcal{A}^\alpha(q|\mathfrak{z}) \mathcal{A}^\alpha(q|\mathfrak{z}').
\end{equation}
\noindent As described in Appendix \ref{app:quasistaticsurface}, each eigenmode $|j_\nu)=\sum_\mathfrak{z} J_\mathfrak{z}^\nu |j_\mathfrak{z})$ is obtained as linear combination of basis elements, where the $J_\mathfrak{z}^\nu$ form a column vector over $\mathfrak{z}$.  Assembling these column vectors into a matrix $\mathbf{J}$, the multi-modal reflectances $\beta_{\mu\nu}$ are matrix elements of $\bm{\beta}=i\omega \,\mathbf{J}^T \mathbf{B} \mathbf{J}$.  When $\mathbf{B}$($=\mathbf{B}^\mathrm{QS}$) is computed with the quasi-electrostatic scalar potential ($\alpha=0$) only and where the surface reflectance is taken hypothetically to unity, the orthogonality relation for eigenmodes can be expressed as:  $i\omega \left[\mathbf{J}^T \mathbf{B}^\mathrm{QS} \mathbf{J}\right]_{\mu\nu}=\delta_{\mu\nu}=(j_\mu|\hat{\mathcal{E}}_S^\mathrm{QS}|j_\nu)$.  We emphasize again the appearance of $i\omega$ in these expressions merely persists the convenient convention for normalization of eigenmodes introduced in Sec. \ref{sec:eigenmodesbasis}.

\section{EigenProbe Imaging: Multi-modal reflectivity from inhomogeneous materials} \label{app:eigenprobeimaging}
\begin{figure}[tbp]
\centering
\includegraphics[width=0.7 \textwidth]{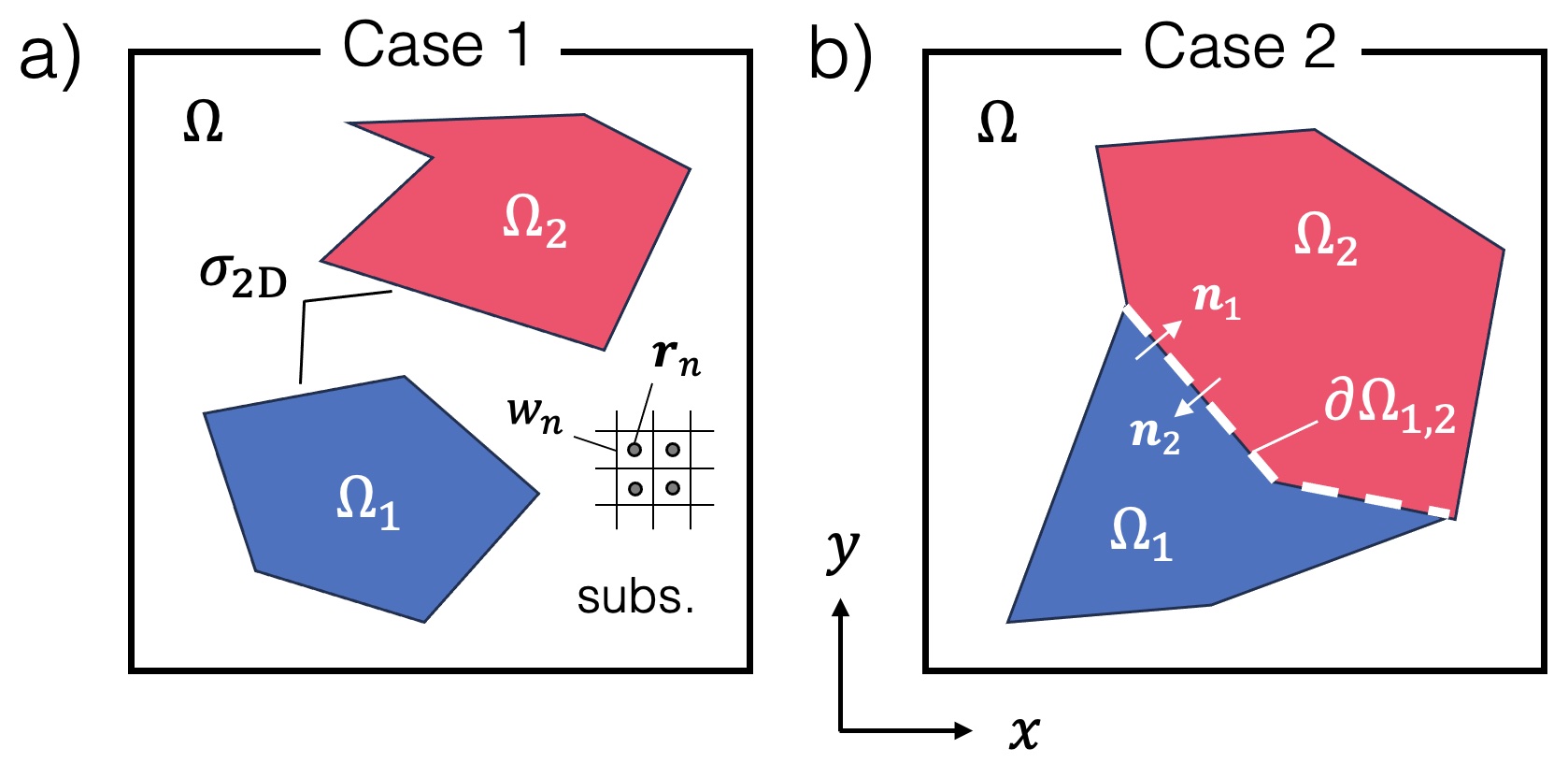}
\caption[Domain configurations for \textit{EigenProbe imaging}.]{\label{fig:domains}  {\bf Domain configurations of \textit{EigenProbe imaging.}} a) Non-overlapping regions $\Omega_{1,2}$ describe non-contiguous 2D materials in domain $\Omega$ atop a quasi-infinite substrate.  Grid nodes ${\bm r}_n$ and weights $w_n$ are used to compute linear transformations of basis functions. b) Overlapping domains require special treatment at their mutual edge $\partial \Omega_{1,2}$ (see text); edge-outwards normal vectors ${\bm n}_{1,2}$ are indicated.}
\end{figure}

Here we elaborate on predictions of inhomogeneous probe-sample interactions and nanoscopic imaging discussed in Sec. \ref{sec:imaging} of the main text.  This description expands upon methods introduced previously.\cite{jing_terahertz_2021,xu_deep_2021} We begin by describing how Eqs. \ref{eq:Rmatrix}-\ref{eq:genreflectanceimaging} are obtained from the simple sample configuration shown in Fig. \ref{fig6}a, in which the probe situates at a lateral coordinate ${\bm r}_\mathrm{probe}$ over an inhomogeneous (quasi-)2D material placed atop a layered substrate.  For simplicity, we consider the substrate optical response described by momentum-($q$-)resolved Fresnel coefficient $r_\mathrm{subs}(q)$, in absence of the surficial 2D material.  The overall goal is to quantify the scattered (``reflected") field from the surface in response to the scalar potential or ``eigenfield" $\Phi_\nu({\bm r} - {\bm r_\mathrm{probe}})$ evaluated in the $z=0$ plane of the surface from a single probe-gap eigenmode emanating from the nanoscopy probe.  We will consider only the quasi-electrostatic optical response of the inhomogeneous material, although generalizing our treatment to the full electrodynamic regime is foreseeable.  We define a convenient bra-ket notation in which linear transformations $\hat{S}$ on functions $\Phi({\bm r})$ defined in the $z=0$ plane are described by
\begin{equation} \label{eq:lineartransformation}
    \hat{S}=\sum_{jk} {\bm G_{jk}} |q_j\rangle\langle q_k|
\end{equation}
\noindent where ${\bm G}$ is the associated transformation matrix and the $|q_j\rangle$ across $j$ correspond to a basis of functions $\Phi_j({\bm r})$.  The basis elements are chosen orthonormal over the lateral (2-dimensional) domain of interest $\Omega$ which resides in the $z=0$ plane.  Here $\langle q_j| q_k\rangle = \delta_{jk}$ describes an inner product that could be computed by \textit{e.g.} $\int_\Omega d^2{\bm r}\, \Phi_j({\bm r})^* \Phi_k({\bm r})$, but in practice could be implemented in any way that preserves the action of linear transformations $\hat{S}$.  (Note this bra-ket notation demands a Hermitian inner product, and is unrelated to the parenthetical bra-ket notation used in the main text.)  As sketched in Fig. \ref{fig:domains} convenient numerical implementation replaces the integral with a weighted summation over a grid of ``node" coordinates ${\bm r}_n$ chosen to sample the domain $\Omega$, in which each weight $w_n$ describes the size of a local ``patch" that approximates the area element $d^2 {\bm r}$.  When the domain $\Omega$ is necessarily made finite, the basis functions could be chosen with periodic boundary conditions, and for subsequent concreteness, a basis of plane waves is convenient $\Phi_j({\bm r}) \propto e^{i {\bm q_j}\cdot {\bm r}}$ equality is fixed by normalization to ensure the basis is orthonormal.

We proceed next to describe the scattering response $|\Phi_\mathrm{\alpha}\rangle= -\hat{R}_\alpha |\Phi_\mathrm{ext}\rangle $ with a ``reflectivity" $\hat{R}_\alpha$ from individual parts of the inhomogeneous system (where $\alpha=$``2D" or ``subs" for the 2D material or substrate, respectively) to an external field $\Phi_\mathrm{ext}$, which we will later select as an eigenfield from the nanoscopy probe.  The scattering responses $-\hat{R}_\alpha$ are now akin to the response functions $\hat{S}_{P,E}$ that can be composited according to Eq. \ref{eq:gcomposite}.  Thanks to its assumed translational invariance, the reflectivity of the substrate is ``diagonal" in our basis of planewaves, whence:
\begin{equation} \label{eq:substratereflectivity}
\hat{R}_\mathrm{subs} = \sum_j |q_j\rangle r_\mathrm{subs}(q_j) \langle q_j|.
\end{equation}
\noindent Here $j$ labels all possible wave-vectors ${\bm q_j}$ but $q_j=|{\bm q_j}|$. If we describe the 2D material by a local inhomogeneous 2D conductivity $\sigma_{2D}({\bm r})$ (analogous to that for graphene)\cite{Fei2012} then the combination of Ohm's law, Coulomb's law, and charge conservation provides a self-consistency relation for its scattered field $\Phi_\mathrm{2D}$:
\begin{equation}
    -\left[\int_{z=0} d^2 {\bm r}' \frac{1}{|{\bm r}'-{\bm r}|} \nabla' \cdot \frac{\sigma_\mathrm{2D}({\bm r}')}{i\omega} \nabla'\right]\Big(\Phi_\mathrm{ext}({\bm r}')+\Phi_\mathrm{2D}({\bm r}')\Big)=\Phi_\mathrm{2D}({\bm r}).
\end{equation}

\noindent In the plane wave basis, the leading convolution by the Coulomb kernel can be represented with the linear transformation $\hat{V}=\sum_i |q_i\rangle 2\pi/q_i \langle q_i|$, where $1/q_i$ is a ``blurring" kernel.  We now suppose that, wherever nonzero, $\sigma_\mathrm{2D}$ is piecewise homogeneous within sub-domains $\Omega_\zeta \in \Omega$ labeled by index $\zeta$, within each of which $\sigma_{2D}/i\omega \equiv (2\pi q_{p,\zeta})^{-1}$, and $q_{p,\zeta}$ denote local plasmon wave-vectors.  Whether positive, negative, or complex-valued, these need not correspond to realizable propagating plasmon modes, but in the latter case, $\mathrm{Im}\,q_p>0$ when $\mathrm{Re}\,\sigma_\mathrm{2D}>0$, as when the conducting medium is lossy.  We can then identify that:
\begin{gather}
    -\nabla \cdot \frac{\sigma_{2D}({\bm r})}{i\omega} \nabla \Phi({\bm r}) = \sum_\zeta (2\pi q_{p,\zeta})^{-1} \left(-\nabla^2_\zeta + \delta({\bm r}-\partial\Omega_\zeta)\, \mathbf{n}_\zeta \cdot \nabla \right) \Phi({\bm r}) \label{eq:2Dreflectivity1} \\
    \text{so that}\quad \frac{\hat{V}}{2\pi} \sum_\zeta \frac{\hat{L}_\zeta}{q_{p,\zeta}} \left( |\Phi_\mathrm{ext}\rangle+|\Phi_\mathrm{2D}\rangle \right) = |\Phi_\mathrm{2D}\rangle \label{eq:2Dreflectivity2} \\
    \text{and}\quad  -\hat{R}_\mathrm{2D} =
    \frac{\hat{V} \sum_\zeta \frac{\hat{L}_\zeta}{2\pi q_{p,\zeta}}}
    {1-\hat{V} \sum_\zeta \frac{\hat{L}_\zeta}{2\pi q_{p,\zeta}}}. \label{eq:2Dreflectivity3}
\end{gather}
\noindent In Eq. \ref{eq:2Dreflectivity1}, $\nabla^2_\zeta$ signifies the Laplacian evaluated strictly in domain $\Omega_\zeta$, $\delta(\dots)$ is a 1D Dirac-delta function nonzero only along the edge $\partial \Omega_\zeta$ of the domain, and ${\bm n_\zeta}$ is the local outward normal vector along the edge, as sketched in Fig. \ref{fig:domains}. The sum of these acting on the scalar potential describes surface and edge charge contributed by the 2D material in domain $\Omega_\zeta$, and Eq. \ref{eq:2Dreflectivity2} provides a shorthand that combines these into a differential operator $\hat{L}_\zeta$, which should be understood in the sense of Eq. \ref{eq:lineartransformation} where matrix elements describe action on the basis elements as
\begin{gather} \label{eq:Loperator1}
\hat{L}_\zeta = \sum_{j,k} \left( {\bm L}_{jk}^{\Omega_\zeta}+{\bm L}_{jk}^{\partial \Omega_\zeta} \right)|q_i\rangle \langle q_j| \quad \text{where} \\
{\bm L}_{jk}^{\Omega_\zeta}\equiv -\int_{\Omega_\zeta} \hspace{-5 pt} d^2{\bm r}\,\Phi_{q_j}^* \nabla^2 \Phi_{q_i} \quad
\text{and} \quad
{\bm L}_{jk}^{\partial\Omega_\zeta} \equiv +{\bm n}_\zeta \cdot \int_{\partial \Omega_\zeta} \hspace{-7 pt}d{\bm r}\, \Phi_{q_i}^* \nabla \Phi_{q_j}. \label{eq:Loperator2}
\end{gather}
\noindent The linear transformations ${\bm L}_{jk}^{\Omega_\zeta}$ and ${\bm L}_{jk}^{\partial \Omega_\zeta}$ map a ``driving" potential to surficial and edge charge densities, respectively.  The inverse in Eq. \ref{eq:2Dreflectivity3} is similarly understood through an inverse matrix ${\bm R}={\bm D}^{-1}$ where the quotient matrix $\bm{D}$ includes the ${\bm L^{\Omega_\zeta,\partial \Omega_\zeta}}$ and should be computed in basis $|q_j\rangle$, which is necessarily (at least approximately) \textit{complete} on $\Omega$.

There is reason to supplement the complete \textit{e.g.} plane wave basis $|q_j\rangle$ with additional orthonormal bases $|Q_{\zeta,j}\rangle$ associated to real-space functions $\Phi_{\zeta,j}({\bm r})$ that are respectively complete (only) on each $\Omega_\zeta$ and elsewhere zero.  It is convenient to choose these basis elements as eigenfunctions of the Laplacian on each domain with eigenvalue $Q_{\zeta,j}$ as $\nabla^2_\zeta \Phi_{\zeta,j}=-Q_{\zeta,j}^2 \Phi_{\zeta,j}$ with Neumann boundary conditions on $\partial \Omega_\zeta$.  Since such a basis is complete on $\Omega_\zeta$ it spans both the domain and the image of the surficial part of $\hat{L}_\zeta$, and the latter part becomes ``diagonal" in this representation $\hat{L}^{\Omega_\zeta}=\sum_j Q^2_{\zeta,j} |Q_{\zeta,j}\rangle \langle Q_{\zeta,j}|$. On the other hand, $\hat{L}^{\partial\Omega_\zeta}$ describes charge accumulated to the edge of each domain $\Omega_\zeta$ is not easily represented in the ``Laplacian basis" on any $\Omega_\zeta$.  However, the image of $\hat{V}\hat{L}^{\partial\Omega_\zeta}$ is straightforwardly represented in the ``full" basis as Eq. \ref{eq:Loperator1}.  Practically speaking, the edge integral Eq. \ref{eq:Loperator2} can be implemented in the following steps:
\begin{enumerate}
    \item \textbf{i)} We define a binary mask $M_{\zeta}$ that is zero in $\Omega_\zeta$ and non-zero in the complement domain $\Omega-\Omega_\zeta$.
    \item \textbf{ii)} We approximate the gradient $\nabla M_{\zeta}$ of this mask (parallel to ${\bm n}_\zeta$ where nonzero along the edge of $\Omega_\zeta$) numerically on the grid ${\bm r_n}$.
    \item \textbf{iii)} Finally, we integrate the inner product of $\nabla M_\zeta$ with $\Phi_j^* \nabla \Phi_k$ over $\Omega$.
\end{enumerate}
When $\Phi_j$ is the basis of plane waves, the integrand is simply equal to $\nabla M_{\zeta} \cdot i{\bm q}_k e^{i({\bm q}_k-{\bm q}_j)}$.  This approach generalizes to both cases shown in Fig. \ref{fig:domains}.  Where domains $\Omega_{\zeta=1,2}$ overlap (\textbf{Case 2}), their mutual edge matrix $\bm{L}^{\partial\Omega_{1,2}}$ should contribute within the summation $\sum_\zeta \hat{L}_\zeta/q_{p,\zeta}$ only once with an overall pre-factor of $\left(q_{p,1}^{-1}-q_{p,2}^{-1}\right)$.

To finally render surficial charge matrices ${\bm L}^{\Omega_\zeta}$ in the $\Omega$-complete basis requires a (non-square) inter-basis projection matrix ${\bm P}_{\zeta,jk}\equiv \langle q_j| Q_{\zeta,k}\rangle$ so that ${\bm L}_\zeta = {\bm P} {\bm Q}^2_\zeta {\bm P}^\dagger+\bm{L}^{\partial\Omega_\zeta}$, where ${\bm P}^\dagger$ is its Hermitian adjoint and ${\bm Q}_\zeta$ is a diagonal matrix of local eigenvalues.  Furthermore, again when the complete basis is provided by planewaves, $\hat{V}$ has the matrix representation ${\bm V}=2\pi/{\bm q}$, where ${\bm q}$ is the diagonal matrix of wave-vectors $q_j$.  \textit{In the uniform case}, Eq. \ref{eq:2Dreflectivity3} then simplifies dramatically:  Since, without edges, $\bm{L}^{\partial\Omega_\zeta}=0$, and all $\Omega_\zeta$ coincide with $\Omega$, their eigenvalues $Q_{\zeta,j}$ match the wave-vectors $q_j$. After obtaining $q_p$ only once summing all conductivities, we recover the surface response considered in Sec. \ref{sec:eigenmodesbasis} that was taken as describing a plasmonic membrane like graphene:
\begin{equation} 
\hat{R} =\sum_q r_p(q) |q\rangle \langle q| \quad \text{with} \quad r_p(q)\equiv \frac{q}{q-q_p}.
\end{equation}

We have now described all ingredients needed to calculate $\hat{R}$, which is given by inserting Eqs. \ref{eq:substratereflectivity} and \ref{eq:2Dreflectivity3} into Eq. \ref{eq:gcomposite}, in which operators $\hat{S}_E$, $\hat{S}_P$, and $\hat{S}_+$ should be replaced with matrices $-{\bm R}_\mathrm{subs}$, $-{\bm R}_\mathrm{2D}$, and $-{\bm R}$, respectively.  Thus we can quite generally compute ${\bm R}$ for an arbitrary distribution of 2D media over a substrate like those shown in Fig. \ref{fig:domains}, and such ${\bm R}$ can be utilized to compute local (generalized) photonic density states according to Eq. \ref{eq:genreflectanceimaging}. In the simplest case of local substrate reflectivity $r_\mathrm{subs}(q)=\beta_\mathrm{subs}$ (constant), the matrices ${\bm R}_\mathrm{subs}$ and ${\bm R}_\mathrm{2D}$ commute in Eq. \ref{eq:gcomposite}.  Then, in the case where a single 2D material sub-domain is described by ${\bm L}$, Eq. \ref{eq:gcomposite} reduces to Eq. \ref{eq:Rmatrix} shown in the main text.

In summary, the general description of local probe-sample scattering requires to 1) describe and appropriately discretize the system in Fig. \ref{fig6}a piecewise according to Fig. \ref{fig:domains}, 2) identify the local plasmon wave-vectors $q_{p,\zeta}$, 3) compute for each domain a basis of functions $\Phi_{\zeta,j}$ that diagonalize the local Laplacian, 4) set out a complete basis on the domain $\Omega$ and compute the change-of-basis matrices ${\bm P}_\zeta$ associated to each 2D material domain, 5) (optionally) evaluate domain-edge contributions to the full series of domain response matrices ${\bm L}_\zeta$, and finally 6) apply Eq. \ref{eq:gcomposite} to composite ${\bm R}$ from ${\bm R}_\mathrm{subs}$ and ${\bm R}_\mathrm{2D}$.  Our implementation accomplishes 2) using the open source finite element solver \textit{FEniCS} \cite{alnaes_fenics_2015}.  A typical computation in 5) involves multiplying and inverting matrices of typical rank $N\sim1000$, equal to the number of basis functions (\textit{e.g.} planewaves) used in the calculation. To calculate elements of these (generally dense) matrices and to invert them are speed-limited by core utilization and can both benefit enormously from GPU-accelerated parallelization.

Notably, highly resolved features of the field scattered by the inhomogeneous sample may scarcely contribute to the inner product represented by Eq. \ref{eq:genreflectanceimaging}, since such features may be sampled only through high-order $\nu$ eigenfield $\Phi_\nu$ of the probe.  Their contribution to the probe-scattered field scales with $|\rho_\nu|^{-1}$ that decreases exponentially with increasing $\nu$.  Therefore, predicted images like Figs. \ref{fig6}b-c can converge rapidly with $N$, provided that largest wave-vectors of the global function basis exceed the largest operative plasmon wave vector $q_p$.  Most importantly, ${\bm R}$ need only be computed once for each predicted image, since only the eigenfield excitations need to be updated for every probe location ${\bm r}_\mathrm{probe}$.  Furthermore, when the global function basis comprises planewaves, this update amounts only to updating ``phase factors" for each component wave in the excitation eigenfield, which is accomplished simply through repeated multiplication by a van der Monde matrix.  Meanwhile, ``imaging spectroscopies" of the sort shown in Figs. \ref{fig6}e-f require recomputation of ${\bm R}$ to describe each reconfiguration of the sample.  Future work will demonstrate how even this calculation can be remarkably accelerated beyond what might be expected from Eq. \ref{eq:gcomposite}.

\section{Barycentric rational approximations for demodulated probe-sample scattering} \label{app:barycentric}

Here we provide rational approximations for the demodulated probe-scattered field that is obtained by combining Eq. \ref{eq:PSScatteringlocal} with Eq. \ref{eq:demodulation}, with $\nu\le20$ and including more than $100$ nodes $d_k$ for the probe-gap distance.  These approximations are suitable to estimate the probe-scattered field $E_\mathrm{scat,n}$ at demodulation harmonic $n$ in terms of the (local) surface reflectivity $\beta$.  The approximations are obtained by the \textit{AAA Algorithm}\cite{AAA_algorithm} with $N=19$ distinct poles and take the following form:
\begin{equation}
     E_\mathrm{scat,n}(\beta) \approx g_n + \sum_{j=1}^N \frac{\mathcal{R}_n^{(j)}}{\beta-\rho_n^{(j)}}
\end{equation}
\noindent Here $\rho_n^{(j)}$ and $\mathcal{R}_n^{(j)}$ denote the $j$th pole and residue of the approximation, respectively, and $g_n$ is a complex-valued gain selected so that $E_\mathrm{scat,n}(\beta=0)=0$ in accord with the assertion that $E_\mathrm{scat}(\beta=0)$ in Eq. \ref{eq:PSScatteringlocal} is independent of probe-sample distance $d$ in the limit that $\nu$ is large.  The following tables provide the estimation for two exemplary choices of probe tapping amplitude $A$ relative to probe radius $a$:  $A/a=2$ and $A/a=3$.  Evaluations for different choices of harmonic $n$ can be taken in ratio as predictors for \textit{e.g.} the self-normalized signal $\xi(\beta)=E_{n=3}(\beta)/E_{n=2}(\beta)$.

\newpage

\begin{table}[h!]
\caption[Barycentric rational approximation for $A/a=2$]{Barycentric rational approximation for the demodulated probe-scattered field corresponding to the standard probe geometry in Fig. \ref{fig:3}b and tapping amplitude relative to probe radius $A/a=2$.  Rows correspond to incrementing $j$.}
\footnotesize
\begin{tabular}{||c|c|c|c||c|c|c|c||}
\hline
\hline
$\mathrm{Re}\,g_{n=1}$ & $\mathrm{Re}\,g_{n=2}$ & $\mathrm{Re}\,g_{n=3}$ & $\mathrm{Re}\,g_{n=4}$ &
$\mathrm{Im}\,g_{n=1}$ & $\mathrm{Im}\,g_{n=2}$ & $\mathrm{Im}\,g_{n=3}$ & $\mathrm{Im}\,g_{n=4}$  \\
\hline
   0.113931 & 0.0232191 & 0.00962331 & 0.234993 & -0.23312 & 0.11057 & 0.0809765 & 0.0684128 \\
   \hhline{|=|=|=|=|=|=|=|=|}
$\mathrm{Re}\,\rho_{n=1}$  & $\mathrm{Re}\,\rho_{n=2}$ & $\mathrm{Re}\,\rho_{n=3}$ & $\mathrm{Re}\,\rho_{n=4}$ &
$\mathrm{Im}\,\rho_{n=1}$  & $\mathrm{Im}\,\rho_{n=2}$ & $\mathrm{Im}\,\rho_{n=3}$ & $\mathrm{Im}\,\rho_{n=4}$ \\
\hline
266.954 & 144.759 & 171.192 & 122.523& 9.95711 & -10.2349 & 0.291681 & -52.3677 \\
101.703 & 93.0116 & 96.763 & 85.6551& -40.6836 & -28.2851 & -30.3629 & -70.2083 \\
71.119 & 61.9175 & 67.0626 & 70.0042& -20.2417 & -15.3375 & -19.9269 & -19.6199 \\
52.8259 & 40.7397 & 46.2036 & 44.1741& -13.1162 & -12.1944 & -10.1201 & -6.86335 \\
36.9163 & 32.5716 & 31.5727 & 35.2859& -11.2183 & -4.24427 & -9.18091 & -13.0463 \\
29.4954 & 21.3808 & 25.6805 & 25.9215& -10.4768 & -8.11862 & -7.78015 & -4.82782 \\
27.5832 & 15.0733 & 17.6766 & 22.1851& -3.69908 & -5.53612 & -7.58757 & -9.31306 \\
15.9718 & 9.87888 & 12.9353 & 13.6829& -1.07067 & -4.14013 & -4.3206 & -5.15188 \\
13.1563 & 7.97151 & 8.64806 & 9.19449& -5.08581 & -3.18124 & -3.45434 & -3.94877 \\
9.66685 & 5.55004 & 5.7492 & 4.51719& -3.37153 & -1.80984 & -2.05624 & -3.20092 \\
6.75158 & 4.25184 & 4.88413 & 5.50294& -1.37545 & -1.1275 & -1.38412 & -1.50104 \\
5.17525 & 3.35751 & 3.53726 & 3.67014& -1.70404 & -0.730621 & -1.00551 & -1.03633 \\
3.6712 & 3.00402 & 3.03844 & 3.0374& -1.38769 & -0.407613 & -0.54473 & -0.659187 \\
3.10234 & 2.62195 & 2.34938 & 2.98697& -0.49605 & -0.698632 & -0.587935 & -0.24945 \\
2.01429 & 2.03658 & 2.42581 & 2.15836& -0.742263 & -0.227129 & -0.328322 & -0.528125 \\
2.11423 & 1.87057 & 1.90929 & 2.12311& -0.0904534 & -0.211256 & -0.203892 & -0.19418 \\
1.66489 & 1.74947 & 1.58278 & 1.58282& -0.269773 & -0.143161 & -0.0938599 & -0.0968805 \\
1.5829 & 1.58508 & 1.74115 & 1.67585& -0.104862 & -0.0955973 & -0.115998 & -0.156032 \\
1.85036 & 1.6538 & 1.64526 & 1.71367& -0.0682434 & -0.111156 & -0.135038 & -0.100162 \\
\hhline{|=|=|=|=|=|=|=|=|}
$\mathrm{Re}\,\mathcal{R}_{n=1}$  & $\mathrm{Re}\,\mathcal{R}_{n=2}$ & $\mathrm{Re}\,\mathcal{R}_{n=3}$ & $\mathrm{Re}\,\mathcal{R}_{n=4}$ &
$\mathrm{Im}\,\mathcal{R}_{n=1}$  & $\mathrm{Im}\,\mathcal{R}_{n=2}$ & $\mathrm{Im}\,\mathcal{R}_{n=3}$ & $\mathrm{Im}\,\mathcal{R}_{n=4}$ \\
\hline
-282.258 & 48.9357 & 76.3611 & 461.376& -198.478 & 187.983 & 128.153 & -831.921 \\
424.34 & 32.3681 & 45.1457 & -536.02& 170.591 & -159.579 & -142.337 & 691.046 \\
-130.855 & -28.7858 & -198.465 & 77.1275& -125.452 & -93.8902 & 12.6695 & 104.537 \\
-36.2565 & -111.159 & 18.8498 & -2.934& -74.6625 & 145.891 & 19.5365 & 2.74167 \\
-158.229 & -0.63548 & 96.1101 & 298.373& -59.1635 & -1.09124 & 61.2845 & -52.4769 \\
59.2867 & 121.049 & 95.0261 & -5.6165& 166.909 & 56.0703 & -3.36819 & -1.1995 \\
-0.78585 & -30.972 & -157.619 & -158.768& 1.53805 & -137.441 & -56.0234 & 71.1235 \\
0.0206821 & 16.9332 & 6.75526 & -108.996& 0.00813555 & 28.7989 & -47.9224 & 45.6106 \\
132.607 & -81.0017 & 25.1894 & 144.275& -122.971 & 0.979729 & 119.676 & -99.4186 \\
-56.0387 & 48.1206 & 0.945734 & -72.9646& 0.620059 & 31.7289 & -59.0824 & 16.0141 \\
0.0933913 & 11.1763 & 19.1198 & 25.2144& 1.35664 & -1.11825 & 9.70145 & 0.89294 \\
-41.5329 & -7.75891 & -18.4087 & -0.191047& 48.973 & -3.50578 & -3.18466 & 17.0221 \\
49.6975 & 1.37793 & -8.62956 & 17.1856& 8.29726 & -0.701914 & -1.94671 & 13.813 \\
-3.05939 & -7.93191 & 10.2012 & 0.00221876& -3.82848 & -12.3389 & 9.0582 & -0.0648674 \\
-8.81125 & 0.735943 & -0.47054 & -8.44741& -7.88081 & 0.336315 & -0.522577 & -8.0567 \\
-0.000310569 & 0.753694 & 0.303408 & 0.241486& -0.0211644 & 2.09597 & -0.60677 & -0.157339 \\
-1.79928 & -0.338397 & -0.12845 & 0.20849& -1.9804 & 1.09745 & -0.658241 & 0.749939 \\
-0.484891 & 0.105902 & -0.14003 & -0.466734& -0.956433 & 0.798433 & -0.325825 & 1.46258 \\
-0.0325597 & -0.253739 & 0.16653 & 0.193709& 0.0155645 & 0.563826 & -1.13469 & -0.139334 \\
\hline
\hline
\end{tabular}
\end{table}

\newpage

\begin{table}[h!]
\caption[Barycentric rational approximation for $A/a=3$]{Barycentric rational approximation for the demodulated probe-scattered field corresponding to the standard probe geometry in Fig. \ref{fig:3}b and tapping amplitude relative to probe radius $A/a=3$.  Rows correspond to incrementing $j$.}
\footnotesize
\begin{tabular}{||c|c|c|c||c|c|c|c||}
\hline
\hline
$\mathrm{Re}\,g_{n=1}$ & $\mathrm{Re}\,g_{n=2}$ & $\mathrm{Re}\,g_{n=3}$ & $\mathrm{Re}\,g_{n=4}$ &
$\mathrm{Im}\,g_{n=1}$ & $\mathrm{Im}\,g_{n=2}$ & $\mathrm{Im}\,g_{n=3}$ & $\mathrm{Im}\,g_{n=4}$  \\
\hline
   0.113931 & 0.0232191 & 0.00962331 & 0.234993 & -0.23312 & 0.11057 & 0.0809765 & 0.0684128 \\
   \hhline{|=|=|=|=|=|=|=|=|}
$\mathrm{Re}\,\rho_{n=1}$  & $\mathrm{Re}\,\rho_{n=2}$ & $\mathrm{Re}\,\rho_{n=3}$ & $\mathrm{Re}\,\rho_{n=4}$ &
$\mathrm{Im}\,\rho_{n=1}$  & $\mathrm{Im}\,\rho_{n=2}$ & $\mathrm{Im}\,\rho_{n=3}$ & $\mathrm{Im}\,\rho_{n=4}$ \\
\hline
152.524 & 166.305 & 178.256 & 177.507& -35.1367 & -37.9905 & -13.0778 & -29.3948 \\
67.6482 & 86.4453 & 106.192 & 54.0939& -25.5817 & -35.0865 & -33.9439 & -80.4892 \\
51.1135 & 64.9383 & 66.9524 & 71.7524& -12.946 & -11.8754 & -23.0408 & -30.6689 \\
36.1831 & 42.3987 & 46.858 & 52.3823& -9.89961 & -12.8257 & -14.3111 & -13.2163 \\
25.7026 & 27.7517 & 37.3031 & 36.4888& -7.16406 & -7.08087 & -0.774079 & -8.99182 \\
21.2019 & 18.8452 & 31.479 & 24.5921& -4.06965 & -5.60464 & -7.27898 & -7.59197 \\
12.217 & 16.6862 & 23.8443 & 23.4136& -9.07917 & -4.80437 & -4.86516 & -1.22909 \\
14.171 & 11.8785 & 21.8089 & 15.1486& -3.62853 & -5.3624 & -6.41319 & -3.51596 \\
9.24369 & 8.49884 & 14.9606 & 9.73346& -4.34623 & -2.68918 & -5.08768 & -5.00486 \\
5.64727 & 5.30453 & 8.4748 & 7.41804& -1.49846 & -2.01253 & -3.24174 & -3.83999 \\
4.45979 & 5.5354 & 5.97618 & 5.64396& -2.09253 & -0.324259 & -2.10896 & -1.38792 \\
4.41754 & 3.15057 & 6.25055 & 4.01596& -0.149363 & -1.43268 & -1.00098 & -1.29398 \\
3.00227 & 3.70619 & 3.74716 & 3.28157& -0.546559 & -0.276697 & -1.19613 & -0.397021 \\
2.58287 & 3.05978 & 3.05483 & 3.0406& -0.811495 & -0.501621 & -0.471994 & -0.619265 \\
2.19346 & 2.20582 & 2.48496 & 2.36951& -0.334286 & -0.373741 & -0.700551 & -0.558684 \\
2.0095 & 2.00212 & 2.04503 & 2.32529& -0.258368 & -0.229624 & -0.248733 & -0.321699 \\
1.58779 & 1.58712 & 1.80502 & 1.8187& -0.100889 & -0.0995057 & -0.139403 & -0.165611 \\
1.81754 & 1.8036 & 1.58882 & 1.58756& -0.166274 & -0.172537 & -0.0982343 & -0.0983387 \\
1.69411 & 1.69944 & 1.68253 & 1.68599& -0.119751 & -0.127551 & -0.141044 & -0.133485 \\
\hhline{|=|=|=|=|=|=|=|=|}
$\mathrm{Re}\,\mathcal{R}_{n=1}$  & $\mathrm{Re}\,\mathcal{R}_{n=2}$ & $\mathrm{Re}\,\mathcal{R}_{n=3}$ & $\mathrm{Re}\,\mathcal{R}_{n=4}$ &
$\mathrm{Im}\,\mathcal{R}_{n=1}$  & $\mathrm{Im}\,\mathcal{R}_{n=2}$ & $\mathrm{Im}\,\mathcal{R}_{n=3}$ & $\mathrm{Im}\,\mathcal{R}_{n=4}$ \\
\hline
338.457 & 361.862 & 121.267 & -202.627& -133.121 & -174.707 & -200.243 & -234.039 \\
-317.398 & -342.586 & -135.475 & -108.659& 5.05161 & 91.7244 & 103.917 & 125.597 \\
23.0165 & -4.1119 & -50.8088 & 349.414& 48.6297 & -4.13673 & 113.411 & 55.2926 \\
47.8663 & 108.766 & 169.936 & 17.207& 76.4176 & 143.496 & -22.6848 & -50.0765 \\
105.802 & 7.33449 & 0.0192446 & -6.0032& 25.4089 & -72.605 & 0.0128418 & -37.8541 \\
-5.98895 & -4.30558 & -16.9073 & -112.697& -2.88591 & -112.959 & -25.067 & 88.2641 \\
35.4137 & 55.595 & -6.90752 & -0.0274107& -229.686 & -27.3459 & -3.87145 & 0.00715897 \\
8.59205 & -96.7029 & -56.2959 & -1.3676& 2.60247 & 131.577 & -81.9078 & -9.08906 \\
-145.669 & -12.5049 & -4.01445 & 185.384& 108.856 & -10.4092 & 70.026 & -17.6207 \\
-9.98266 & 89.9777 & 42.001 & -149.665& -5.5352 & -1.70182 & 40.6073 & -31.5973 \\
85.3427 & 0.00922995 & -7.58004 & 8.93844& 53.1166 & -0.0123551 & -56.1136 & 3.87579 \\
0.00467023 & -42.719 & 0.777338 & 3.53922& 0.000218412 & -12.0305 & 0.185485 & 23.522 \\
-8.0087 & -0.036238 & -15.8015 & -0.471073& -0.367258 & -0.00104897 & 9.21448 & -0.224739 \\
-12.7874 & 4.02217 & -2.82289 & 11.3641& -16.9929 & 1.94822 & -0.858945 & 0.827609 \\
-2.43031 & 1.07428 & 9.18017 & -8.03177& -1.94751 & 1.81688 & 7.25225 & -7.41944 \\
0.0449347 & 0.262535 & 0.355801 & 1.26654& -3.17491 & 1.45046 & -0.635439 & -0.403642 \\
-0.171565 & 0.161262 & -0.240876 & -0.0310175& -0.842935 & 0.788712 & -0.486705 & 0.620668 \\
0.254597 & 0.201449 & -0.0782241 & 0.112427& -1.38175 & 1.60306 & -0.764898 & 0.754172 \\
0.312098 & -0.547348 & 0.136312 & -0.259976& -0.551356 & 0.600538 & -1.01837 & 0.652901 \\
\hline
\hline
\end{tabular}
\end{table}


\clearpage

\bibliography{apssamp}

\end{document}